\documentclass[12pt]{article}
\pdfoutput=1
\usepackage{graphicx}
\usepackage{amsmath, amssymb}
\usepackage{slashed}
\usepackage{xcolor}
\usepackage{array,arydshln}
\usepackage{cite}
\usepackage[italicdiff]{physics}
\usepackage[pdftex,
  colorlinks=true,
citecolor=green!60!blue]{hyperref}
\usepackage{hyperref}

\begin{document}

\begin{titlepage}

\def\thefootnote{\fnsymbol{footnote}}

\begin{center}

\hfill KANAZAWA-21-01\\
\hfill June, 2021

\vspace{0.5cm}
{\Large\bf Sommerfeld-enhanced dark matter searches with dwarf
  spheroidal galaxies}

\vspace{1cm}
{\large }Shin'ichiro Ando$^{\it (a,b)}$, 
{\large } Koji Ishiwata$^{\it (c)}$

\vspace{1cm}

{\it $^{(a)}${GRAPPA Institute, University of Amsterdam, 1098 XH
  Amsterdam, \\The Netherlands}}

\vspace{0.2cm}

{\it $^{(b)}${Kavli Institute for the Physics and Mathematics of the Universe (Kavli IPMU, WPI), University of Tokyo, Kashiwa, Chiba 277-8583, Japan}}

\vspace{0.2cm}

{\it $^{(c)}${Institute for Theoretical Physics, Kanazawa University,
  Kanazawa 920-1192, Japan}}

\vspace{1cm}

\abstract{We study observable signals from dark matter that
  self-annihilates via the Sommerfeld effect in dwarf spheroidal galaxies
  (dSphs). Since the effect of the Sommerfeld enhancement depends on
  the velocity of dark matter, it is crucial to determine the profile
  of dSphs to compute the J-factor, i.e., the line-of-sight integral
  of density squared. In our study we use the prior distributions of
  the parameters for satellite density profiles in order to determine
  the J-factor, making most out of the recent developments in the
  N-body simulations and semi-analytical modeling for the structure
  formation.  As concrete models, we analyze fermionic dark matter that
  annihilates via a light scalar and Wino dark matter in
  supersymmetric models.  We find that, with the more realistic prior
  distributions that we adopt in this study, the J-factor of the most
  promising dSphs is decreased by a factor of a few, compared
  with earlier estimates based on non-informative
  priors. Nevertheless, the Cherenkov Telescope Array should be able
  to detect the thermal Wino dark matter by pointing it toward best
  classical or ultrafaint dSphs for 500 hours.}

\end{center}
\end{titlepage}

\section{Introduction}
\label{sec:intro}    
\setcounter{equation}{0}

What dark matter (DM) is made of is one of the greatest mysteries in
contemporary particle physics, astrophysics, and cosmology. Although
its nature, such as mass and interaction with other particles, is
unknown, weakly interacting massive particles (WIMPs) are one of the
most popular DM candidates. In the WIMP hypothesis, the current
abundance of DM measured by the Planck
Collaboration~\cite{Akrami:2018vks} is explained by the thermal
freeze-out scenario in the early Universe.  This means that the WIMP
DM possibly annihilates into the standard-model (SM) particles, which
may be detected as high-energy messengers such as gamma rays on the
Earth.

While lots of particle physics models for WIMP candidates are proposed
to account for DM, Wino or Higgsino is one of the well-motivated
candidates among them. They are superpartners of the $W$ and Higgs
bosons in supersymmetric models. Motivated by the results of the Large
Hadron Collider, supersymmetric models where supersymmetry breaks at
very-high-energy scales have been revisited as an attractive framework
up to the grand unification scale. Split
supersymmetry~\cite{Giudice:2004tc,Giudice:2011cg,Arvanitaki:2012ps,ArkaniHamed:2012gw},
high-scale supersymmetry~\cite{Hall:2009nd}, spread
supersymmetry~\cite{Hall:2011jd}, and pure gravity
mediation~\cite{Ibe:2011aa,Ibe:2012hu} are the examples.  In the
high-scale supersymmetry, {\it pure} Wino or {\it pure} Higgsino is
the DM candidate. Due to their pure nature, they interact with the SM
particles via the gauge interactions. As a consequence, their
scattering or annihilation cross sections are in principle determined
with small theoretical uncertainties for a given mass. The scattering
cross section of the Wino or Higgsino with nucleons can be determined
at the next-to-leading order level of quantum chromodynamics in a
consistent manner~\cite{Hisano:2015rsa}.\footnote{See
Refs.\,\cite{Hisano:2004pv,Hisano:2010fy,Hisano:2010ct,Hisano:2011cs,Hisano:2012wm,Hill:2011be,Hill:2013hoa,Hill:2014yka,Hill:2014yxa}
for earlier work.} On the other hand, the annihilation process is
associated with the so-called Sommerfeld enhancement that leads to
non-trivial behavior of the annihilation cross section as a function
of mass~\cite{Hisano:2003ec,Hisano:2004ds,Hisano:2006nn}. As a
consequence, 2.7--3.0\,TeV mass is predicted to yield the right relic
abundance of the Wino as DM~\cite{Hisano:2006nn} (See
Ref.\,\cite{Beneke:2016ync} for recent developments.)  The Sommerfeld
effect is an issue not just for the Wino or Higgsino, but it can
happen for DM candidates that, for example, interact with a light
mediator via a Yukawa interaction (see Ref.\,\cite{ArkaniHamed:2008qn}
for a review). For such DM candidates, accurate predictions of
Sommerfeld-enhanced signals is important in order to detect DM
indirectly using high-energy gamma rays and other cosmic messengers.

Detecting gamma rays from dwarf spheroidal galaxies (dSphs) is one of the most promising avenues in the indirect DM searches. While dSphs have
abundant DM component, they contain less baryonic matter compared
with other galaxies. Consequently, they are an ideal object in order to detect
gamma-ray signals from DM with small astrophysical
backgrounds. Recently, lots of new ultrafaint dSphs have been
discovered~\cite{ultrafaintdSph}. Since they are faint with
much less amount of baryons, they might be more suitable for the DM
searches. This fact, however, makes it difficult to determine density
profiles of the ultrafaint dSphs due to the lack of stellar kinematics
data. 

A Bayesian approach is often adopted for estimations of density
profile parameters.  Representative examples are a scale radius $r_s$
and a characteristic mass density $\rho_s$ of the often adopted
Navarro-Frenk-White (NFW) profile~\cite{Springel:2008cc}.  The
Bayesian approach has to adopt the {\it prior} distribution of these
parameters, which are combined with the likelihood to obtain the {\it
  posterior} probability distribution function.  While uninformative,
uniform priors for both $\log r_s$ and $\log \rho_s$ are usually
adopted in the literature (e.g.,~\cite{Geringer-Sameth:2014yza}), more
realistic priors based on theories of the structure formation, called
satellite priors, are proposed~\cite{Ando:2020yyk}. The authors showed
substantial impact on DM searches using dSphs. As a result of adopting
the satellite priors, the constraints on the annihilation cross
section for 10-1000~GeV DM are found to be relaxed by a factor of 2--7
compared with those obtained with uninformative, log-uniform
priors~\cite{Hoof:2018hyn}. It is therefore crucial to adopt more
realistic priors that take into account the observed satellite
dynamics in order to constrain the DM cross section.

In this work, we extend the approach of Ref.~\cite{Ando:2020yyk} and
adopt the satellite priors to predict the rates of WIMP DM
annihilation that is enhanced by the Sommerfeld effect. To see the
impact of the satellite priors on the observable signatures, we study
two Majorana DM models: (1) a model with a light scalar mediator and
(2) the Wino DM in supersymmetric scenarios with masses around 3~TeV.
In the former model,
Refs.\,\cite{Boddy:2017vpe,Lu:2017jrh,Bergstrom:2017ptx,Petac:2018gue}
have already done intensive analysis, but using uninformative flat
priors. Since the annihilation cross section, $\sigma v $, in this
case has velocity dependence, one has to take the internal velocity
structure of the dSphs into account. The models developed in
Ref.~\cite{Ando:2020yyk} specify the internal density structure of
each given dSph, from which we can compute the velocity distribution
function (or the phase-space density of DM in the dSph). We find, for
the first time, that in comparison with the case of uninformative
priors, adopting the informative satellite priors \textit{decreases}
the expected annihilation rate by a factor of 2.7 and 1.4 using models
with $V_{50}=10.5$~km~s$^{-1}$ and 18~km~s$^{-1}$, respectively, where
$V_{50}$ is a parameter corresponding to the threshold of the maximum
circular velocity of the subhalo above which satellite galaxies are
assumed to form (see Sec.\,\ref{sec:method} for more details), for one
of the most promising ultrafaint dSphs, Reticulum~II. In addition, we
find that those factors are almost independent of a parameter
$\epsilon_\phi$ defined later in Eq.\,\eqref{eq:epsilons}, {\it i.e.},
ratio of DM mass and a mediator mass times the coupling that mediates
the DM annihilation, except for a region where the annihilation is
enhanced significantly.

For the latter Wino model, the Galactic center would be the most
promising object to detect gamma rays from the DM
annihilation~\cite{Rinchiuso:2018ajn,Rinchiuso:2020skh}.\footnote{See
also \cite{Hryczuk:2019nql} for constraints on the lightest
superparticle in the pMSSM analysis using signals from the Galactic
center. In addition, see Ref.\,\cite{Lefranc:2016fgn} that compares
the sensitivity of observing the Galactic center and the selected
dSphs, Draco and Triangulum II. }  On the other hand, dSphs are also
considered to be one of the most important targets for the DM search
with the Cherenkov Telescope Array (CTA)
observatory~\cite{CTA,CTAConsortium:2018tzg}. Yet, as far as we are
aware, no dedicated study on the CTA sensitivity estimates for the
Wino DM searches has been performed. We, therefore, show sensitivity
projection for the CTA North for several classical and ultrafaint
dSphs.  Even though the velocity dependence can be safely ignored for
the Wino masses much heavier than the gauge boson mass, the resultant
J factor changes, for example, by a factor of 1.8 between adopting
priors with $V_{50}=18$~km~s$^{-1}$ and $10.5$~km~s$^{-1}$ for
Reticulum II. Eventually it is found that both classical and
ultrafaint dSphs are promising targets to detect the Wino DM with
2.7--3~TeV masses, providing a complementary method to the observation
of the Galactic center.

Lastly, we note that we made the codes to generate the satellite prior
distributions of $r_s$, $\rho_s$, etc., for any of both the classical
and ultrafaint dSphs that are discussed in this paper publicly
available at \url{https://github.com/shinichiroando/dwarf_params}.

The rest of the paper is organized as follows. In
Sec.\,\ref{sec:method}, we summarize the satellite priors and how to
determine probability distribution of the profile parameters.
Section\,\ref{sec:flux} introduces formulae to compute J-factor and
shows the numerical results of the light mediator model and the Wino
DM.  Current constraints and future sensitivity by the CTA
collaboration are discussed in Sec.\,\ref{sec:results}, and we
conclude the paper in Sec.\,\ref{sec:conclusion}.

\section{Analysis with satellite priors}
\label{sec:method}    
\setcounter{equation}{0}

In our study, we perform the analysis by using satellite priors, which
we briefly summarize in this section. See
Refs.\,\cite{Ando:2020yyk,Hiroshima:2018kfv} for more details.

We use prior probability distribution function (PDF) for the subhalo
parameters and the observed parameters of dSphs to give the posterior
PDF for the satellite parameters.  Here is a list of quantities
involved in the PDF:
\begin{itemize}
\item subhalo parameters: $\vb*{\theta}=(\rho_s,r_s,r_t)$;
\item observed dSphs data: $\vb*{d}=(\theta_h, \sigma_{\rm los}, D)$.
\end{itemize}
Here, $r_t$ is the truncation radius, and $\theta_h$, $\sigma_{\rm
  los}$ and $D$ are the projected angular half-light radius,
line-of-sight velocity dispersion and the distance,
respectively. In our study, we adopt the NFW profile for the
DM density,
\begin{equation}
  \rho (r) = \frac{\rho_s}{(r/r_s)(1+r/r_s)^2}\,.
\end{equation}
The prior PDF for the subhalo parameters $\vb*{\theta}$ is proportional
to the number of subhalos per volume in the parameter space
$d\vb*{\theta}$, i.e.,
\begin{equation}
    P_{\rm sh}(\vb*{\theta})\propto \frac{d^3N_{\rm sh}}{d \rho_s dr_s dr_t}\,.
\end{equation}

We note that not all of subhalos host
satellite galaxies. To take the satellite formation into account, we adopt a
prescription provided in Ref.\,\cite{Graus_2019}. We parameterize the
probability that a satellite galaxy is formed in a host subhalo as
\begin{align}
  P_{\rm form}(V_{{\rm peak}})=
  \frac{1}{2}\left[1+{\rm erf}
    \left(\frac{V_{{\rm peak}}-V_{50}}{\sqrt{2}\sigma}\right)\right]\,,
\end{align}
where $V_{{\rm peak}}$ is the peak value of the maximum circular
velocity of the subhalo. In our model, it is given as the maximum
circular velocity at the subhalo accretion onto its host, $V_{\rm
  peak} = V_{{\rm max},a}=r_{s,a}(4\pi G\rho_{s,a}/4.625)^{1/2}$,
where $G$ is the Newtonian constant of gravity and the subscript $a$
represents quantities at the time of accretion. $V_{50}$ is another
input parameter.  If $V_{\rm peak}$ exceeds $V_{50}$, then a satellite
galaxy is formed in a host subhalo. $V_{50}=18$~km\,s$^{-1}$ is a
value according to the conventional theory of galaxy formation.  On
the other hand, Ref.\,\cite{Graus_2019} suggests
$V_{50}=10.5$~km\,s$^{-1}$ since $V_{50}=18$~km\,s$^{-1}$
underpredicts the number of dSphs and their radial distributions
compared with observations.  In our analysis, we adopt
$V_{50}=10.5$~km\,s$^{-1}$ and $18$~km\,s$^{-1}$ for ultrafaint dSphs
and $25$~km\,s$^{-1}$ for classical dSphs, and
$\sigma=2.5$~km\,s$^{-1}$ based on Ref.\,\cite{Graus_2019}. Therefore,
we construct the satellite prior PDF, $P_{\rm sat}(\vb*{\theta})$, as
\begin{align}
  P_{\rm sat}(\vb*{\theta})=
  P_{\rm sh}(\vb*{\theta}) P_{\rm form}(V_{{\rm peak}})\,.
\end{align}

According to the Bayes' theorem, the posterior PDF $P_{\rm
  sat}(\vb*{\theta}|\vb*{d})$ for the subhalo parameters is given by
\begin{align}
  P_{\rm sat}(\vb*{\theta}|\vb*{d}) \propto P_{\rm sat}(\vb*{\theta})
  {\cal L}(\vb*{d}|\vb*{\theta}),
\end{align}
where ${\cal L}(\vb*{d}|\vb*{\theta})$ is
likelihood
function of obtaining data $\vb*{d}$ for a dSph given the model parameters
$\vb*{\theta}$. The likelihood is given by
\begin{align}
  &{\cal L}(\vb*{d}|\vb*{\theta})=
  \prod_{x=\{\theta_h, \sigma_{\rm los}, D\}}
  \frac{1}{\sqrt{2\pi \sigma_x^2}}
  \exp\left[-\frac{(x-x_{\rm obs})^2}{2\sigma_x^2}\right]\,,
\end{align}
where $x_{\rm obs}$ is the
central value of the observation and $\sigma_x$ is the measurement
uncertainty of $x$.
For these values, we use the results summarized in
Tables~I and II of Ref.\,\cite{Ando:2020yyk}.

$P_{\rm sh}(\vb*{\theta})$ is obtained as follows. The input parameters
are $(z_a,m_a,c_{{\rm vir},a})$, where $m_a$ is the mass of a halo
that accreted onto its host halo and $z_a$ is the redshift at the
accretion. $c_{{\rm vir},a}$ is the virial concentration parameter at
the accretion, for which we adopt log-normal distribution with
standard deviation $\sigma_{\log c}=0.13$~\cite{Ishiyama:2011af} while
its mean value is obtained by Ref.\,\cite{Correa:2015dva}. From a set
of the input parameters, characteristic radius $r_{s,a}$ and density
$\rho_{s,a}$ at the accretion are obtained. In general $(r_{s}(z)$,
$\rho_{s}(z))$ at redshift $z$ are related to the maximum velocity
$V_{\rm max}(z)$ and radius $r_{\rm max}(z)$. For a profile that is
proportional to $r^{-1}$ in the inner region, it is studied that
$V_{\rm max}(z)$ and $r_{\rm max}(z)$ are determined by the subhalo
mass $m(z)$ for given initial values of $V_{\rm max}(z_a)$, $r_{\rm
  max}(z_a)$, and $m_a$~\cite{Penarrubia:2010jk}. Therefore, once we
know the evolution of $m(z)$ after accretion, we obtain $r_s(z)$ and
$\rho_s(z)$ for a given redshift. After accretion, the halo loses its
mass by tidal stripping that can be described by the following
differential equation,
\begin{align}
  \frac{dm(z)}{dt}=-A\frac{m(z)}{\tau_{\rm dyn}(z)}
  \left[\frac{m(z)}{M(z)}\right]^\zeta\,,
  \label{eq:dmdt}
\end{align}
where $\tau_{\rm dyn}(z)$ is the dynamical
timescale~\cite{Jiang:2014nsa}, and $M(z)$~\cite{Correa:2014xma} is
the host halo mass at the redshift $z$. For parameters $A$ and
$\zeta$, we use the results given in Ref.\,\cite{Hiroshima:2018kfv}
that agree with the results of the N-body simulation. Then, by solving
Eq.\,\eqref{eq:dmdt}, we obtain the subhalo parameters
$(\rho_s,r_s,r_t)$ at $z=0$.

The distribution of mass $m_a$ and redshift $z_a$ of a subhalo at the
accretion, $dN_{\rm sh}/(dz_a dm_a)$, is obtained by the extended
Press-Schechter formalism~\cite{Bond:1990iw} that is calibrated with
the numerical simulations~\cite{Yang:2011rf}.  With the initial
distribution, we simulate the tidal effect for each subhalo to obtain
the present distribution $d^3N_{\rm sh}/(d \rho_s dr_s dr_t)$.

Prior and posterior PDFs of $r_s$ and $\rho_s$ are extensively
discussed and shown in Ref.\,\cite{Ando:2020yyk} for each dSph and for
a chosen value of $V_{50}$ parameter (see Figs.\,3 and 5--10 in
Supplemental Material of Ref.\,\cite{Ando:2020yyk}). For each
parameter set $(r_s,\rho_s,r_t)$ drawn from the posterior PDFs that
can be obtained using publicly available codes at
\url{https://github.com/shinichiroando/dwarf_params}, one can
calculate the J-factors including the Sommerfeld enhancement, which is
the focus of the following section.

\section{Sommerfeld-enhanced gamma-ray flux from dSphs }
\label{sec:flux}    
\setcounter{equation}{0}

We formulate gamma-ray flux from dSphs in two types of Majorana
fermionic DM models; the annihilation is boosted via a light scalar
mediator (Sec.\,\ref{sec:scalarmed}) and the electroweak gauge bosons
especially focusing on the Wino DM (Sec.\,\ref{sec:Wino}). In
Sec.\,\ref{sec:scalarmed} we review how to compute J-factor when the
annihilation cross section is velocity dependent and give numerical
results of some selected dSphs suitable for the DM search that are
computed by the method explained in Sec.\,\ref{sec:method}

\subsection{Light scalar mediator model}
\label{sec:scalarmed}

The Sommerfeld enhancement factor via a light scalar mediator is given
by the wavefunction, which will be defined soon in
Eq.\,\eqref{eq:def_S} (see Ref.\,\cite{ArkaniHamed:2008qn} for a
review).  The wavefunction obeys the following Schr\"{o}dinger
equation where the potential is given by the Yukawa type,
\begin{align}
  \frac{1}{m_{\rm dm}}\frac{d^2\psi(r)}{dr^2}-V(r)\psi(r)
  = -m_{\rm dm}v^2\psi(r)\,,
\end{align}
where $m_{\rm dm}$ is the mass of DM and $v$ is the velocity of each
DM particle, which can be written using the relative velocity $v_{\rm
  rel}$ as $v=v_{\rm rel}/2$, and the potential $V(r)$ is
\begin{align}
  V(r)=-\alpha_y \frac{e^{-m_\phi r}}{r}\,.
\end{align}
Here $\alpha_y$ is the coupling constant of the interaction and
$m_\phi$ is the mass of the scalar mediator. It is legitimate to
introduce dimensionless parameters,
\begin{align}
  \epsilon_v=\frac{v}{\alpha_y}\,,~~~
  \epsilon_\phi=\frac{m_\phi}{\alpha_y m_{\rm dm}}\,,~~~
  x=\alpha_y m_{\rm dm} r\,,
  \label{eq:epsilons}
\end{align}
to get
\begin{align}
  \psi''(x)+\frac{e^{-\epsilon_\phi x}}{x}\psi(x)=-\epsilon_v^2\psi(x)\,,
\end{align}
where $\prime$ means derivative with respect to $x$. Here we have used
the same symbol for wavefunction for simplicity.

The Sommerfeld-enhancement factor is then given by solving the
Schr\"{o}dinger equation under the boundary condition $\psi(0)=1$ and
$\psi'(\infty)=i\epsilon_v\psi(\infty),$\footnote{Namely, it
corresponds to $\psi(x)\to e^{i\epsilon_v x}$ at $x=\infty$.}
\begin{align}
  S^\phi(v_{\rm rel};\alpha_y, \epsilon_\phi) = |\psi(\infty)|^2\,.
  \label{eq:def_S}
\end{align}
Meanwhile we can solve the equation numerically to obtain
$S^\phi(v_{\rm rel})$, there is an approximated formula,
\begin{align}
  S^\phi_{\rm apprx}(v_{\rm rel};\alpha_y,\epsilon_\phi)
  =\frac{(\pi/\epsilon_v)\sinh s}
  {\cosh s-\cos t}\,,
\end{align}
where
\begin{align}
  s=\frac{12 \epsilon_v}{\pi \epsilon_\phi}\,,~~~~
  t=2\pi
  \sqrt{\frac{6}{\pi^2 \epsilon_\phi}-
    \left(\frac{6\epsilon_v}{\pi^2 \epsilon_\phi}\right)^2}\,.
\end{align}
For example, see Ref.\,\cite{Lu:2017jrh} for a comparison between the
numerical results with this approximation. As presented there, the
approximated formula agrees with the numerical results at the 10\%
level, which is of sufficient accuracy for the purpose of our present
analysis. In the following numerical analysis of the light mediator
model, we will use $S^\phi_{\rm apprx}(v_{\rm
  rel};\alpha_y,\epsilon_\phi)$, and parameterize the total
annihilation cross section as
\begin{align}
  \sigma v= (\sigma v)_0 
 S^\phi_{\rm apprx}(v_{\rm rel};\alpha_y,\epsilon_\phi)\,.
\end{align}
Then the gamma-ray flux is given by
\begin{align}
  \frac{d\Phi_\gamma}{dE}
  =J(\theta)\frac{(\sigma v)_0}{8\pi m_{\rm dm}^2}
  \frac{dN_\gamma}{dE}\,,
  \label{eq:flux}
\end{align}
where $J(\theta)$ is the J-factor which will be given below and
$dN_\gamma/dE$ is the gamma-ray spectrum per annihilation.

When the annihilation cross section is velocity-dependent, we need to
take into account that effect to give the J-factor. Here we briefly
review the formulation to compute the J-factor for velocity-dependent
annihilation cross section. The velocity distribution $f(v,r)$ of the
DM particle at the radius $r$ is given by Eddington’s
formula~\cite{BinneyAndTremaine},
\begin{align}
  f(v,r) =\frac{1}{\sqrt{8}\pi^2 \rho(r)}
  \int^{\infty}_{R({\cal E}(r,v))}
  dr'\frac{P(r')}{\sqrt{{\cal E}(r,v)-\Psi(r')}}\,,
\end{align}
where
\begin{align}
  {\cal E}(r,v)&=-\frac{v^2}{2}+\Psi(r)\,, \\
   P(r)&=
   \frac{d\rho(r)}{dr}\frac{d^2\Psi(r)}{dr^2}
   \left(\frac{d\Psi(r)}{dr}\right)^{-2}
  -\frac{d^2\rho(r)}{dr^2}\left(\frac{d\Psi(r)}{dr}\right)^{-1}\,.
\end{align}
$\Psi(r)$ is the gravitational potential and $R(X)$ is the inverse
function of $\Psi(r)$. For the NFW profile $\Psi(r)$ is given
analytically as
\begin{align}
  \Psi(r)&= 4\pi G \rho_s r_s^2 \frac{\ln(1+r/r_s)}{r/r_s}\,,
\end{align}
where $\rho_s$ and $r_s$ are
the parameter of the NFW profile. By definition, $f(v,r)$ satisfies,
\begin{align}
  \int^\infty_0 dv \, 4\pi v^2 f(v,r) = 1\,.
\end{align}
Using $f(v,r)$, the velocity distribution of the two-body DM state
is derived in terms of the relative velocity $v_{\rm rel}$ as
\begin{align}
  F(v_{\rm rel},r) =
  \int_0^1dz \, 4\pi \int_0^\infty dv_{\rm cm}v^2_{\rm cm} f(v_+,r)f(v_-,r)\,,
\end{align}
where
\begin{align}
  v_{\pm} =\sqrt{v_{\rm cm}^2+v_{\rm rel}^2/4\pm v_{\rm cm}v_{\rm rel}z}\,.
\end{align}

Finally the J-factor is obtained by the line of sight (l.o.s) integral,
\begin{align}
  J(\theta)&=\int_{{\rm l.o.s.}(\theta)}d\ell\int_0^\infty dv_{\rm rel}\,
  4\pi v_{\rm rel}^2 F(v_{\rm rel},r) \rho^2(r) S(v_{\rm rel};\vb*{X})\,,  
\end{align}
where $S(v_{\rm rel};\vb*{X})$ is the factor that describes the
velocity dependence of the annihilation cross section and $\vb*{X}$
stands for the other parameters in a certain DM model. In the present
case, $S(v_{\rm rel};\vb*{X})=S^\phi_{\rm apprx}(v_{\rm
  rel};\alpha_y,\epsilon_\phi)$.  On the other hand, it is a good
approximation to give the line of sight integral in cylindrical
coordinate with radius $R$ and height $z$, which relate to $r$ as
$r^2=R^2+z^2$,
\begin{align}
  \int_{{\rm l.o.s.}(\theta)}d\ell = \frac{1}{D^2}
  \int_0^{D\theta}dR
  \int_0^{r_t}dz\,
  4\pi R\,,
\end{align}
Here $D$ and $\theta$ are the distance to the dSph and the
maximum polar angle from the center of the dSph that will be given in
the later discussion.

\begin{figure}
  \begin{center}
    \includegraphics[width=4.9cm]{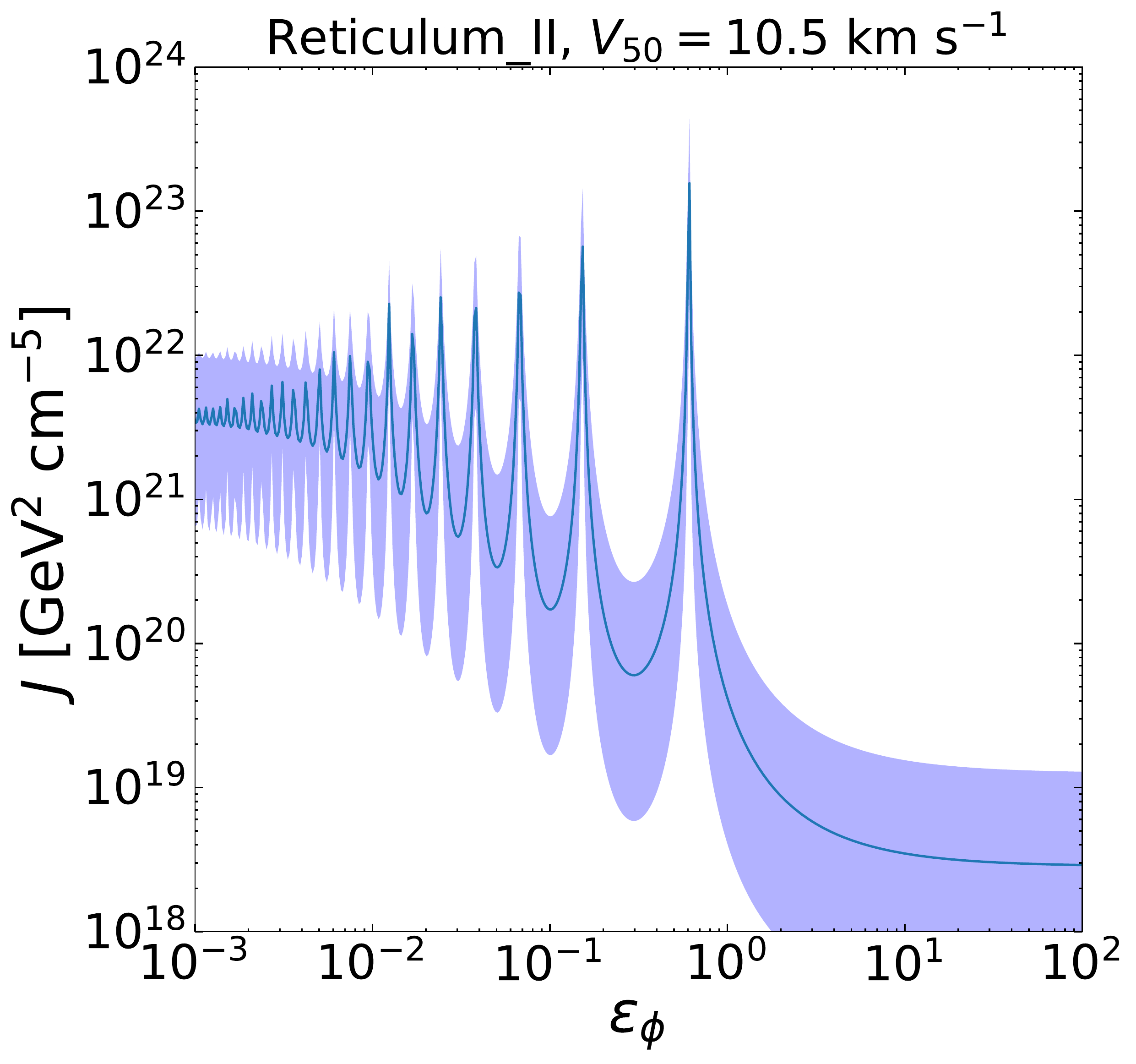}
    \includegraphics[width=4.9cm]{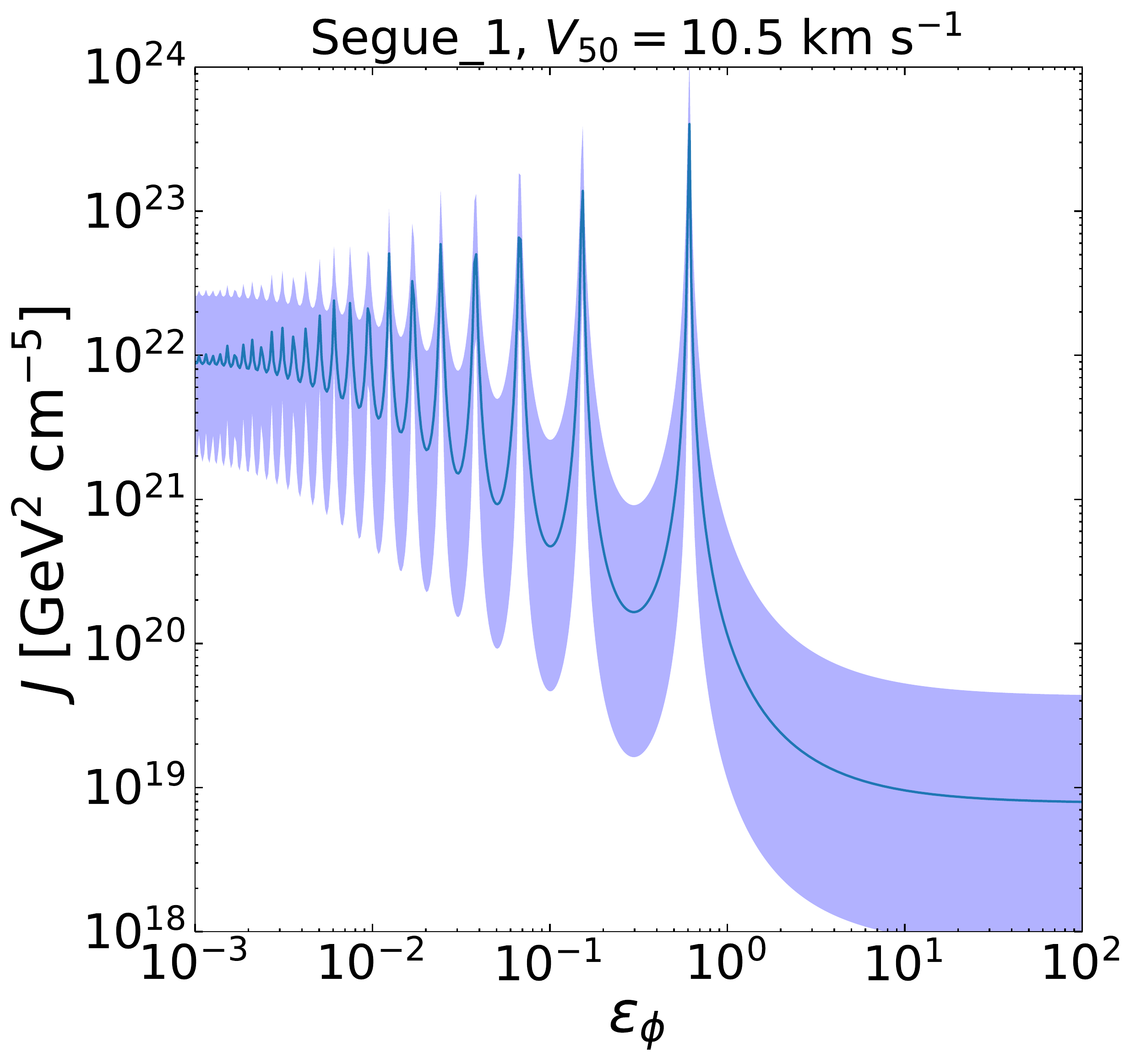}
    \includegraphics[width=4.9cm]{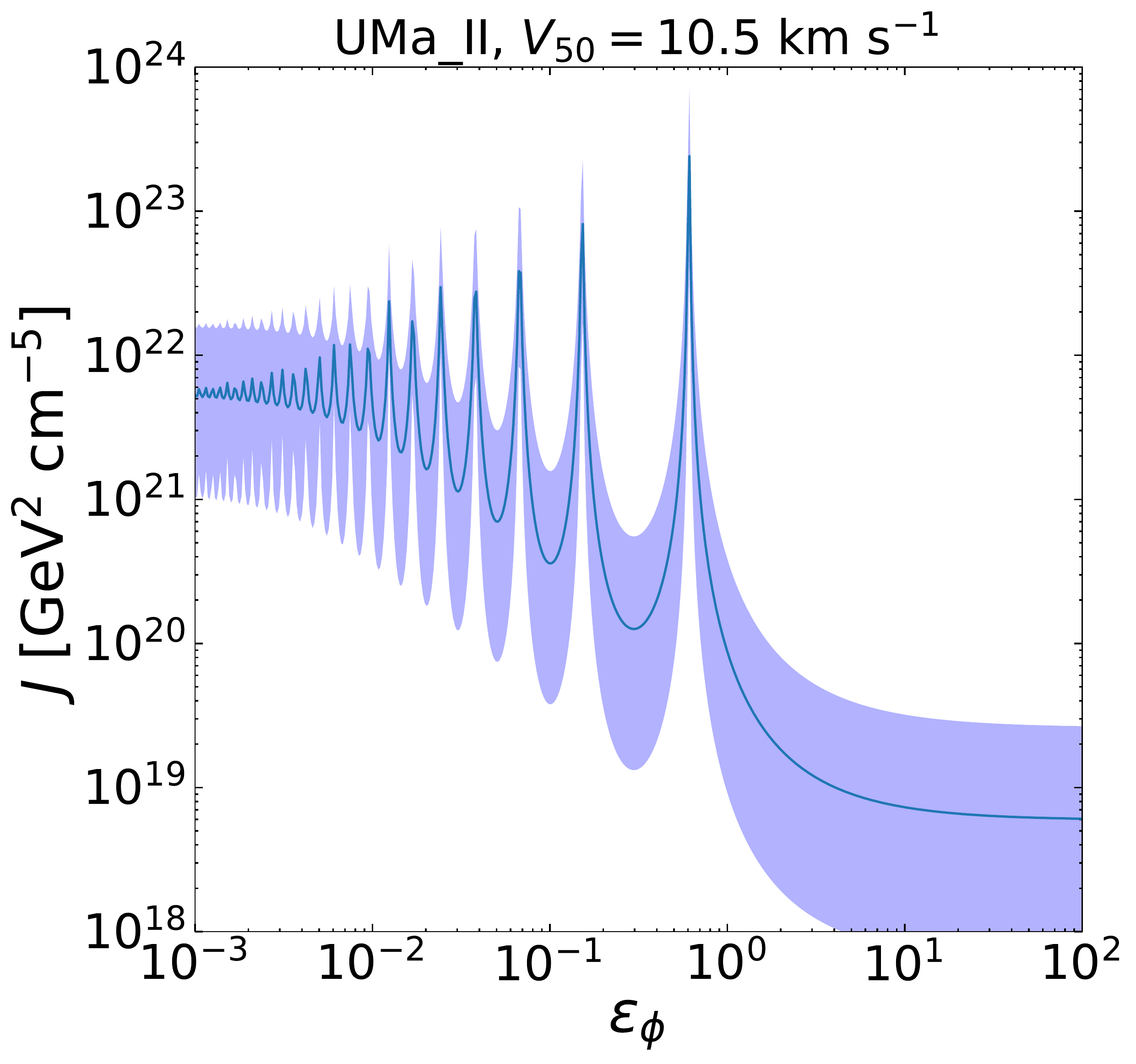}
    
    \includegraphics[width=4.9cm]{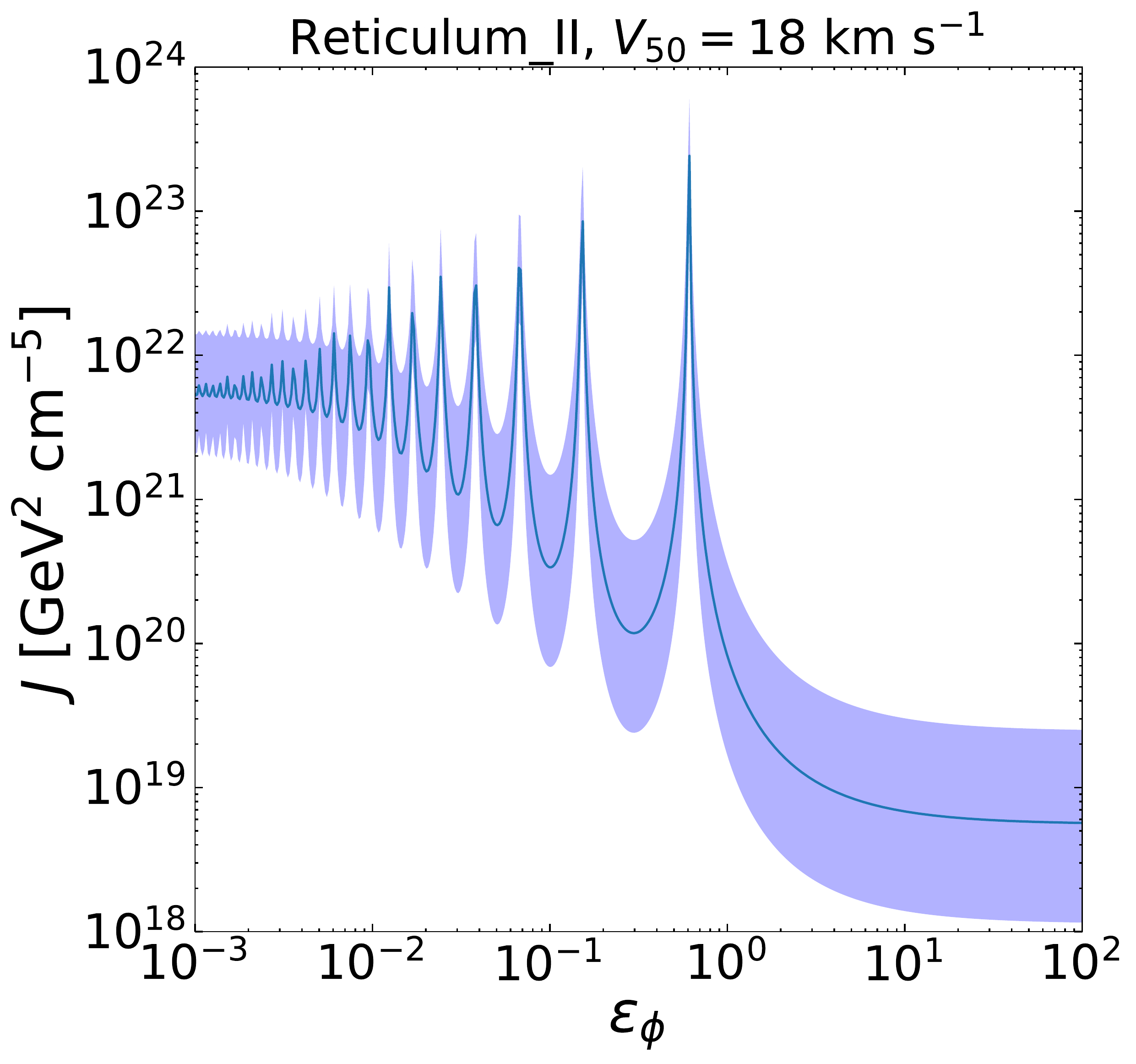}
    \includegraphics[width=4.9cm]{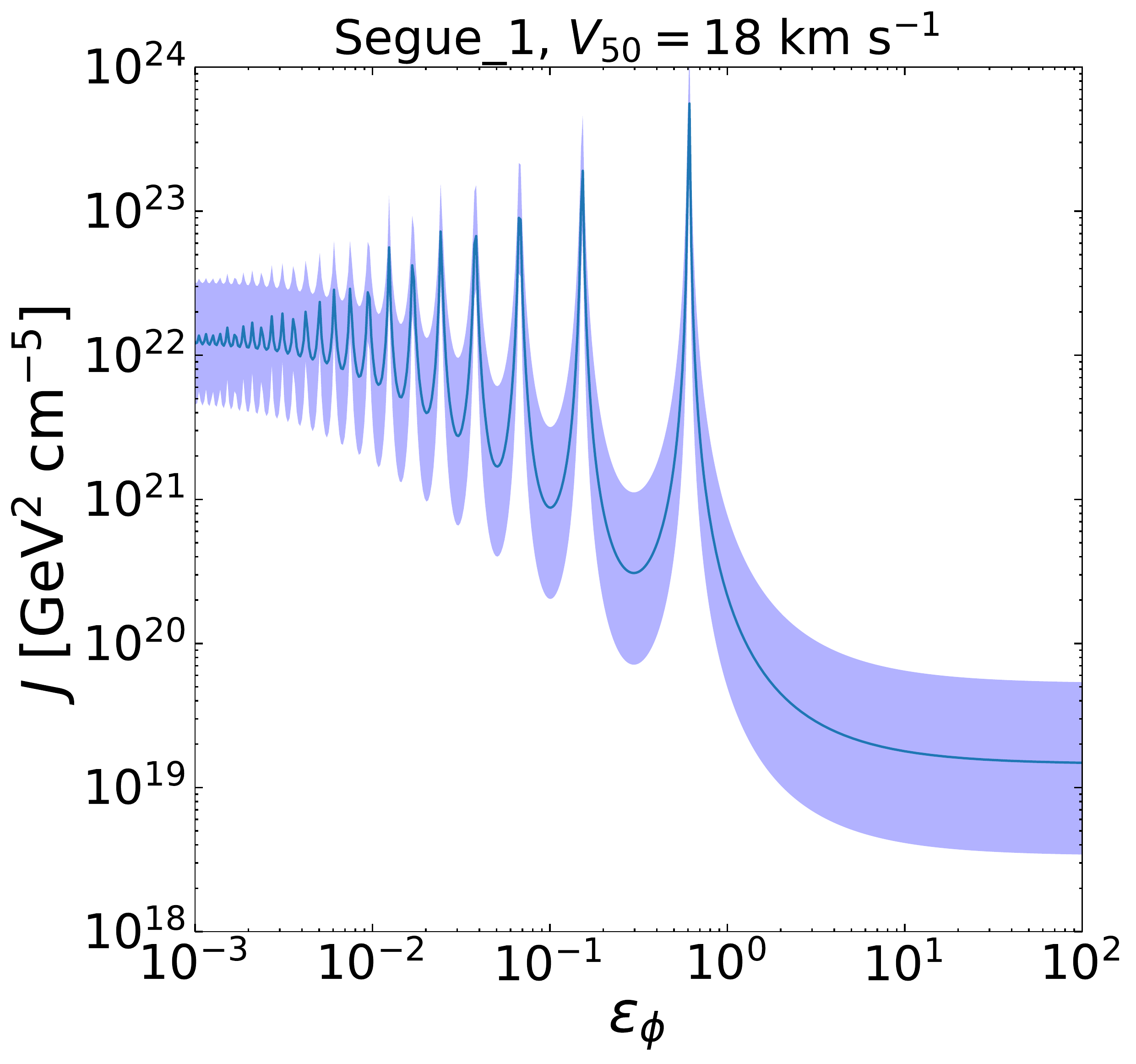}
    \includegraphics[width=4.9cm]{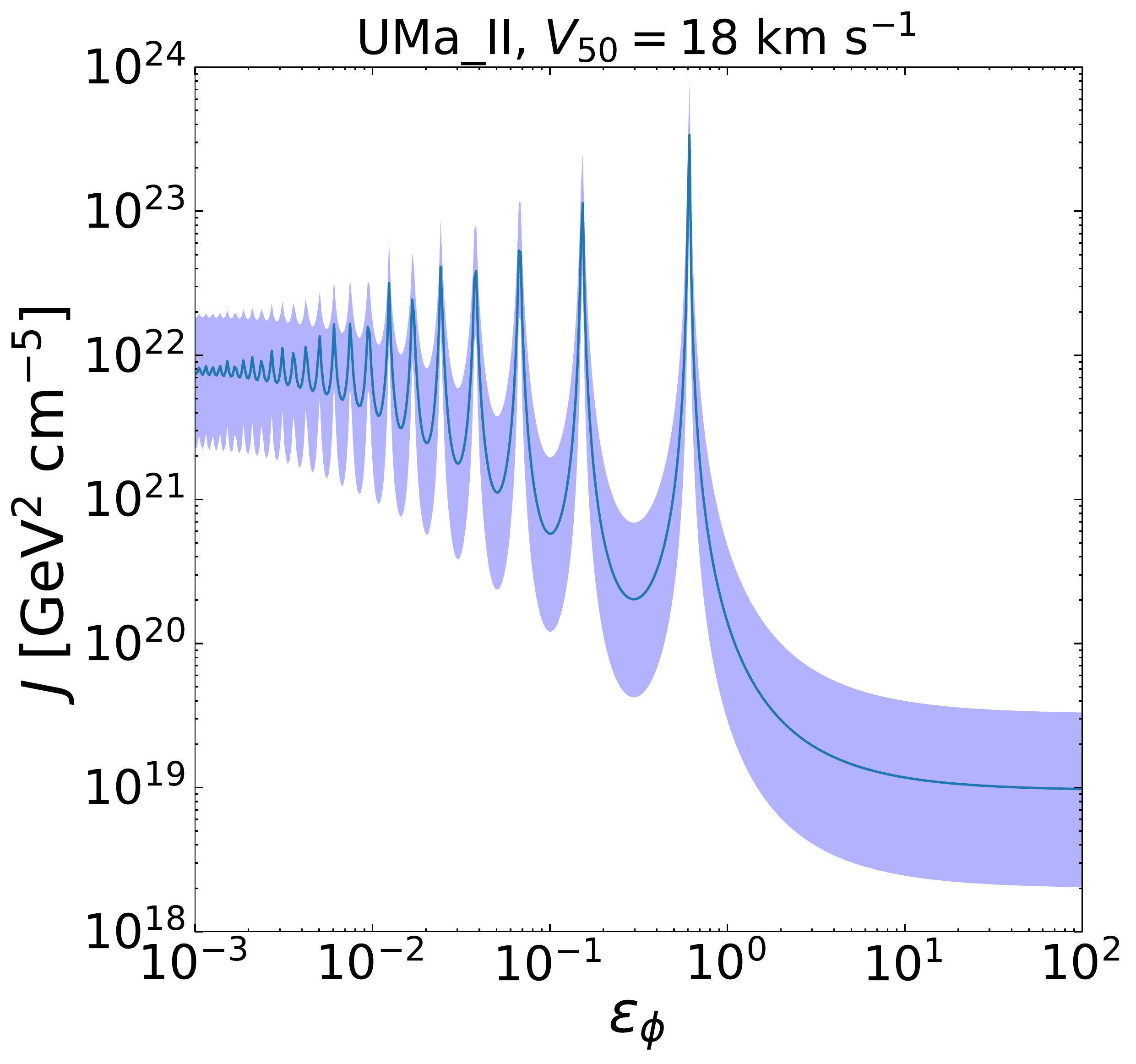}
    
    \includegraphics[width=4.9cm]{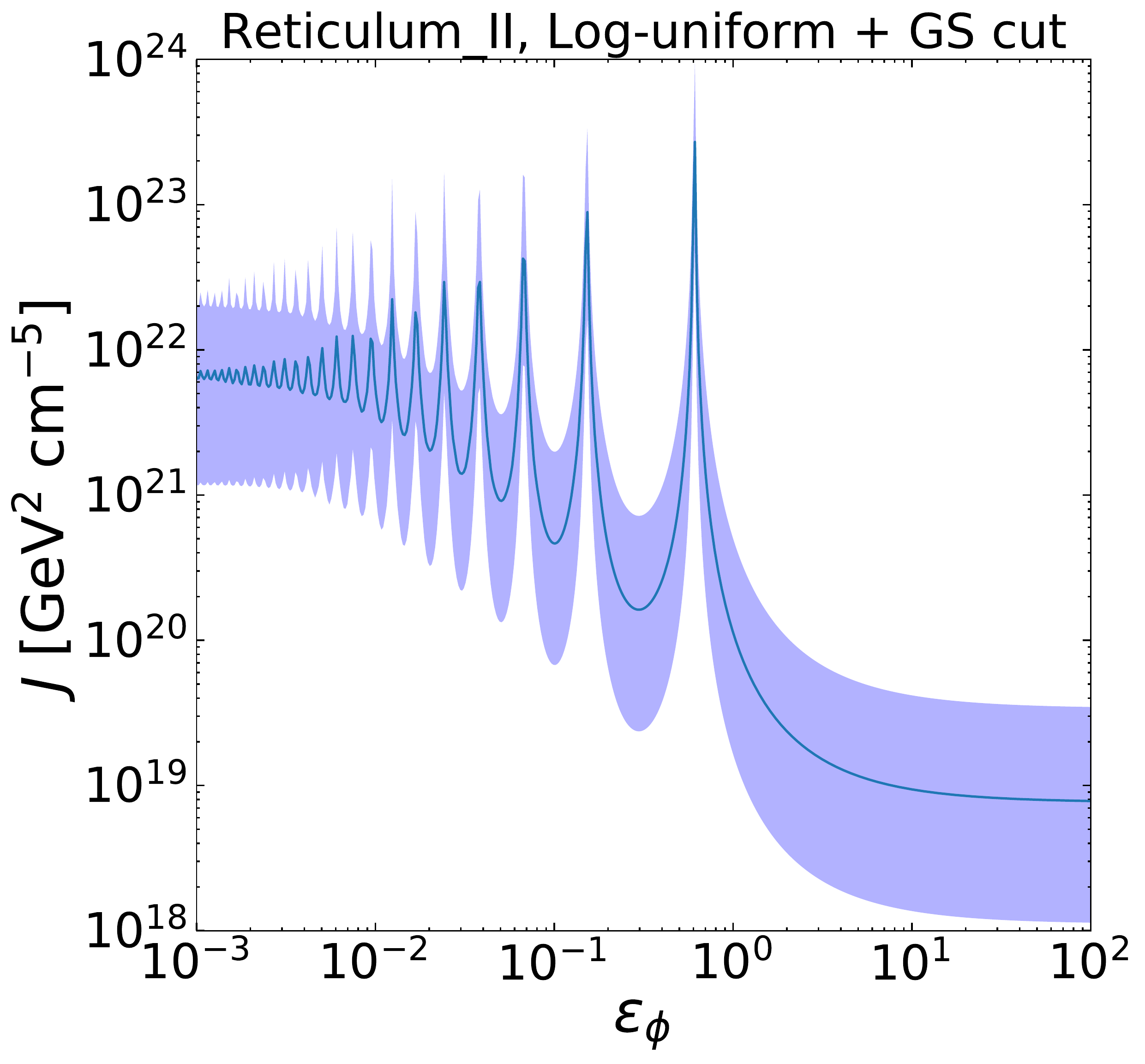}
    \includegraphics[width=4.9cm]{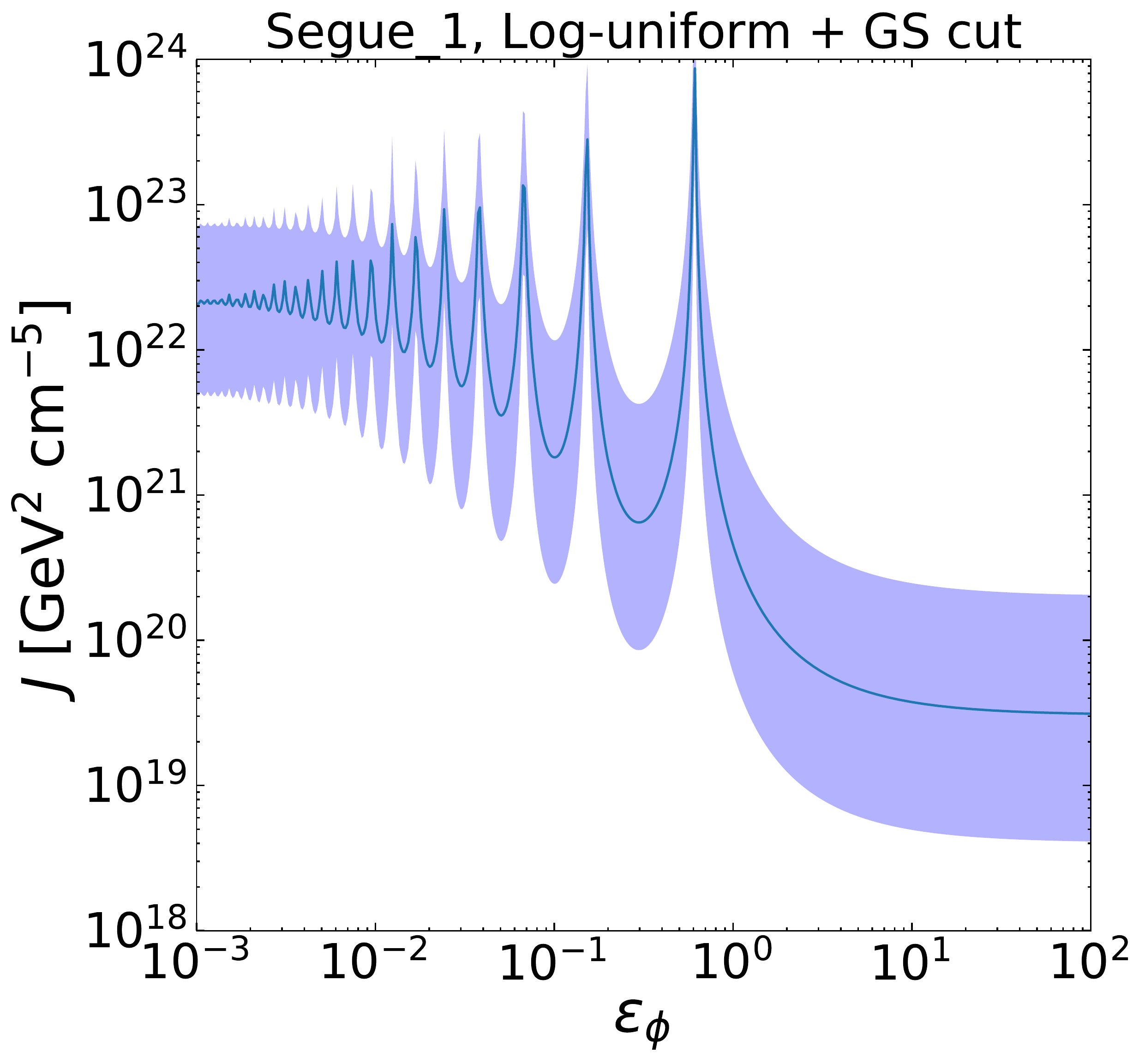}    
    \includegraphics[width=4.9cm]{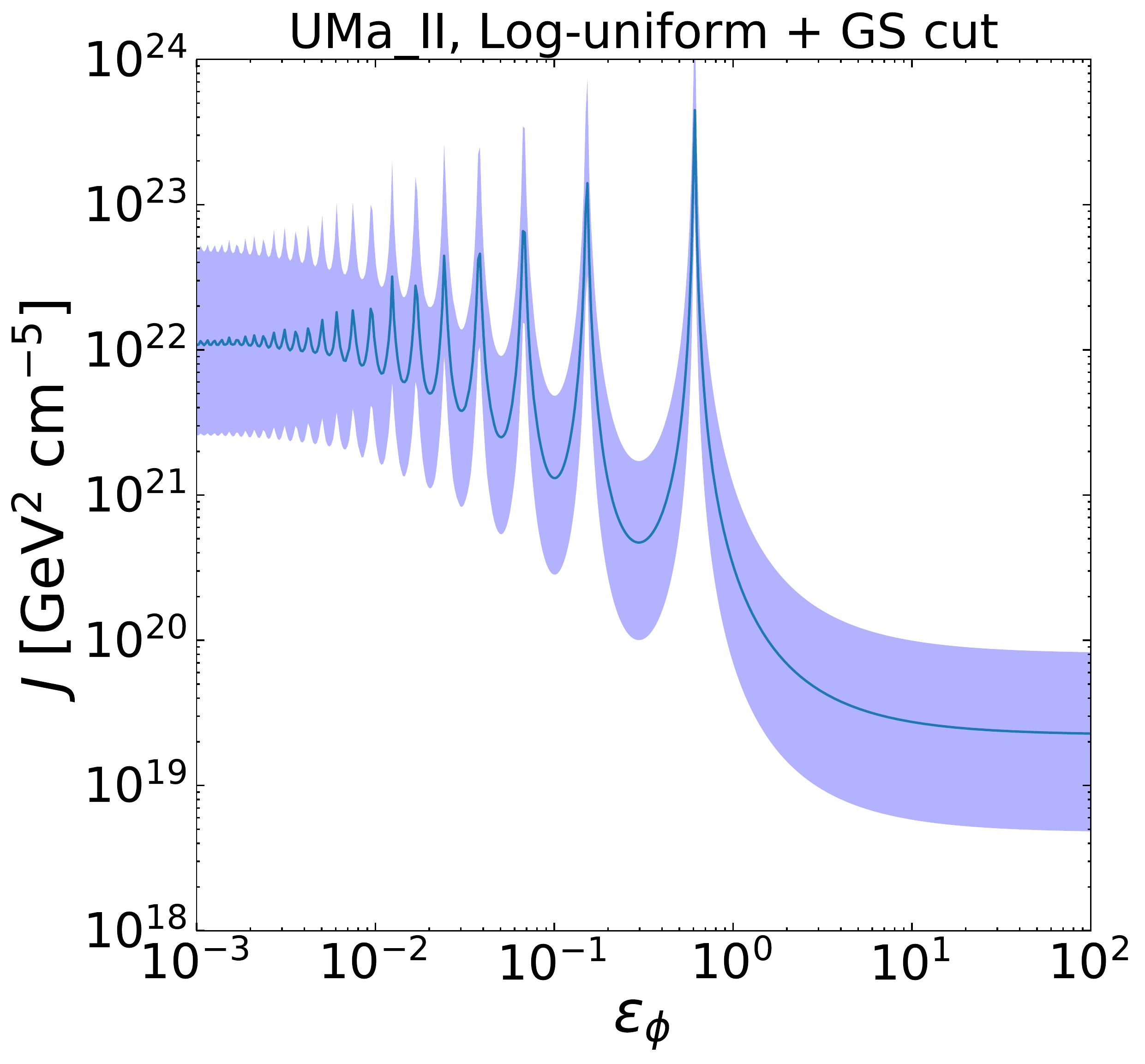}
    
     \caption{J-factor calculated as function of
       $\epsilon_\phi=m_\phi/(\alpha_y m_{\rm dm})$ with
       $\theta=0.5^\circ$ for ultrafaint dSphs, Reticulum II, Segue 1,
       and Ursa Major II (from left to right). $\alpha_y=10^{-2}$ and
       $V_{50}$ is taken to 10.5 km~s$^{-1}$ (top) and 18 km~s$^{-1}$
       (middle). Results by using the log-uniform prior with GS15
       cut~\cite{Geringer-Sameth:2014yza} is plotted as a reference
       (bottom). Line shows the median values and shaded region
       corresponds to 95\% credible intervals (see text for detail).}
  \label{fig:J_ultrafaints_selected}
 \end{center}
\end{figure}

\begin{figure}
  \begin{center}
    
    \includegraphics[width=4.9cm]{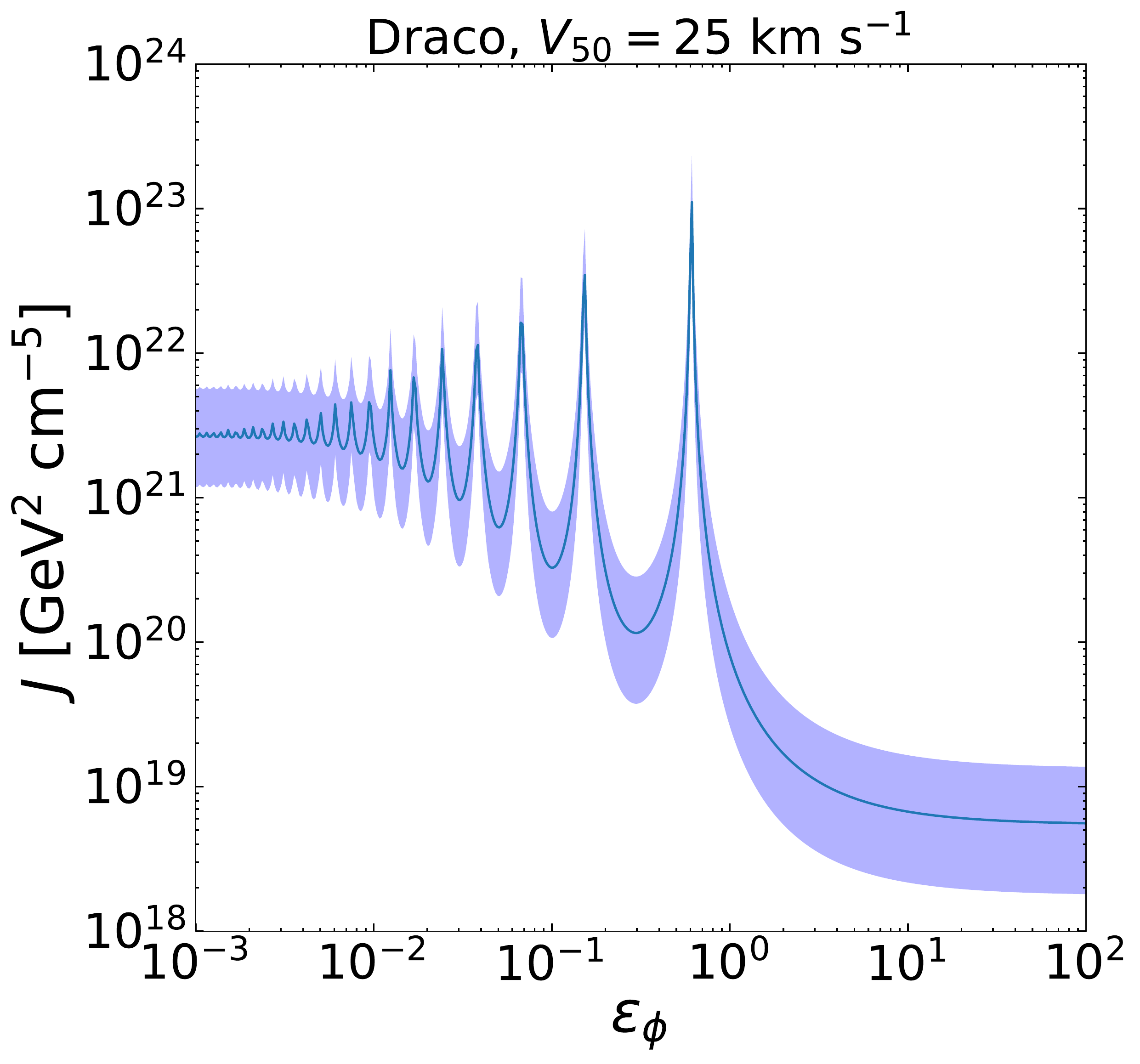}
    \includegraphics[width=4.9cm]{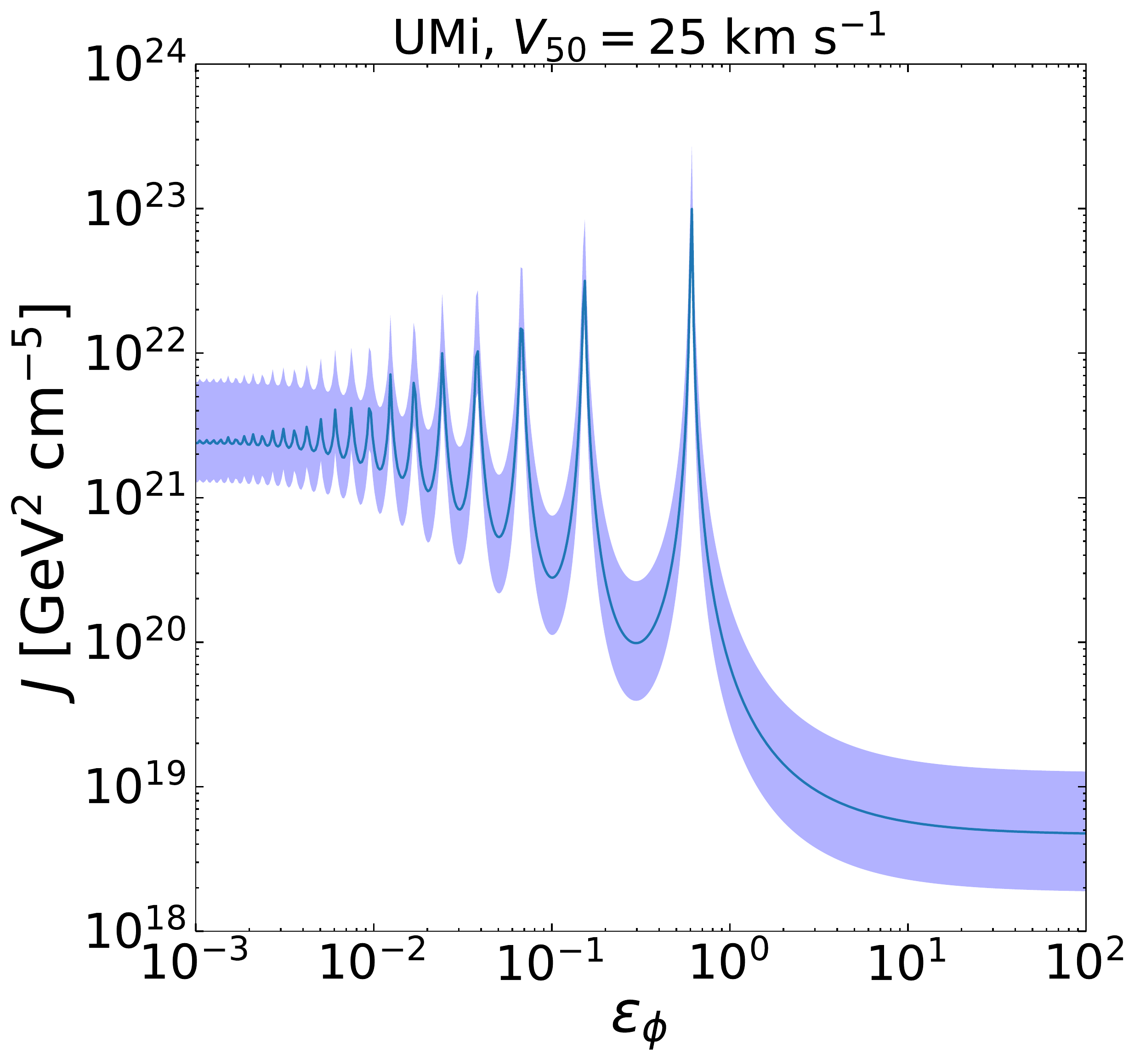}
    
    \includegraphics[width=4.9cm]{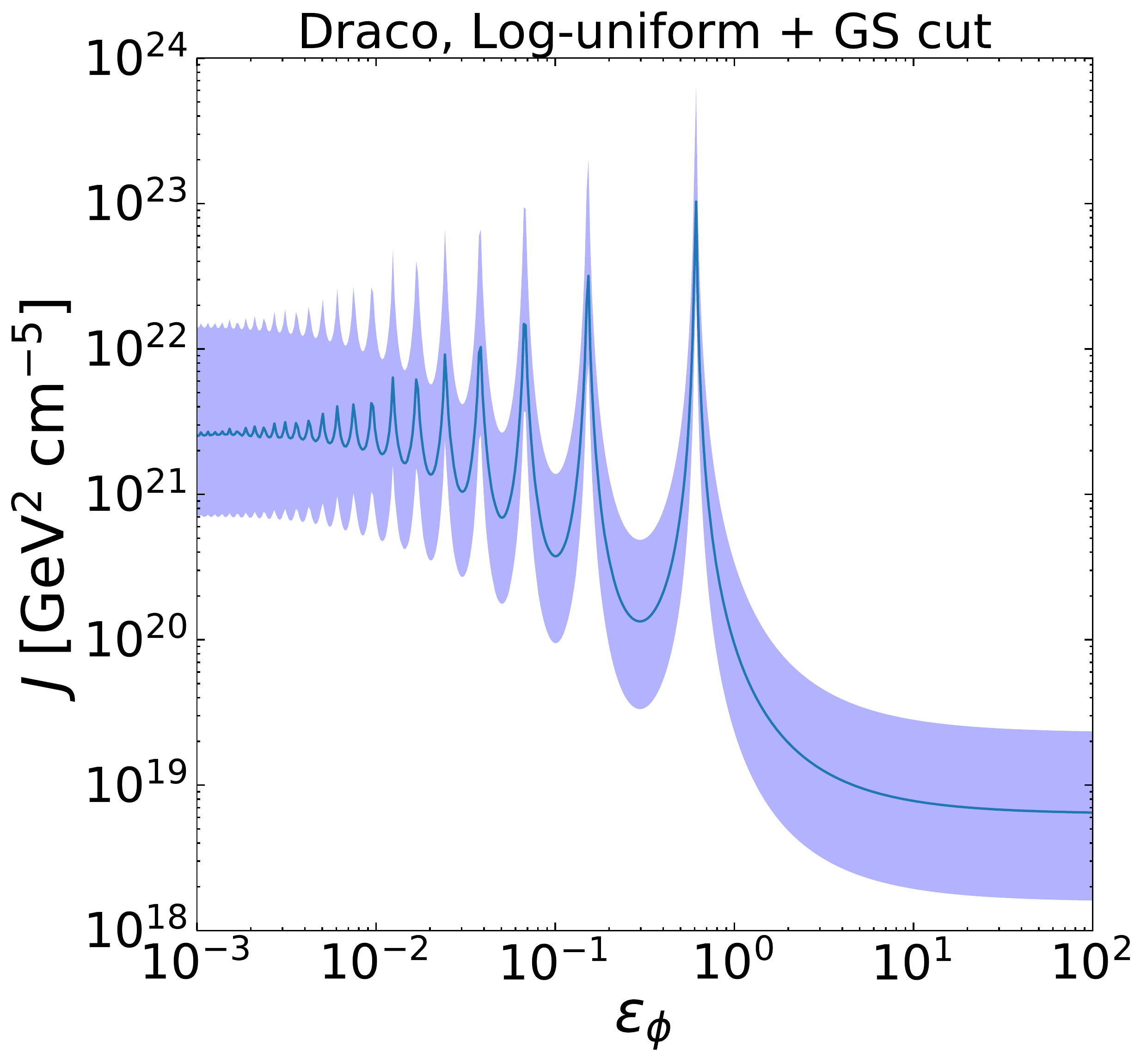}
    \includegraphics[width=4.9cm]{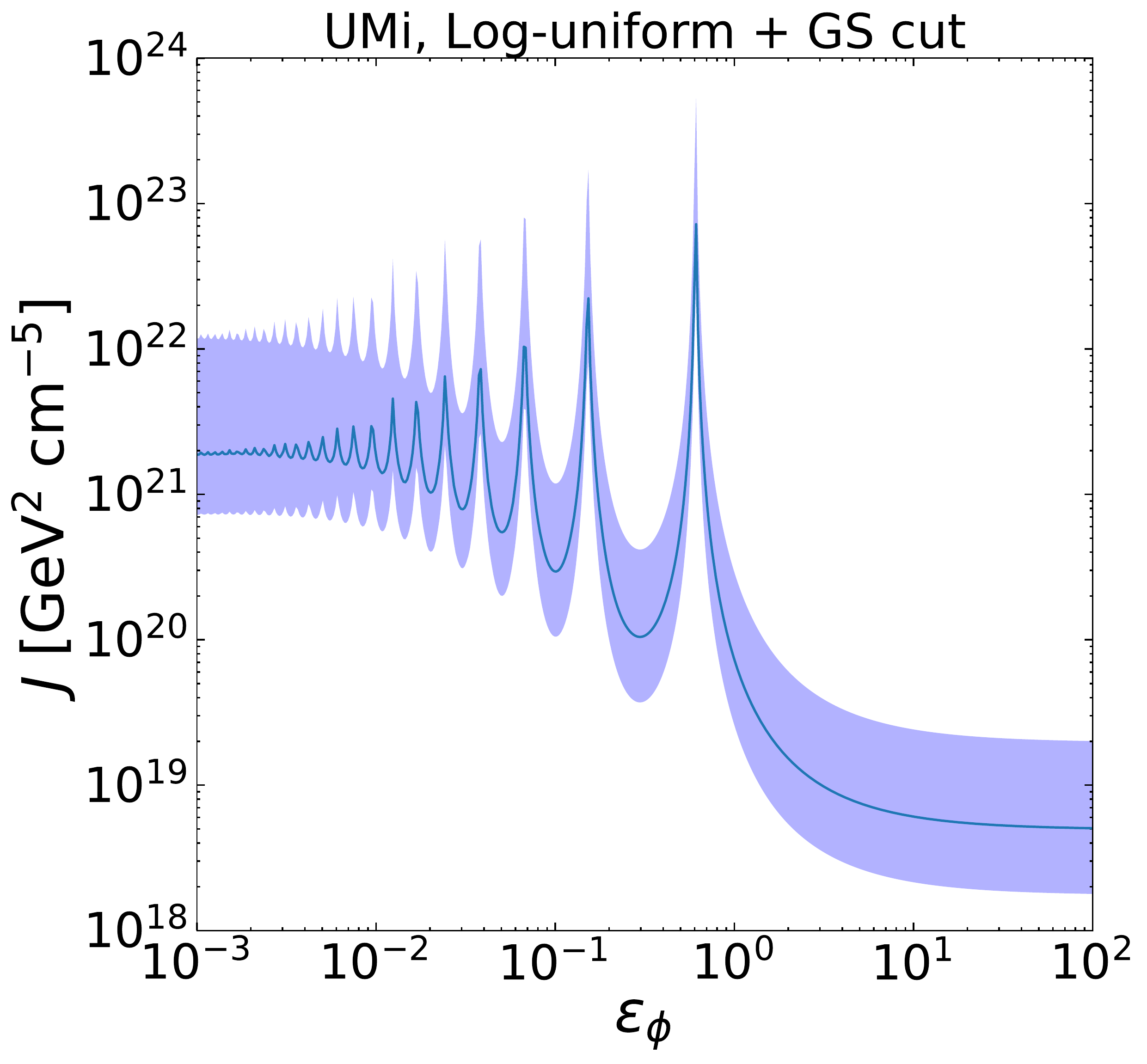}

    \caption{Same as Fig.\,\ref{fig:J_ultrafaints_selected} but
      J-factor calculated as function of $\epsilon_\phi$ for classical
      dSphs, Draco (left) and Ursa Minor (right). $V_{50}=$25
      km~s$^{-1}$ (top) is taken and result using log-uniform + GS
      cut prior is also shown (bottom).}
  \label{fig:J_classicals_selected}
 \end{center}
\end{figure}

\begin{figure}
  \begin{center}
    \includegraphics[width=4.9cm]{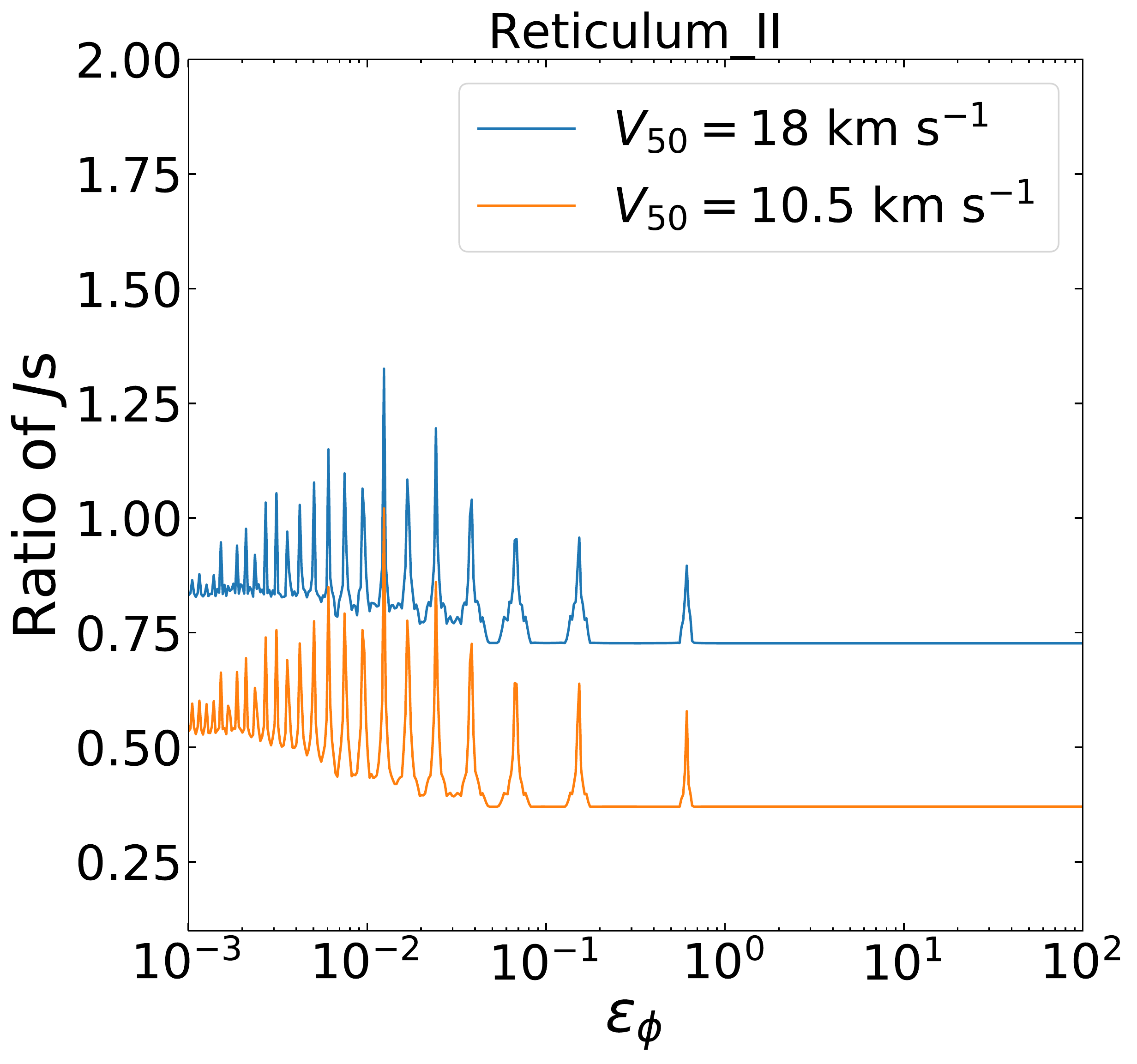}
    \includegraphics[width=4.9cm]{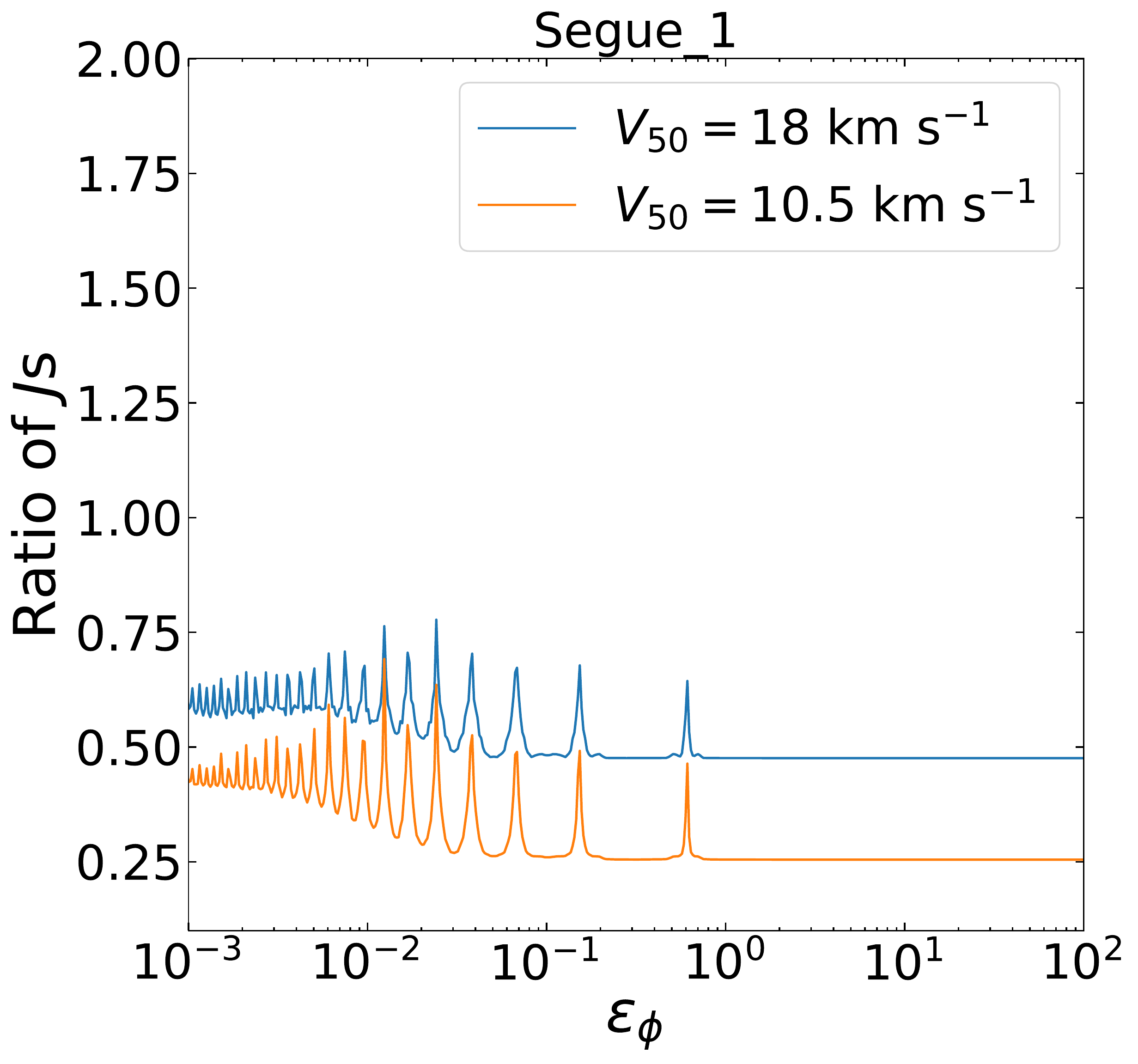}
    \includegraphics[width=4.9cm]{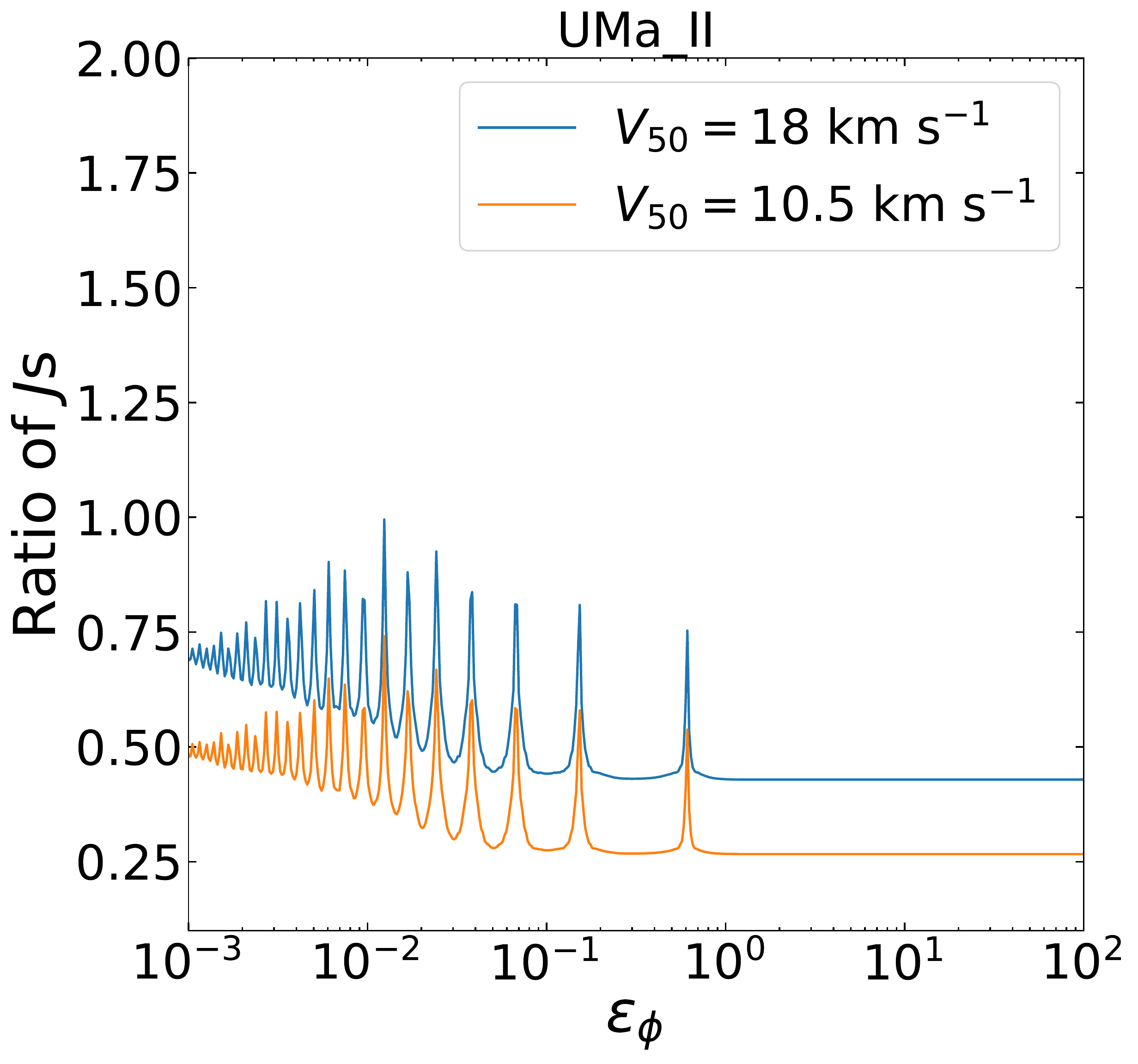}
     \caption{J-factor calculated by using $V_{50}=$10.5 km~s$^{-1}$
       and 18 km~s$^{-1}$ that are normalized by J-factor with
       log-uniform prior with GS15 cut. The other parameters are the
       same as Fig.\,\ref{fig:J_ultrafaints_selected}.}
  \label{fig:compareJs_ultrafaints}
 \end{center}
\end{figure}

\begin{figure}
  \begin{center}
    \includegraphics[width=4.9cm]{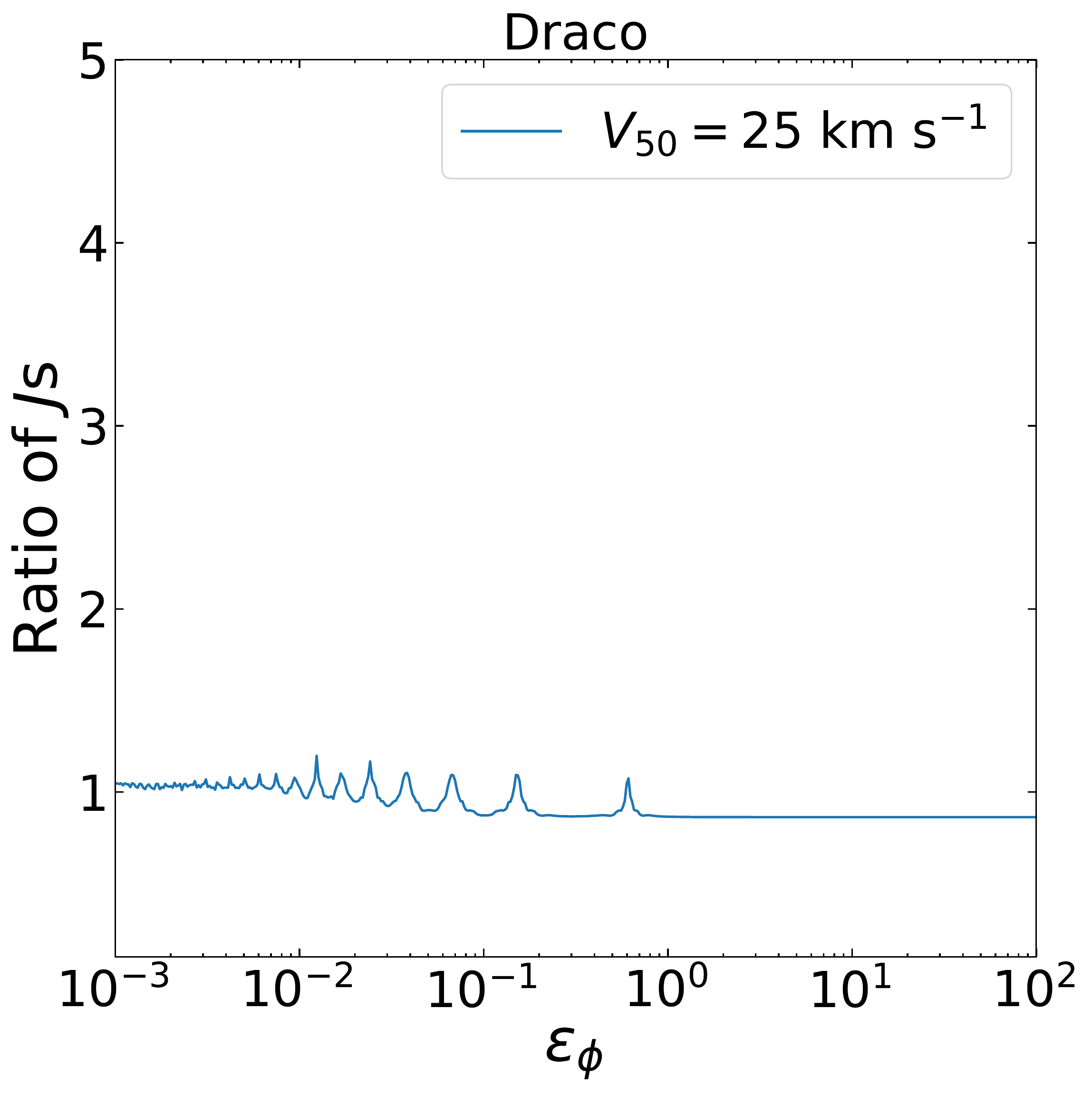}
    \includegraphics[width=4.9cm]{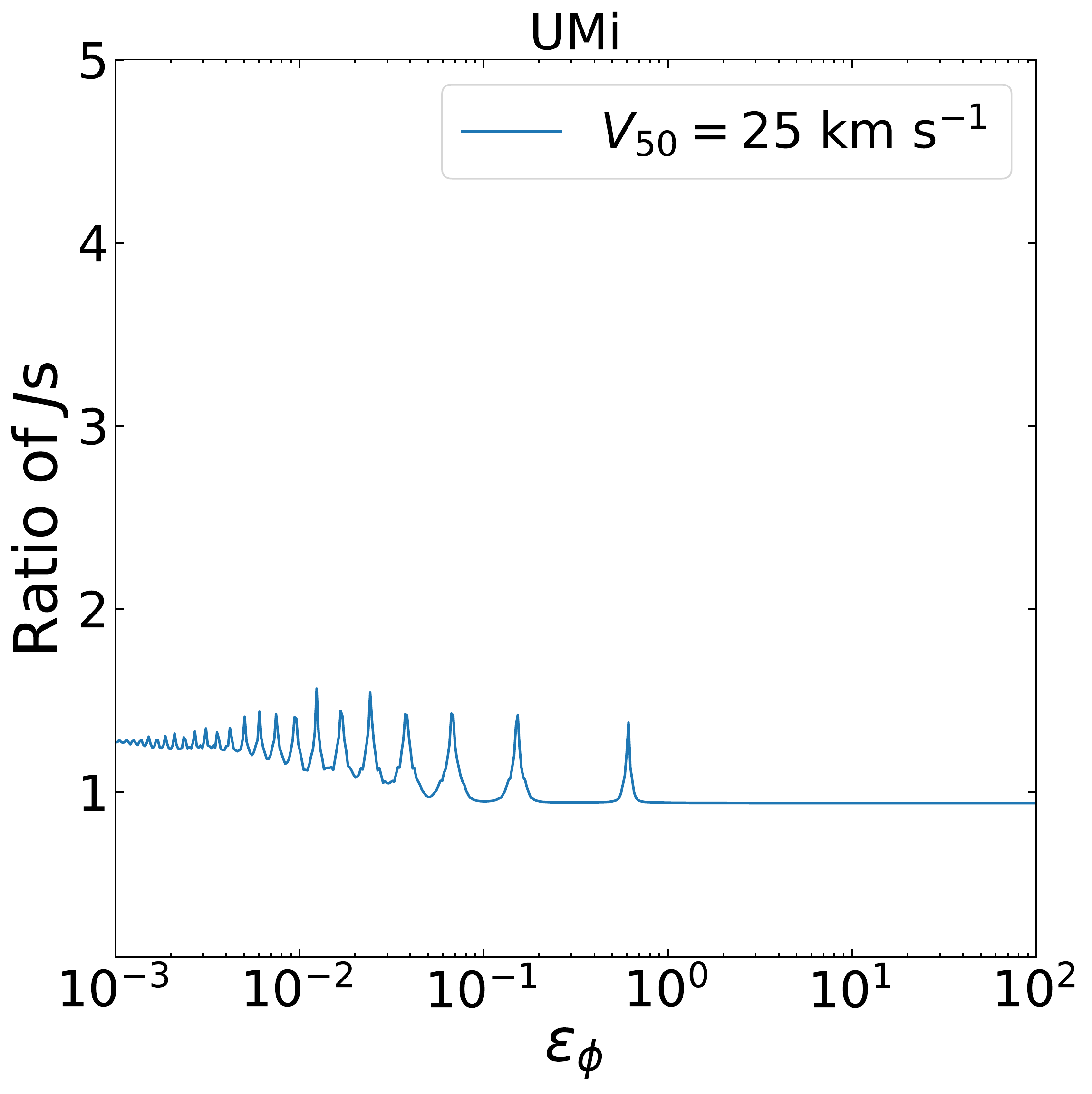}
     \caption{Same as Fig.\,\ref{fig:compareJs_ultrafaints} but for classical 
     dSphs, Draco and Ursa Minor, and using 
     $V_{50}=$25 km~s$^{-1}$.}
  \label{fig:compareJs_classicals}
 \end{center}
\end{figure}

Fig.\,\ref{fig:J_ultrafaints_selected} shows J-factors of a few
selected ultrafaint dSphs that are found to have relatively large
J-factor. Here $\theta=0.5^\circ$ and $V_{50}=$10.5 km~s$^{-1}$ (left)
and 18 km~s$^{-1}$ (middle) are adopted.  As a comparison, the results
by log-uniform priors for both $\rho_s$ and $r_s$ with abrupt cut
according to cosmological argument (GS15
cut)~\cite{Geringer-Sameth:2014yza} is shown for comparison (right).
Shaded band shows 95\% credible intervals, which are determined by the
posterior distributions of the J factor for a given value of
$\epsilon_\phi$ and $\alpha_y$.  $\epsilon_\phi \to \infty$ limit
corresponds to the results without the Sommerfeld enhancement that are
consistent with those in Ref.\,\cite{Ando:2020yyk}.  It is seen that
the median and credible region change in different values of $V_{50}$
for a given $\epsilon_\phi$. It is found that they trace the value at
$\epsilon_\phi \to \infty$ limit; if the J-factor of a dSph for a
given $V_{50}$ in this limit is larger compared to another, then the
J-factor with finite $\epsilon_\phi$ is also larger.  I.e., the
differences of J-factor are nearly independent of $\epsilon_\phi$.

To see this explicitly, we compare J-factors computed in the prior
models in Fig.\,\ref{fig:compareJs_ultrafaints}, where J-factors using
$V_{50}=10.5$ km s$^{-1}$ and $18$ km s$^{-1}$ divided by J-factor
using log-uniform prior are shown. It is seen that relative value of
J-factors with different priors are less sensitive to $\epsilon_\phi$,
except for especially enhanced regions.  If DM annihilation happens in
the region, more precise computation might be required to determine
the J-factor. Additionally, it is seen that the J factor with
$V_{50}=18$ km s$^{-1}$ is larger than that with $V_{50}=10.5$ km
s$^{-1}$. However, the ratio between them is differs for different
dSphs. Finally, it is noted that J factor with the log-uniform prior
tends to give the largest value. Since studies in the existing
literature had to rely on this uninformative prior, this result shows
how incorporating a more realistic, satellite prior is important to in
order to determine the J factor, or the gamma-ray flux from dSphs.

For completeness, we show J-factors of some selected classical dSphs
in Fig.\,\ref{fig:J_classicals_selected}, where we adopt $V_{50}=25$
km~s$^{-1}$. Similarly to Fig.\,\ref{fig:compareJs_ultrafaints}, we
show the ratio of J-factors with different prior models in
Fig.\,\ref{fig:compareJs_classicals}. Compared to the previous
ultrafaint dSphs case, the impact of adopting the satellite prior,
i.e., the choice of $V_{50}$, is small for both the Draco and Ursa
Minor, as their density profiles are well determined by rich
kinematics data that lead to the constraining likelihood functions.
We also investigated another classical dSph,
Sagittarius~\cite{Sagittarius}, which indeed yielded the largest
J-factor, particularly because of its proximity. However, it is a
system that is far from thermal equilibrium because of tidal heating
and disk shocking, and our assumptions made in this study may not
apply.  Therefore, even though we find that our satellite priors
greatly help determine the J-factor of Sagittarius by reducing
associated errors, we do not discuss it any further in this study.
More accurate determination of J-factors using the proper priors is
important for hunting DM using dSphs, which will be discussed in
Sec.\,\ref{sec:results}.

\subsection{Wino dark matter model}
\label{sec:Wino}

Wino is the superpartner of $W$ boson. It is $SU(2)_L$ triplet with
hypercharge zero.  Since the triplet consists of neutral and charged
states, we need to solve two-component Schr\"{o}dinger equation,
\begin{align}
    \frac{1}{m_{\rm dm}}\frac{d^2\psi_i(r)}{dr^2}-V_{ij}(r)\psi_j(r)
  = -m_{\rm dm}v^2\psi_i(r)\,,
\end{align}
where
\begin{align}
  V_{ij}(r)=
  \left(
  \begin{array}{cc}
    0 & -\sqrt{2}\alpha_W \frac{e^{-m_W r}}{r} \\
    -\sqrt{2}\alpha_W \frac{e^{-m_W r}}{r}~~~ &
    2\delta m-\frac{\alpha}{r}-\alpha_Wc_W^2\frac{e^{-m_Z r}}{r}
  \end{array}
  \right)\,.
\end{align}
Here $\alpha$ is the fine-structure constant, $\alpha_W =
g_2^2/(4\pi)$ ($g_2$ is the gauge coupling constant of $SU(2)_L$),
$c_W=\cos \theta_W$ ($\theta_W$ is the Weinberg angle), and $m_W$ and
$m_Z$ are the masses of $W$ and $Z$ bosons, respectively. $\delta m$
is the mass difference between charged Wino and neutral Wino. In our study,
we adopt $\delta m = 0.1645$\,GeV~\cite{Ibe:2012sx}. As in the previous case,
we introduce dimensionless parameters,
\begin{align}
  \epsilon_{v}=\frac{v}{\alpha_W}\,,~~~
  \epsilon_{W,Z}=\frac{m_{W,Z}}{\alpha_W m_{\rm dm}}\,,~~~
  x=\alpha_W m_{\rm dm} r\,,
\end{align}
to obtain,
\begin{align}
  \psi''_i(x)+\tilde{V}_{ij}(x)\psi_j(x)=-\epsilon_v^2\psi_i(x)\,,
\end{align}
where $\prime$ means derivative with respect to $x$ and 
\begin{align}
    \tilde{V}_{ij}(x)=\frac{V_{ij}(r)}{\alpha_W^2 m_{\rm dm}}=
  \left(
  \begin{array}{cc}
    0 & -\sqrt{2}\frac{e^{-\epsilon_W x}}{x} \\
    -\sqrt{2}\frac{e^{-\epsilon_W x}}{x}~~~ &
    \frac{2\delta m}{\alpha_W^2m_{\rm dm}}
    -\frac{s_W^2}{x}-c_W^2\frac{e^{-\epsilon_Z x}}{x}
  \end{array}
  \right)\,.
\end{align}
Here $s_W=\sin \theta_W$. Technically, it is useful to rewrite the
equation by introducing $\chi_i(x)$ and $\phi_i(x)$ ($i=1,2$),
\begin{align}
  \psi_i(x) = \chi_i(x) \phi_i(x)\,,
\end{align}
where
\begin{align}
  \phi_1(x)&=e^{i\epsilon_v x}\,, \\
  \phi_2(x)&=e^{-\sqrt{2\delta m/(\alpha_W^2 m_{\rm dm})-\epsilon^2_v}x}\,.
\end{align}
Then the differential equations that $\chi_i(x)$ should obey are
\begin{align}
  &\chi_1''(x)+2i\epsilon_v \chi_1'(x)
  +\sqrt{2}\frac{e^{-\epsilon_W x}}{x}
  \frac{\phi_2(x)}{\phi_1(x)}\chi_1(x)=0\,, \\
  &\chi''_2(x)-\sqrt{\frac{2\delta m}{\alpha_W^2 m_{\rm dm}}
    -\epsilon^2_v}\chi'_2(x)
  +\sqrt{2}\frac{e^{-\epsilon_W x}}{x}
  \frac{\phi_1(x)}{\phi_2(x)}\chi_2(x)
  +\left(\frac{s_W^2}{x}+c_W^2\frac{e^{-\epsilon_Zx}}{x}\right)\chi_2(x)=0\,.
\end{align}
We solve the equations under the boundary condition
$\chi_i'(\infty)=0$ and the initial conditions of (i) $\chi_1(0)=1$,
$\chi_2(0)=0$, and (ii) $\chi_1(0)=0$, $\chi_2(0)=1$. We call the
solutions as $\psi_i^{\rm sol1}(x)$ and $\psi_i^{\rm sol2}(x)$ for the
conditions (i) and (ii), respectively.

The differential cross section for the process where the Wino
annihilates to $\gamma \gamma$ is computed at the leading-log
(LL)~\cite{Baumgart:2017nsr} and the next-LL
(NLL)~\cite{Baumgart:2018yed}.\footnote{See also
Refs.\,\cite{Beneke:2019vhz,Beneke:2020vff} for the recent
developments. } In the following analysis, we adopt the analytical
formulas at the NLL to give the differential cross section. In their
notation, $s_{00}$ and $s_{\pm 0}$ corresponds to $\psi_{1}^{\rm
  sol1}(\infty)$ and $\psi_{1}^{\rm sol2}(\infty)$, respectively. For
the process where the Wino annihilates to $W^+W^-$, we adopt the
formula given in Refs.\,\cite{Hisano:2004ds,Hisano:2006nn}. Namely,
the cross section $\sigma v_{W^+W^-}$ is
\begin{align}
  \sigma v_{W^+W^-} = 2
  \left(A_{\rm som} \Gamma_{WW} A_{\rm som}^\dagger\right)_{11}\,,
\end{align}
where
\begin{align}
  A_{\rm som} =
  \left(
  \begin{array}{cc}
    s_{00} & s_{\pm 0} \\
    0 & 0
  \end{array}
  \right)\,, ~~~~~
  \Gamma_{WW} = \frac{\pi \alpha_W^2}{m^2_{\rm dm}}
   \left(
  \begin{array}{cc}
    1 &\sqrt{2}/2 \\
    \sqrt{2}/2 & 1/2
  \end{array}
  \right)\,.
\end{align}
As a reference, leading order cross section to $\gamma\gamma$ is
obtained as
\begin{align}
  \sigma v_{\gamma \gamma,\,{\rm LO}}
  = 2
  \left(A_{\rm som} \Gamma_{\gamma\gamma} A_{\rm som}^\dagger\right)_{11}\,,
\end{align}
where
\begin{align}
  \Gamma_{\gamma\gamma}=\frac{\pi \alpha^2}{m^2_{\rm dm}}
   \left(
  \begin{array}{cc}
    0 &0\\
    0 & 1
  \end{array}
  \right)\,.
\end{align}

In the Wino annihilation, it is known that the cross section is
numerically independent of the relative velocity when $m_W/m_{\rm dm}
\ll 1$ and this is the parameter space we are interested
in. Therefore, the gamma-ray flux in this case is simply given by
\begin{align}
  \frac{d\Phi_\gamma}{dE}
  =J(\theta)\frac{\sigma v}{8\pi m_{\rm dm}^2}
  \frac{dN_\gamma}{dE}\,,
\end{align}
where $J(\theta)$ is given by taking $S(v_{\rm rel};\vb*{X})=1$. While
the Sommerfeld effect hardly changes the J-factor, the annihilation
cross section is significantly affected by it, which was shown by
Refs.\,\cite{Hisano:2003ec,Hisano:2004ds,Hisano:2006nn,Baumgart:2017nsr,Baumgart:2018yed,Beneke:2019vhz,Beneke:2020vff}. The
photon spectrum by the Wino annihilation is given by
\begin{align}
  \sigma v \frac{dN_\gamma}{dE}
  =\sum_i  (\sigma v)_i \left[\frac{dN_\gamma}{dE}\right]_i\,.
  \label{eq:spectrum_Wino}
\end{align}
Here $i$ shows annihilation mode, $i=\gamma \gamma$, $\gamma Z$,
$W^+W^-$, and $ZZ$. For the later analysis, we re-parameterize
Eq.\,\eqref{eq:spectrum_Wino} as
\begin{align}
  \sigma v \frac{dN_\gamma}{dE}
  =\sum_\alpha  (\sigma v)_\alpha \left[\frac{dN_\gamma}{dE}\right]_\alpha\,,
\end{align}
where $\alpha= \{{\rm line}, {\rm cascade}\}$,\footnote{We have
checked the spectrum is consistent with Figs.\,2 and 5 in
Ref.\,\cite{Rinchiuso:2018ajn}.  Although it is not consistent with
Fig.4 in the literature, we have checked it is just a numerical bug
and their conclusion is not affected. We thank N.~L.~Rodd and T.~Cohen
for confirming this point.}
\begin{align}
  (\sigma v)_{\rm line}\left[\frac{dN_\gamma}{dE}\right]_{\rm line}
  &=2\left(\sigma v_{\gamma\gamma}+\frac{1}{2}\sigma v_{\gamma Z}\right)
  \left[\delta(E-m_{\rm dm})+\frac{dN_\gamma}{dE}\Bigr|_{\rm endpoint}\right]\,,
  \\
  (\sigma v)_{\rm cascade}\left[\frac{dN_\gamma}{dE}\right]_{\rm cascade}
  & =\sigma v_{\gamma Z}\frac{dN_\gamma}{dE}\Bigr|_{Z\,{\rm cascade}}
    +\sum_{i=W^+W^-,ZZ}(\sigma v)_i\left[\frac{dN_\gamma}{dE}\right]_i\,.
\end{align}
Here $dN_\gamma/dE|_{\rm endpoint}$ is the endpoint contributions
computed in Refs.\,\cite{Baumgart:2017nsr,Baumgart:2018yed}. As
presented in the references, they are cascading photons from
monochromatic photons.  Although the gamma-ray spectrum is determined
for fixed $m_{\rm dm}$, we take $\sigma v_{\rm line}$
\begin{align}
  \sigma v_{\rm line}=\sigma v_{\gamma\gamma}+\frac{1}{2}\sigma v_{\gamma Z}\,,
  \label{eq:sigmav_line}
\end{align}
as a free parameter in the later analysis and give the projected
sensitivity limit on that to see the impact of the line and endpoint
gamma-ray spectrum in the DM search.

\section{CTA sensitivity to dark matter by observing dSphs}
\label{sec:results}

Given the flux $d\Phi_\gamma/ dE$ from the DM annihilation,
Eq.~\eqref{eq:flux}, the gamma-ray events in a given energy range
between $E_1$ and $E_2$ are calculated as
\begin{equation}
  N_\gamma(E_1,E_2) = T \int_{E_1}^{E_2} dE_R A_{\rm eff}(E_R)
  \int dE P(E_R|E) \frac{d\Phi_\gamma}{dE},
\end{equation}
where $T$ is the exposure time, $E_R$ is the reconstructed energy, $E$
is the true gamma-ray energy, and $P(E_R|E)$ is the energy dispersion
function that takes the finite energy resolution of the detector into
account.  For the detector specification of the CTA North such as the
effective area $A_{\rm eff}(E_R)$, we adopt information extracted from
\url{www.cta-observatory.org}, and for the exposure time, we assume
500~hours~\cite{CTA,CTAConsortium:2018tzg}.  For the energy resolution
we adopt a flat value of 10\%, i.e., $\Delta E/ E = 0.1$.  The photon
counts in this given energy bin and spatial pixel are then added on
top of other background events caused by both the cosmic ray electrons
and protons, for which we adopt a model given by
Ref.\,\cite{Silverwood:2014yza}.  Then we perform a Poisson likelihood
analysis by combining the information on all the pixels in both the
spatial and energy bins under the null hypothesis that there is no DM
component in the mock data, and obtain the expected upper limits on
the annihilation cross section at 95\% confidence level (CL).

\begin{figure}
  \begin{center}

    \includegraphics[width=4.9cm]{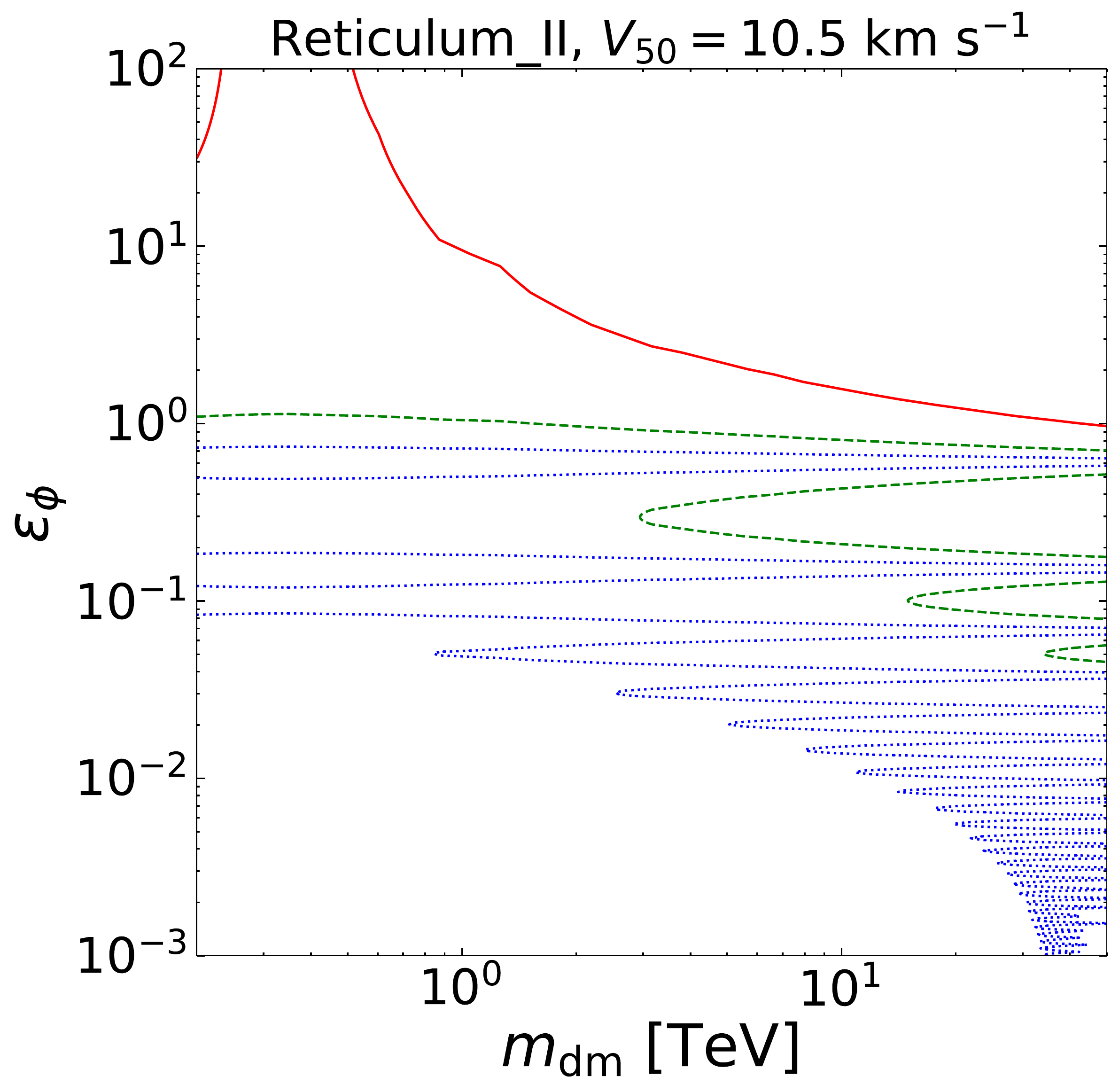}
    \includegraphics[width=4.9cm]{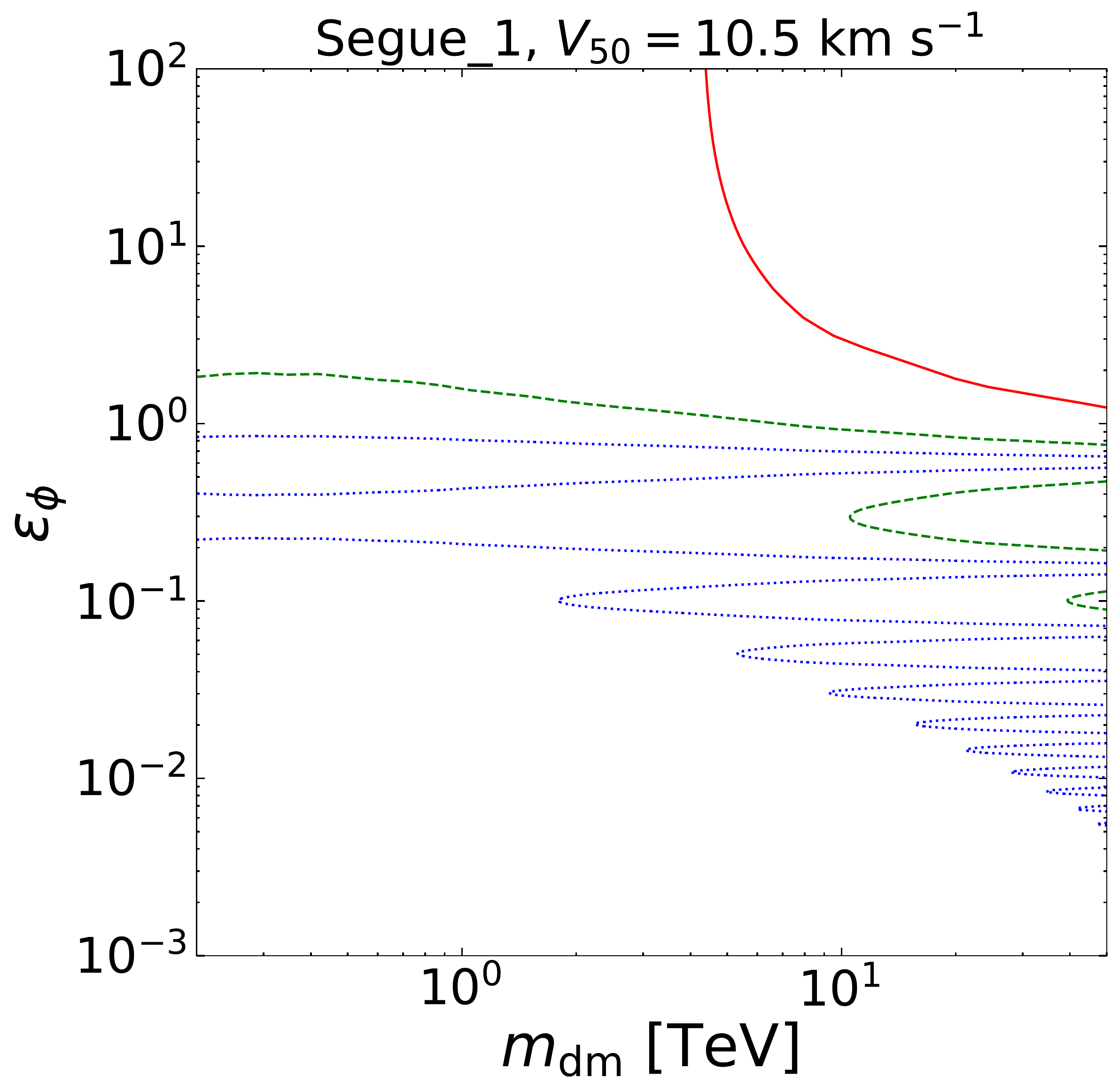}
     \includegraphics[width=4.9cm]{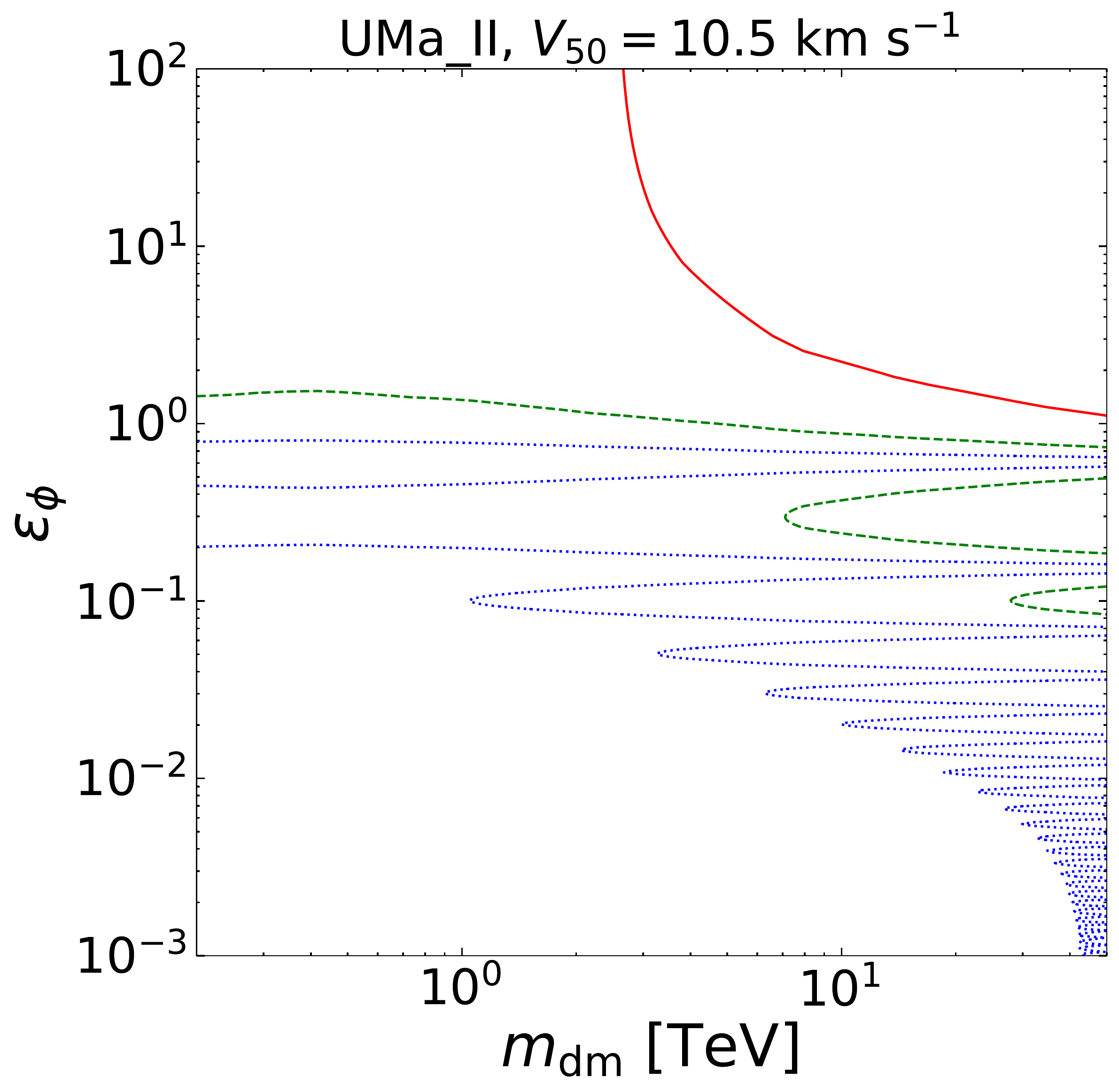}
    
    \includegraphics[width=4.9cm]{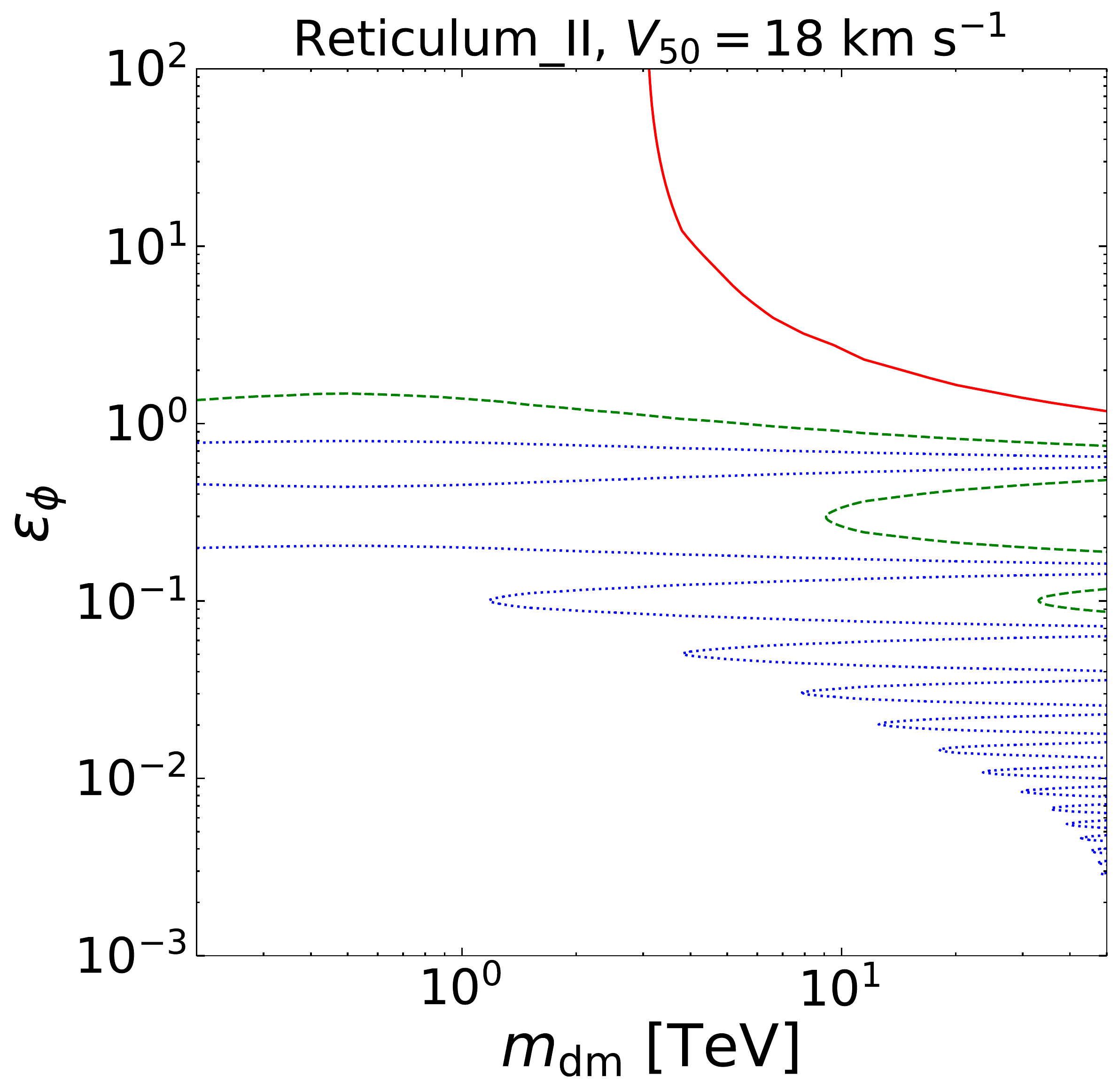}
    \includegraphics[width=4.9cm]{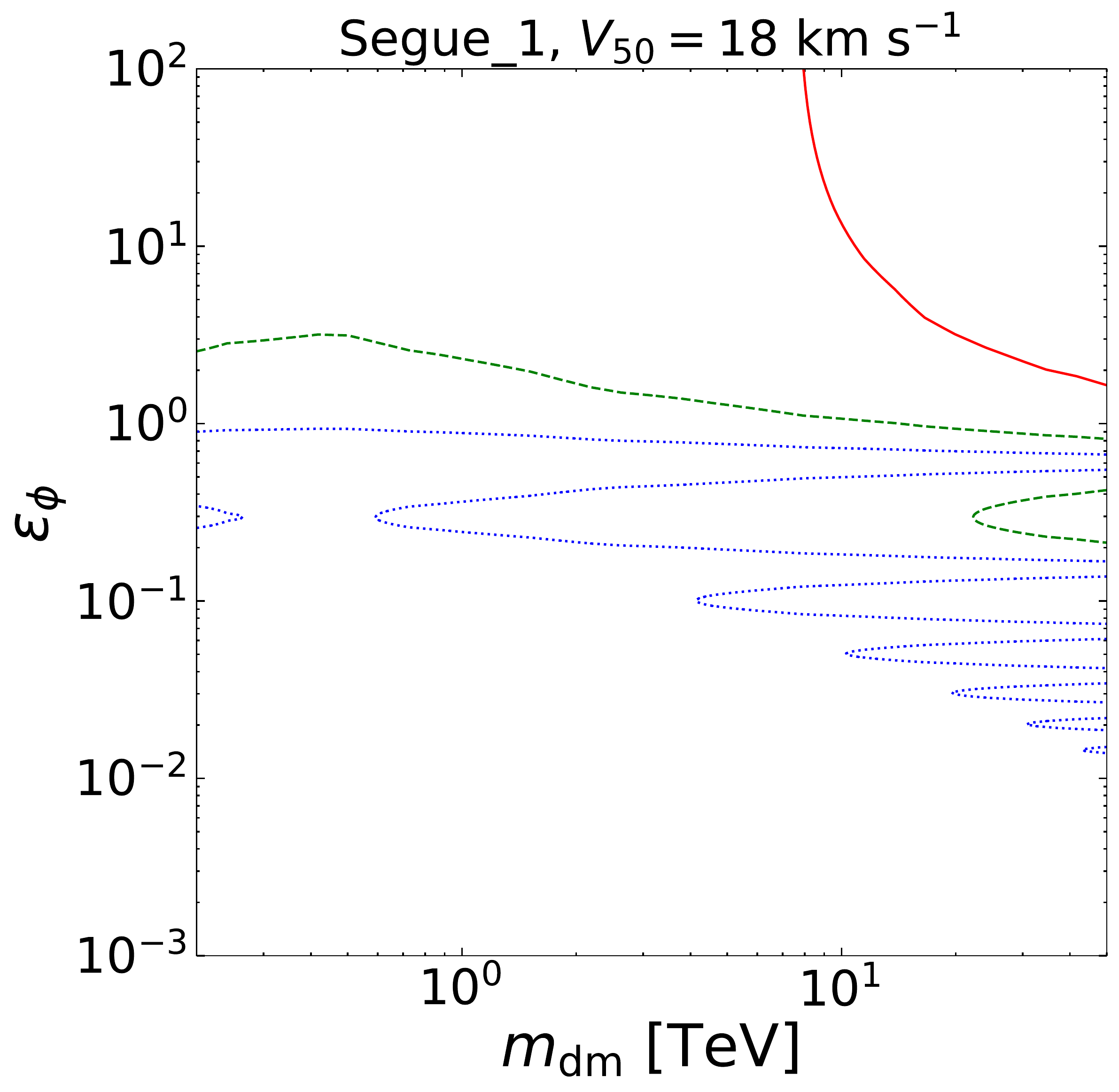}
    \includegraphics[width=4.9cm]{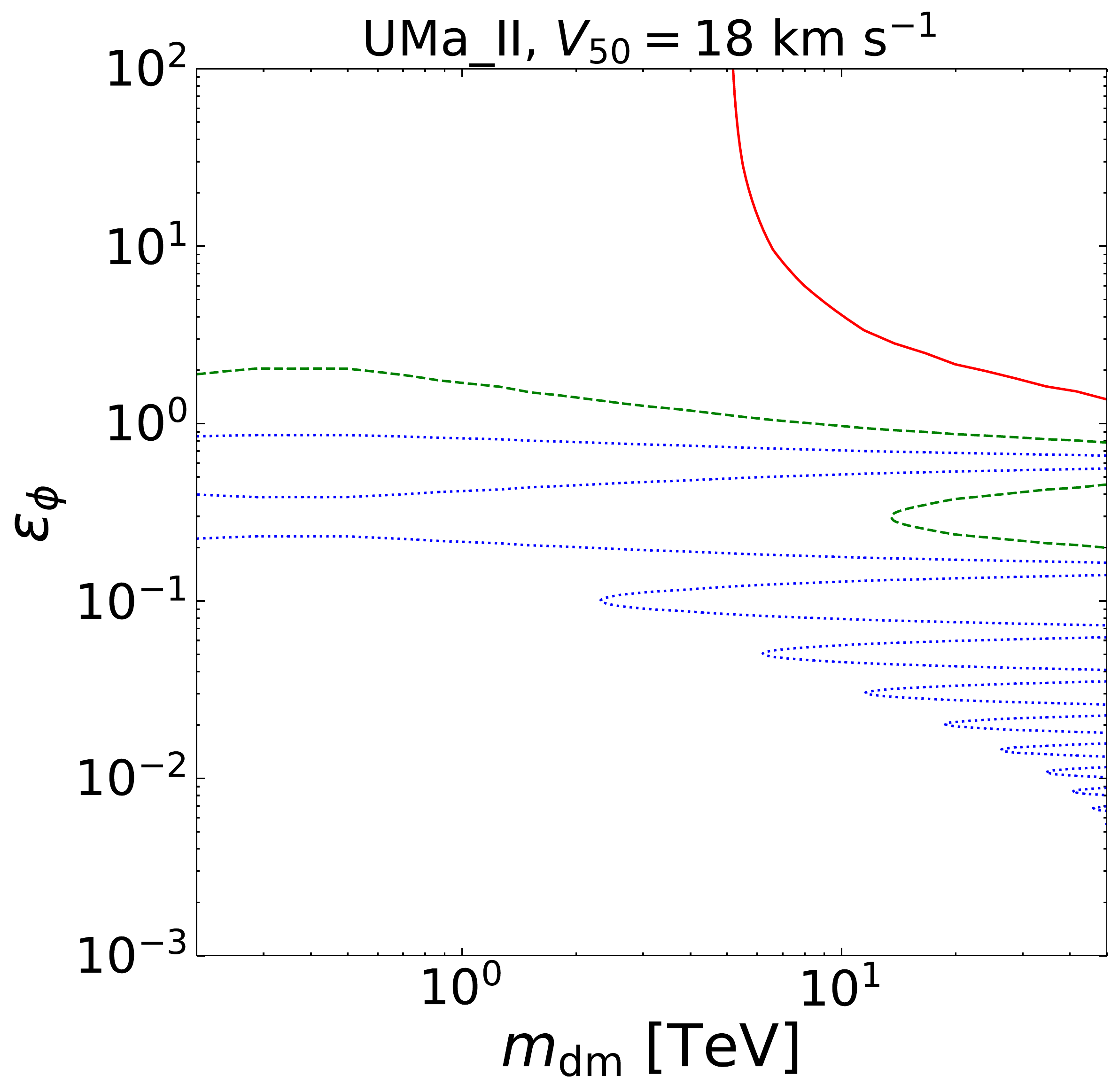}
    
    \includegraphics[width=4.9cm]{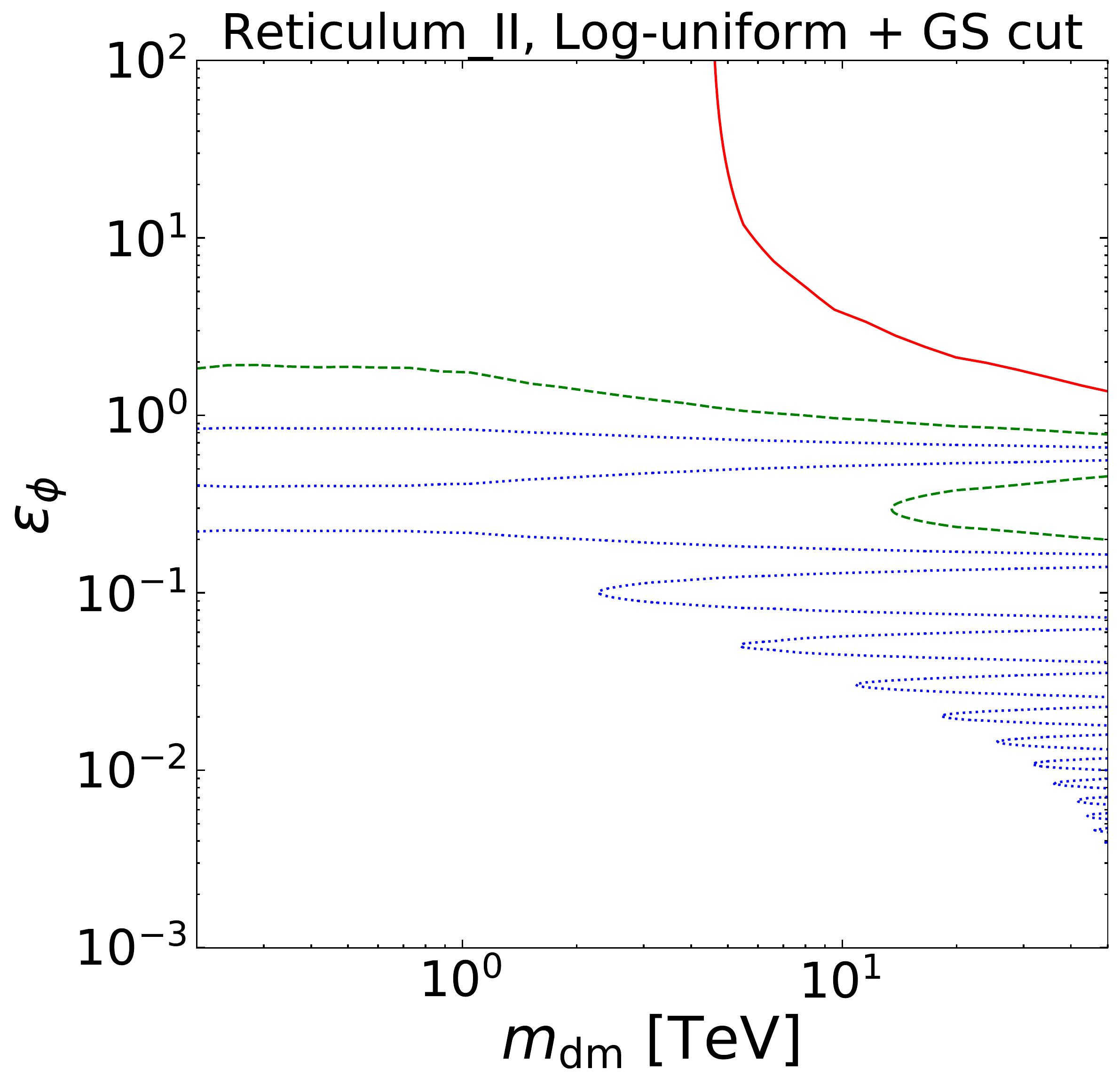}
    \includegraphics[width=4.9cm]{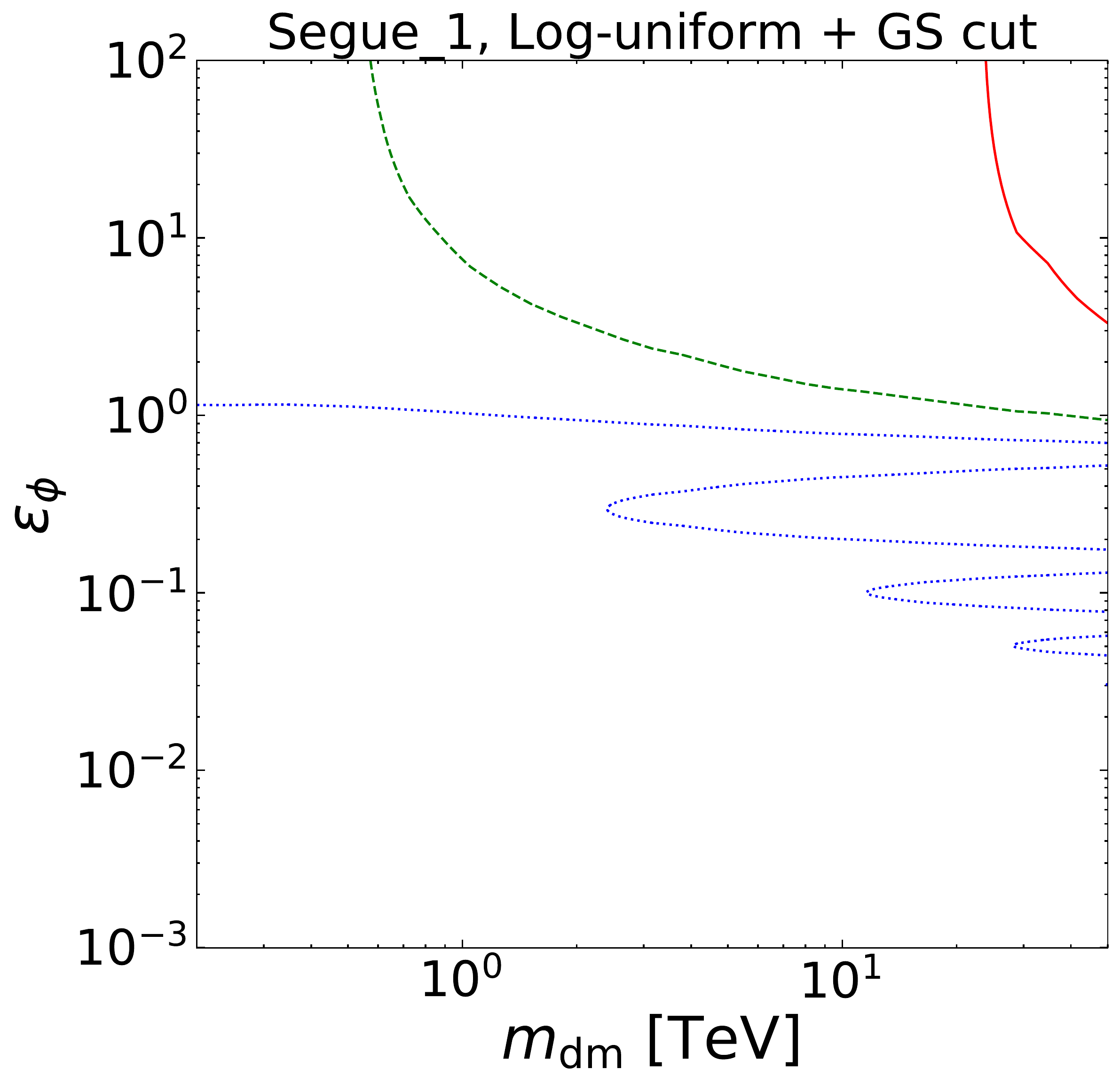}
    \includegraphics[width=4.9cm]{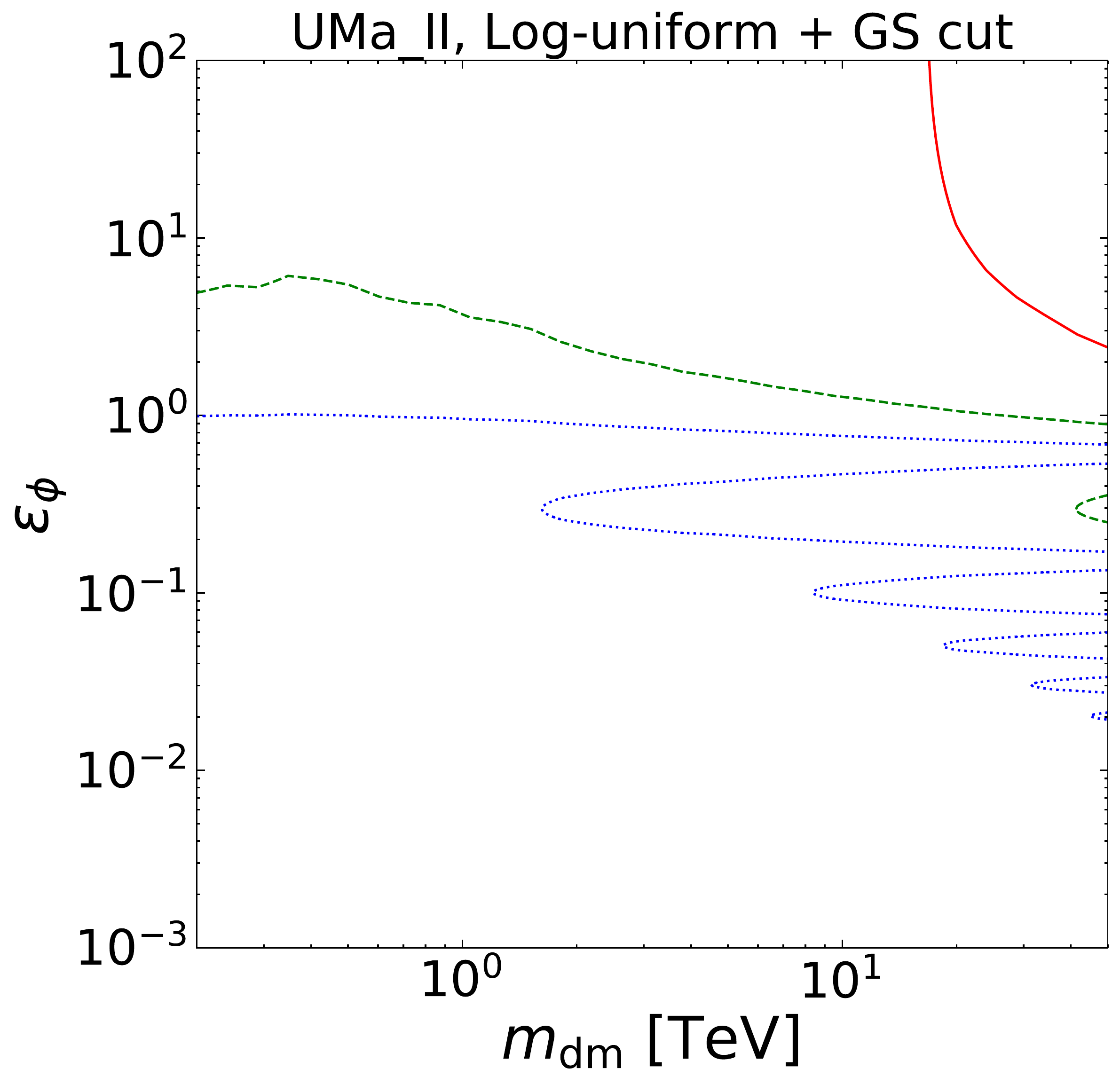}   

    \caption{Expected upper limit contour of the annihilation cross
      section, $\sigma v\,[{\rm cm}^3\,{\rm
          s}^{-1}]=10^{-24}$ (solid red), $10^{-25}$ (dashed green),
      $10^{-26}$ (dotted blue) by CTA North with 500 hour
      experiment. We use median J-factor and the annihilation mode is
      assume to be $b\bar{b}$. Limits are given by observing Reticulum
      II, Segue 1, and Ursa Major II (left to right) and satellite
      prior with $V_{50}=$ 10.5 km s$^{-1}$ (top) and 18 km s$^{-1}$
      (middle) are adopted. As a comparison, result using log-uniform
      prior with GS15 cut is given (bottom).}
  \label{fig:contour_ultrafaints_lightmediator}
 \end{center}
\end{figure}

\begin{figure}
  \begin{center}
    \includegraphics[width=4.9cm]{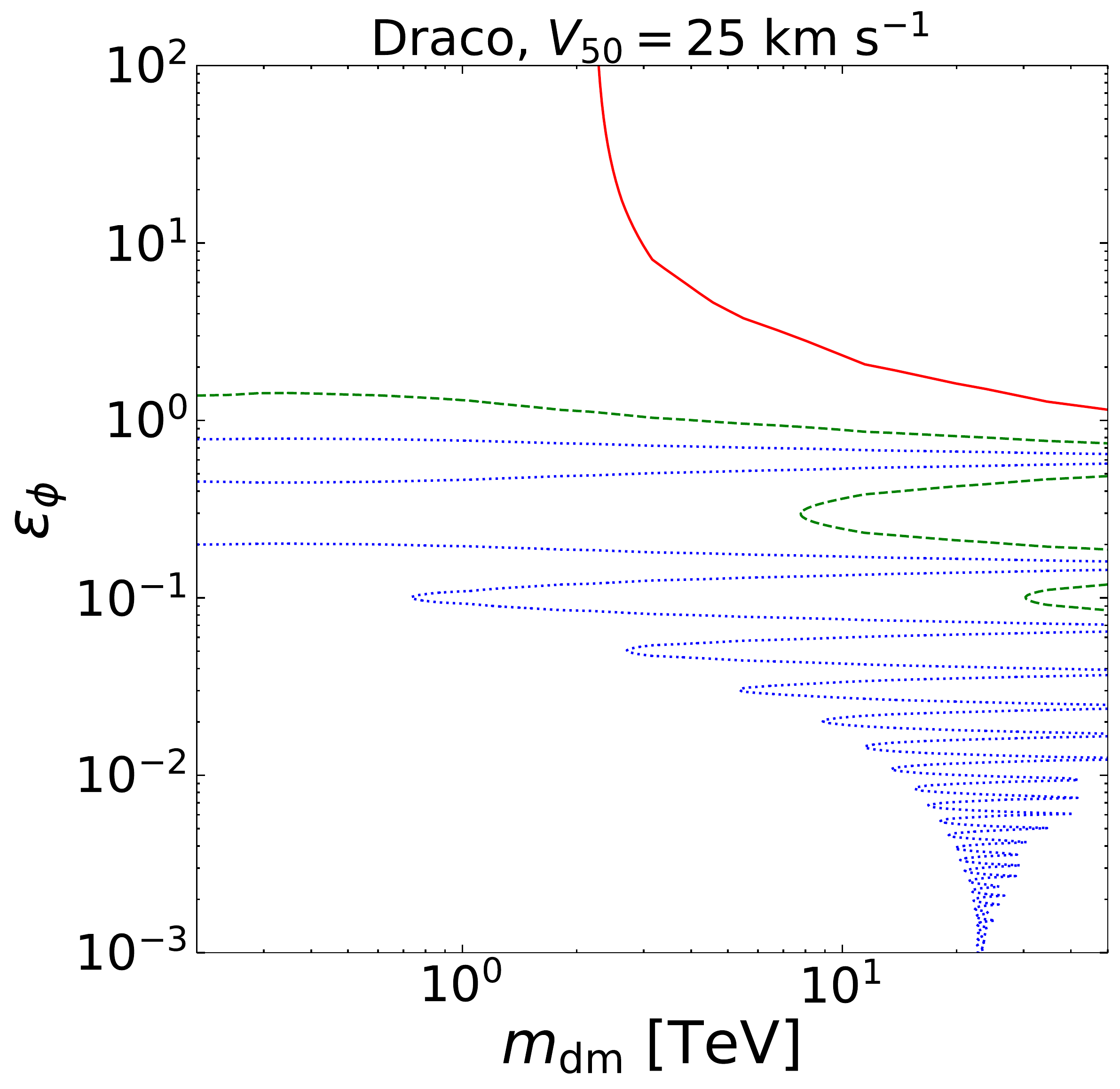}
    \includegraphics[width=4.9cm]{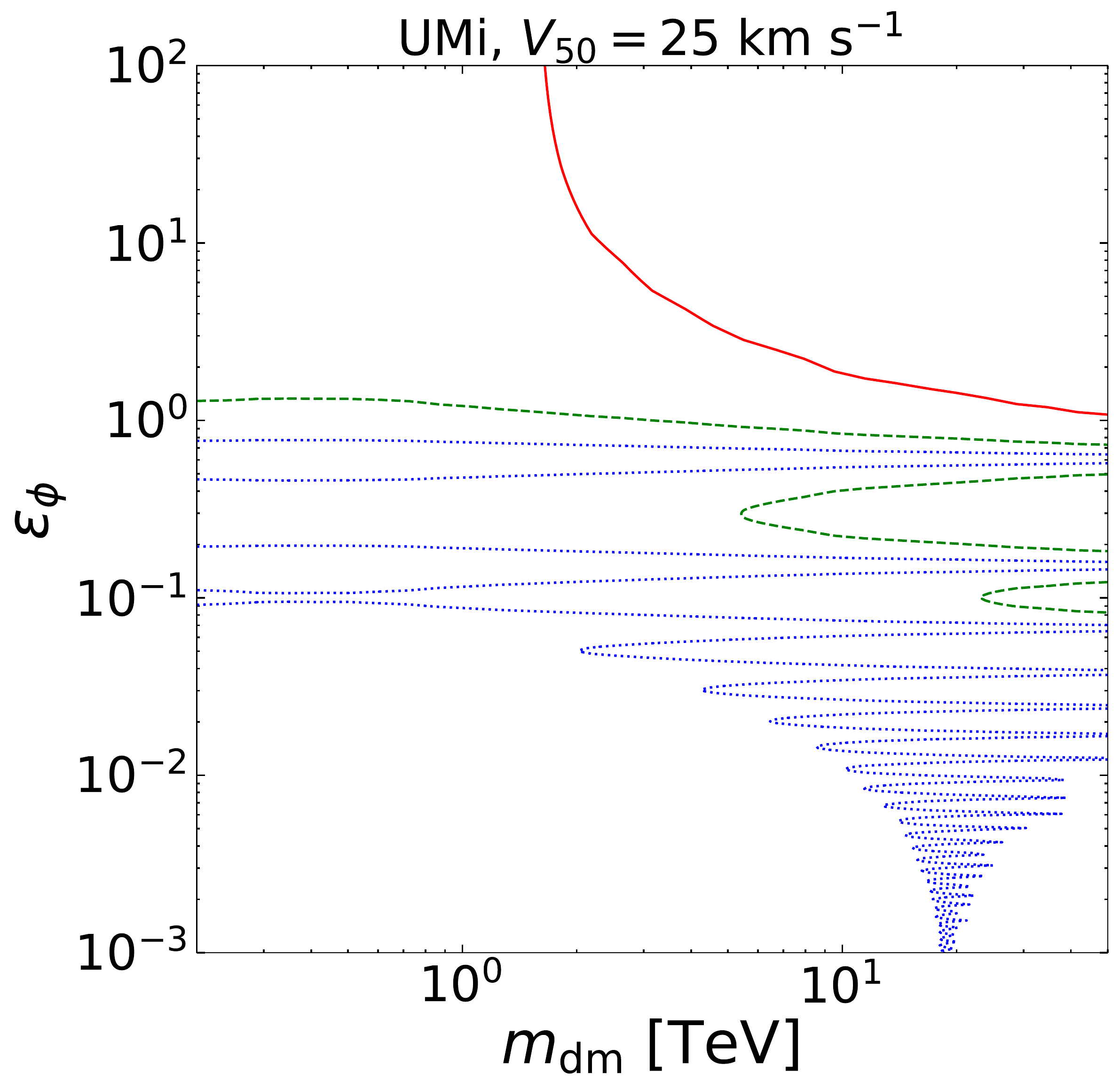}
    
    \includegraphics[width=4.9cm]{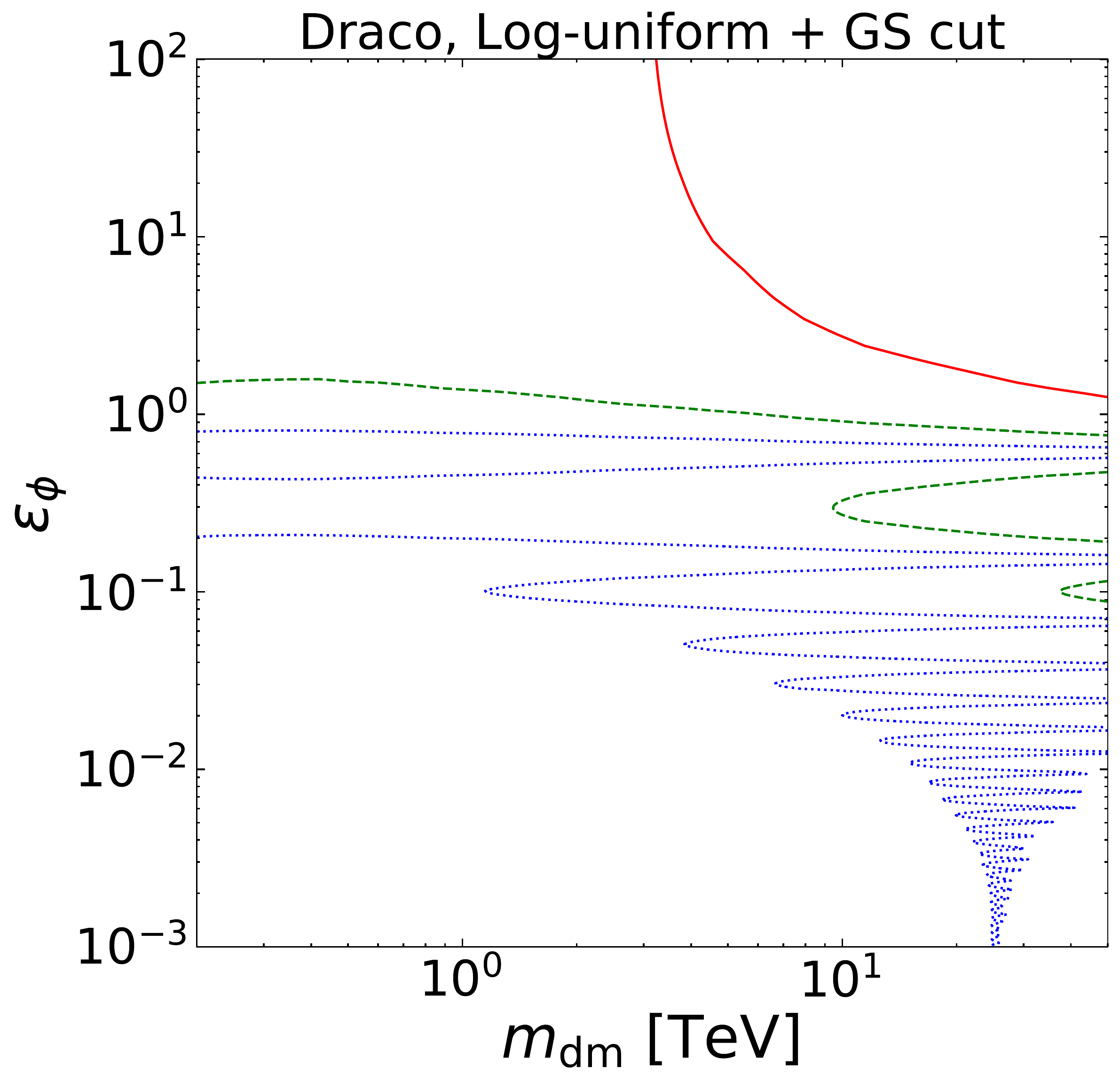}
    \includegraphics[width=4.9cm]{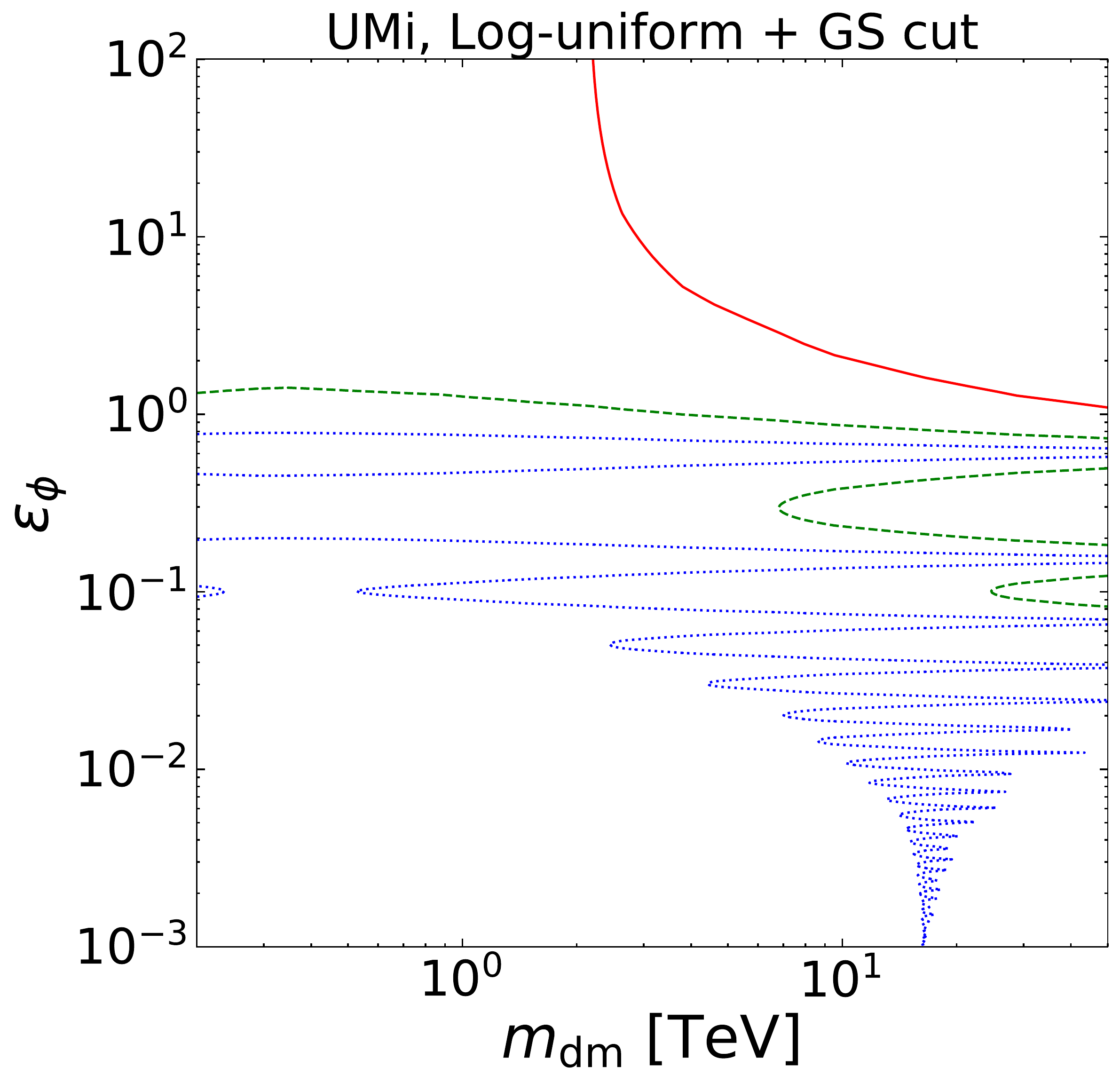}
    \caption{The same as
      Fig.\,\ref{fig:contour_ultrafaints_lightmediator} but analyzed
      for classical dSph, Draco and Ursa Minor. 
      The parameters are the same as
      Fig.\,\ref{fig:J_classicals_selected}.}
    \label{fig:contour_classicals_lightmediator}
 \end{center}
\end{figure}

\begin{figure}
  \begin{center}

    \includegraphics[width=4.9cm]{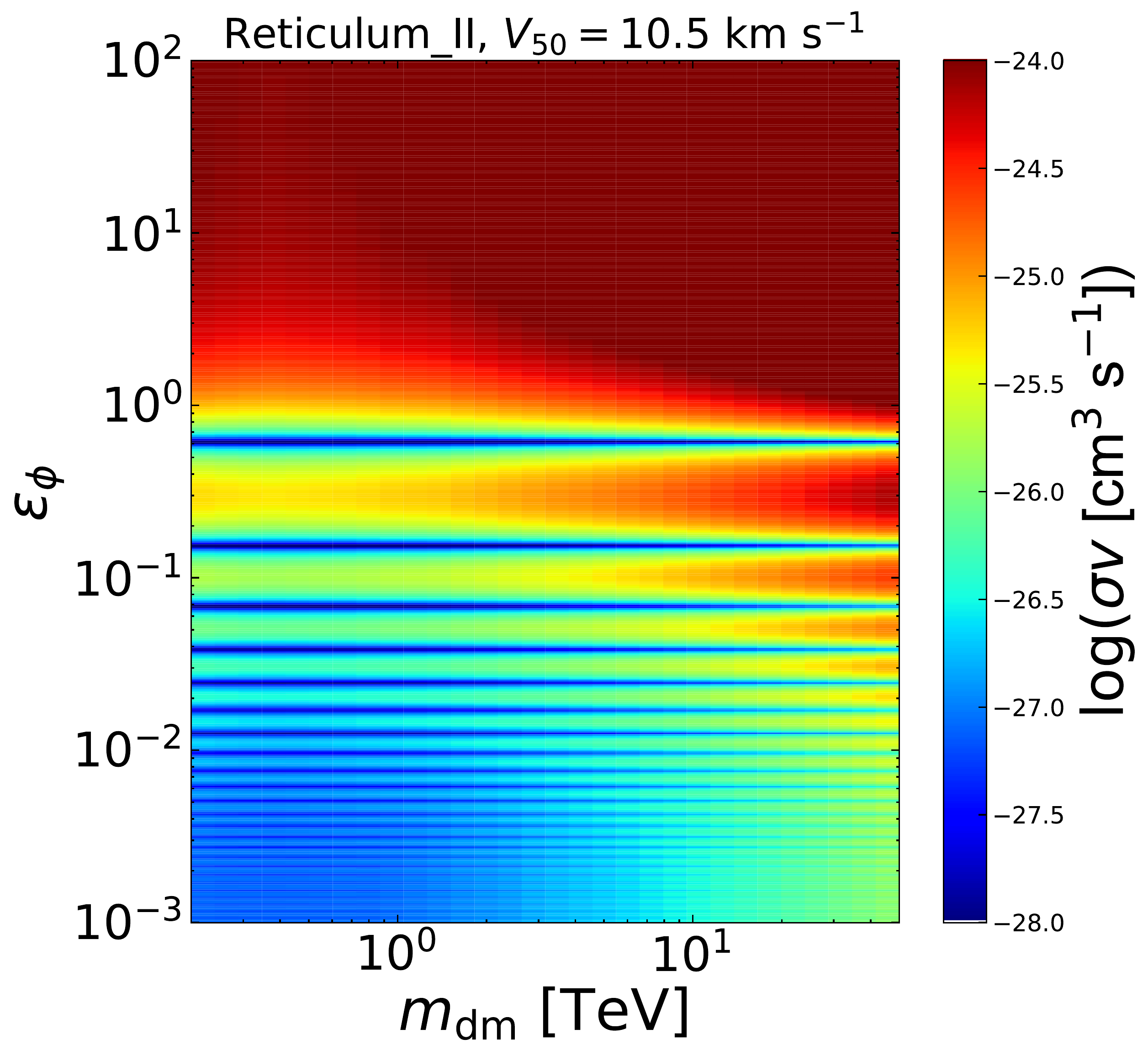}
    \includegraphics[width=4.9cm]{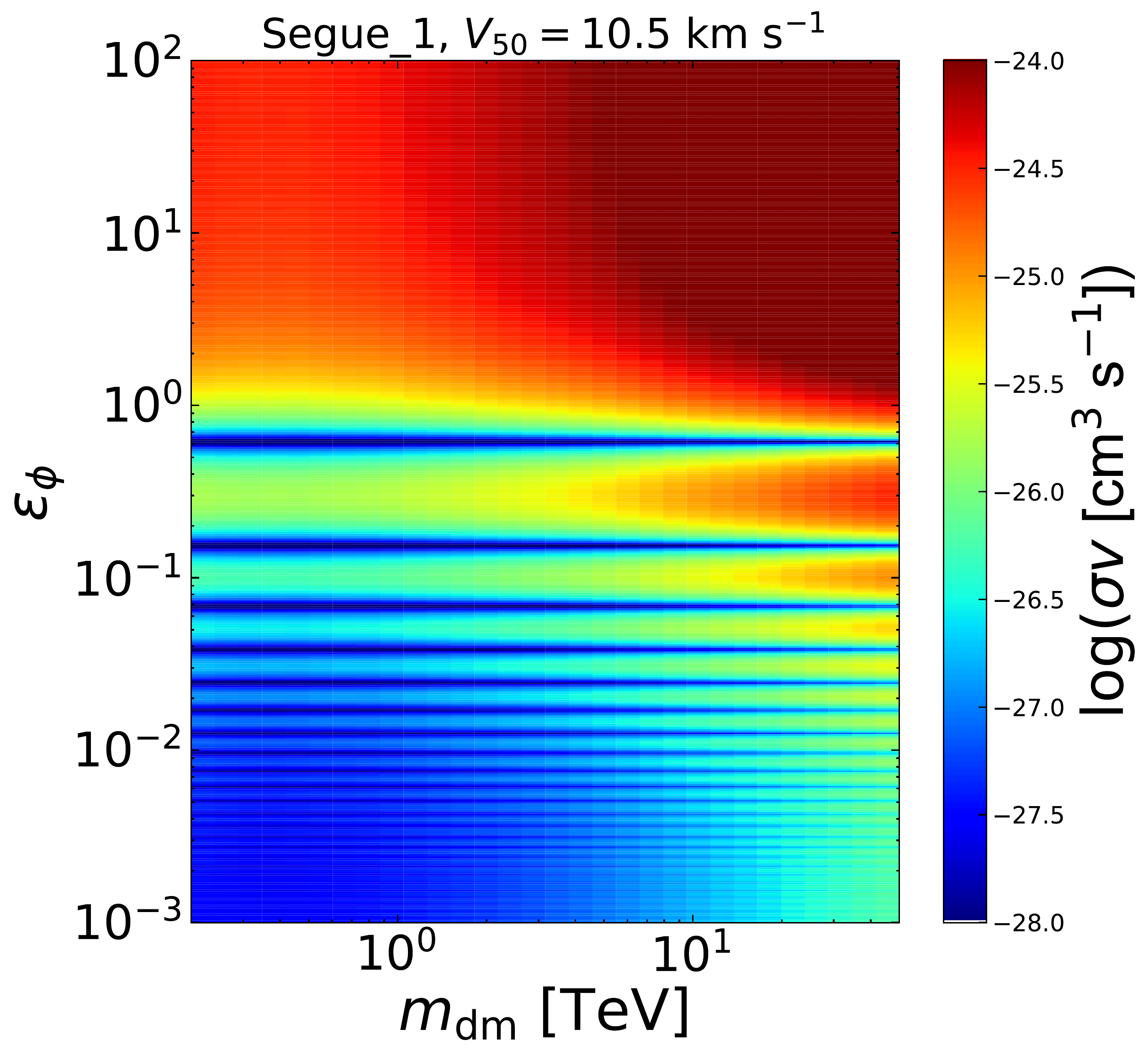}
     \includegraphics[width=4.9cm]{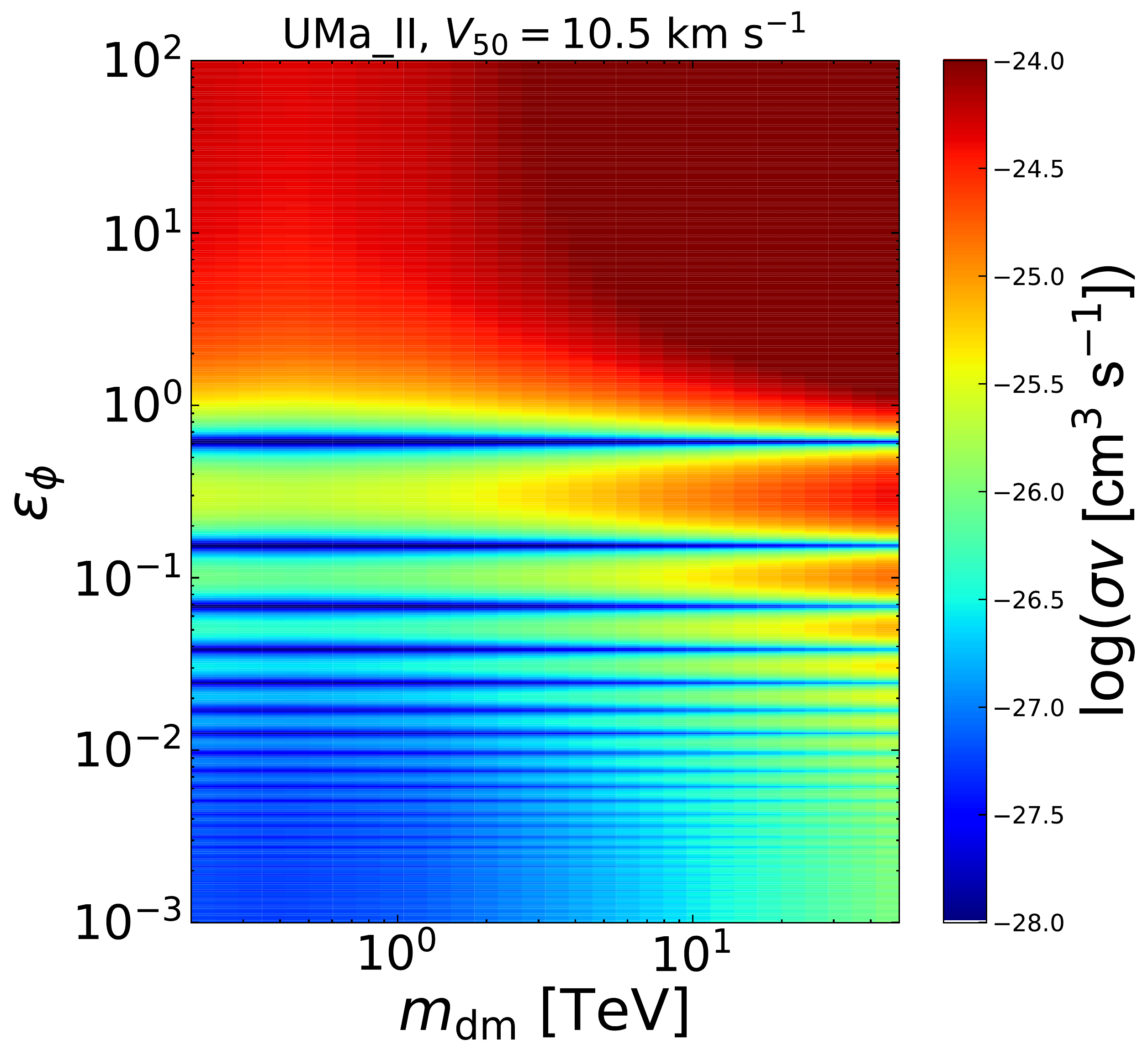}
    
    \includegraphics[width=4.9cm]{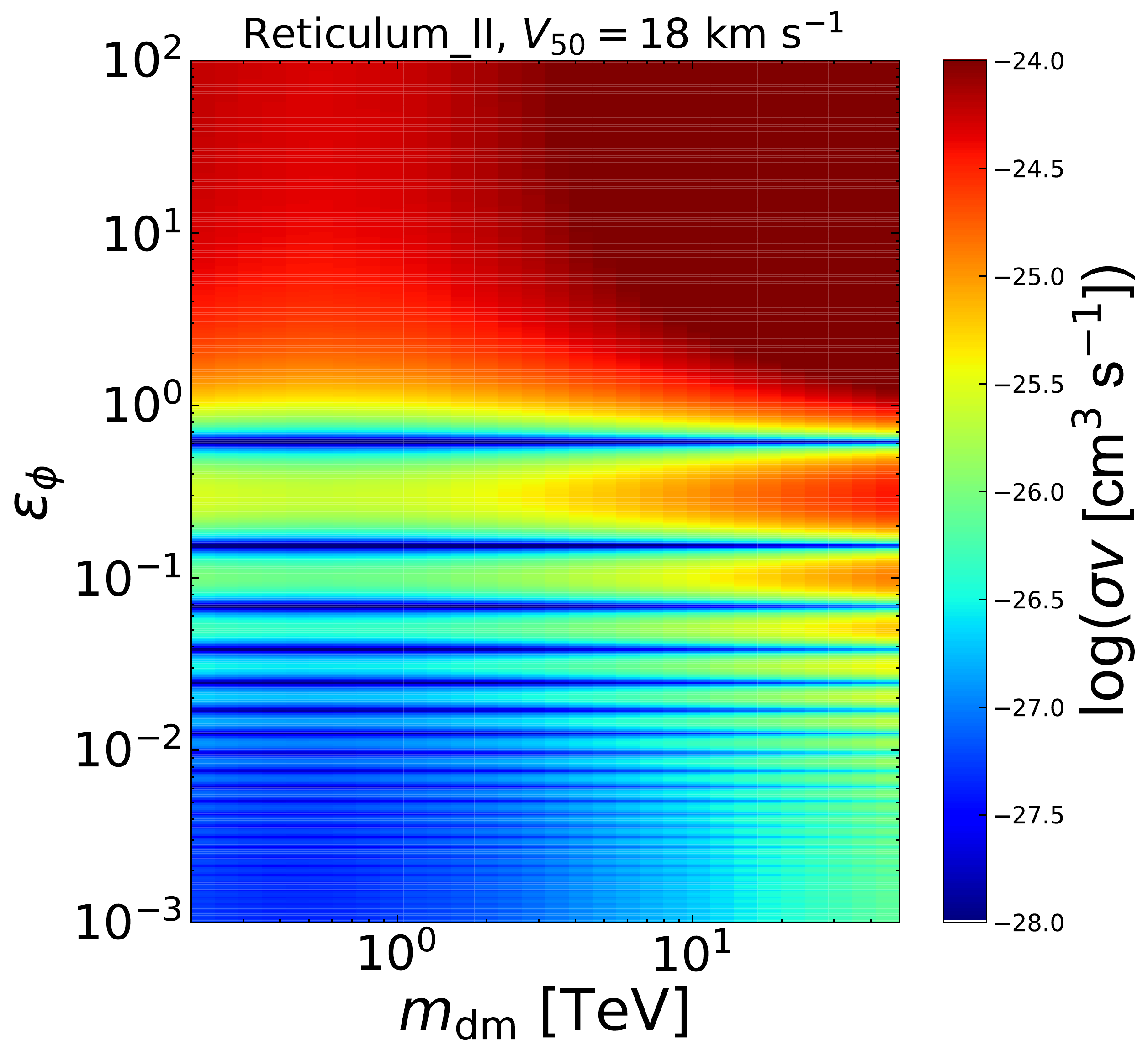}
    \includegraphics[width=4.9cm]{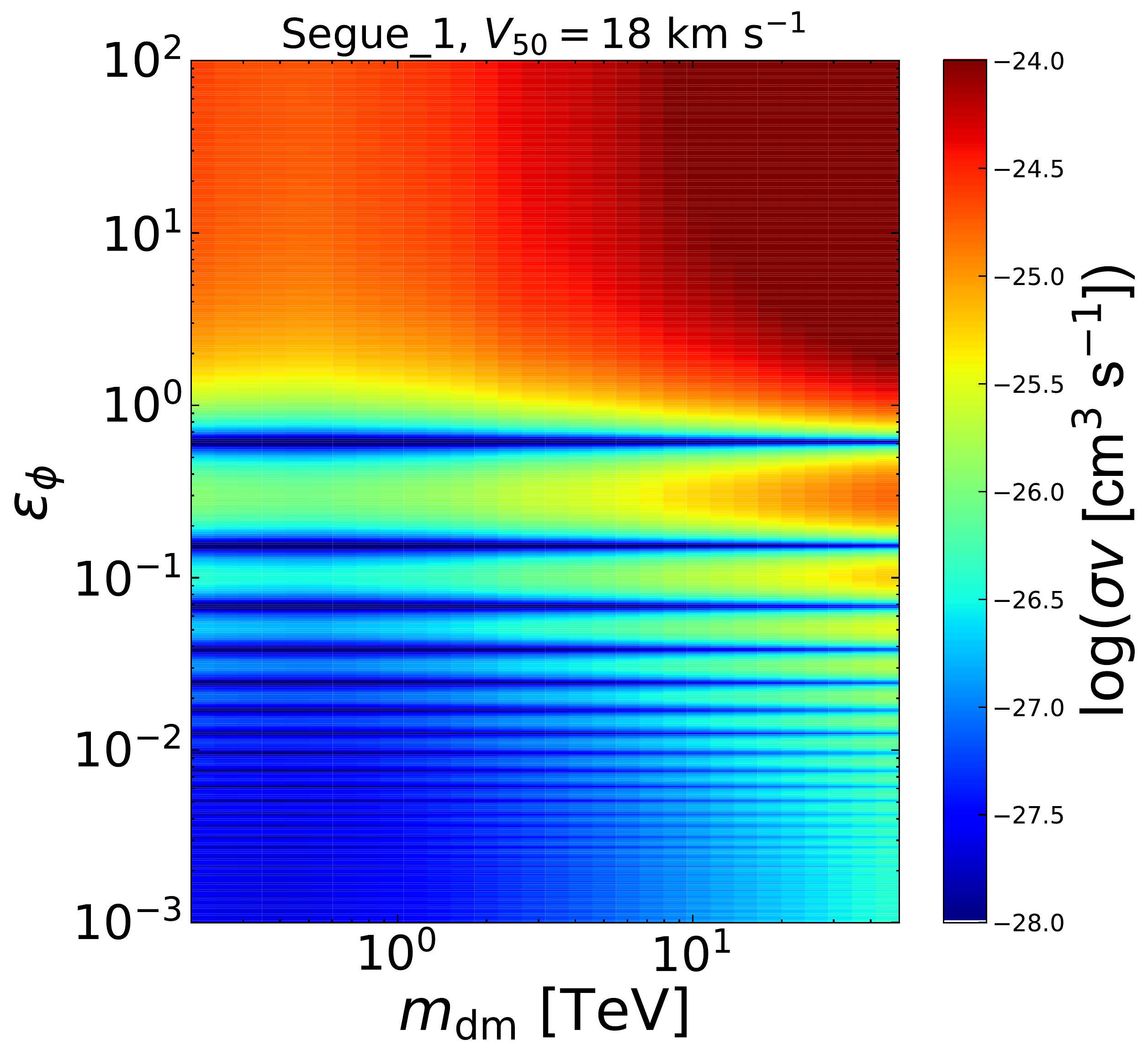}
    \includegraphics[width=4.9cm]{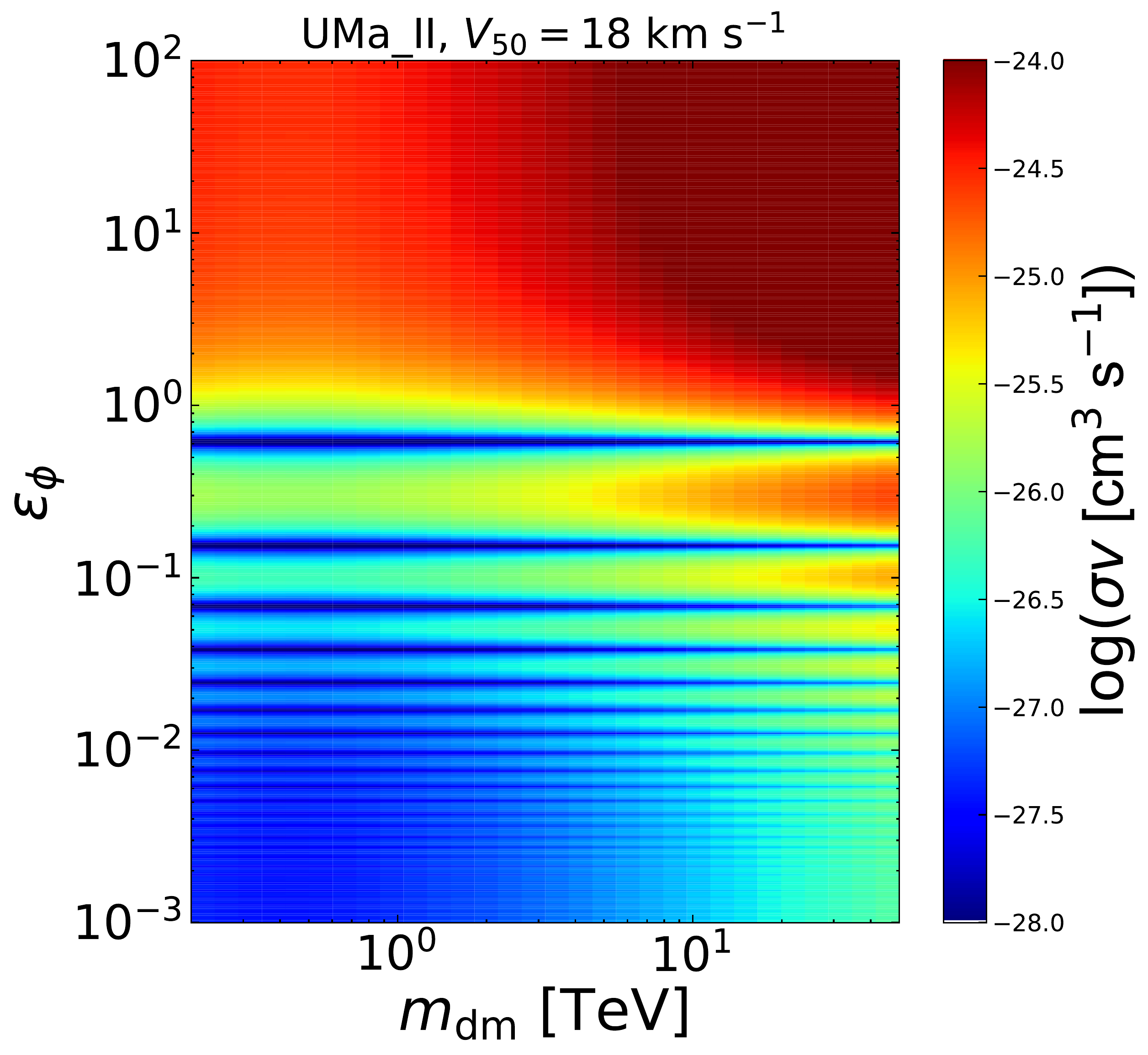}
   
    \includegraphics[width=4.9cm]{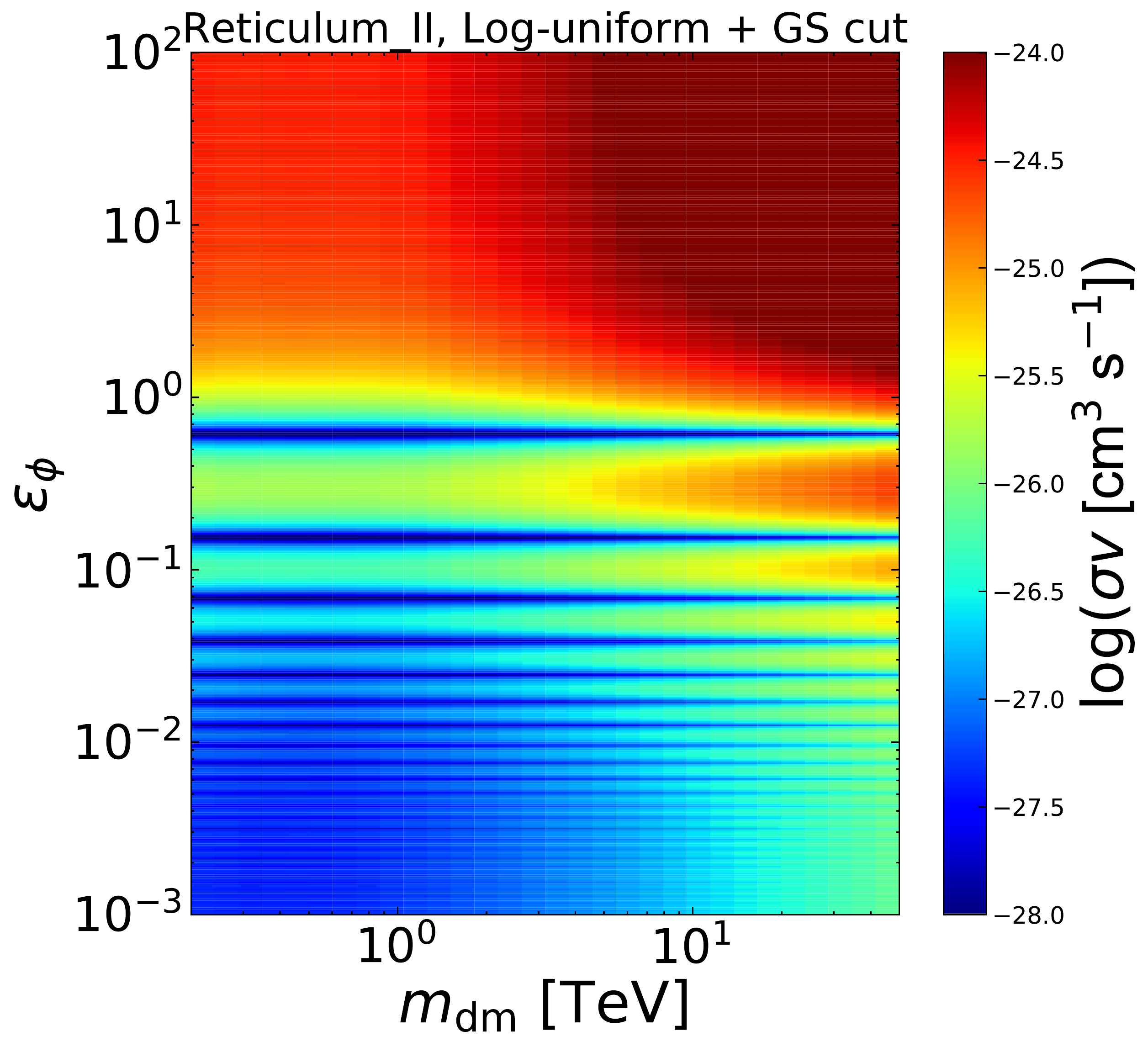}
    \includegraphics[width=4.9cm]{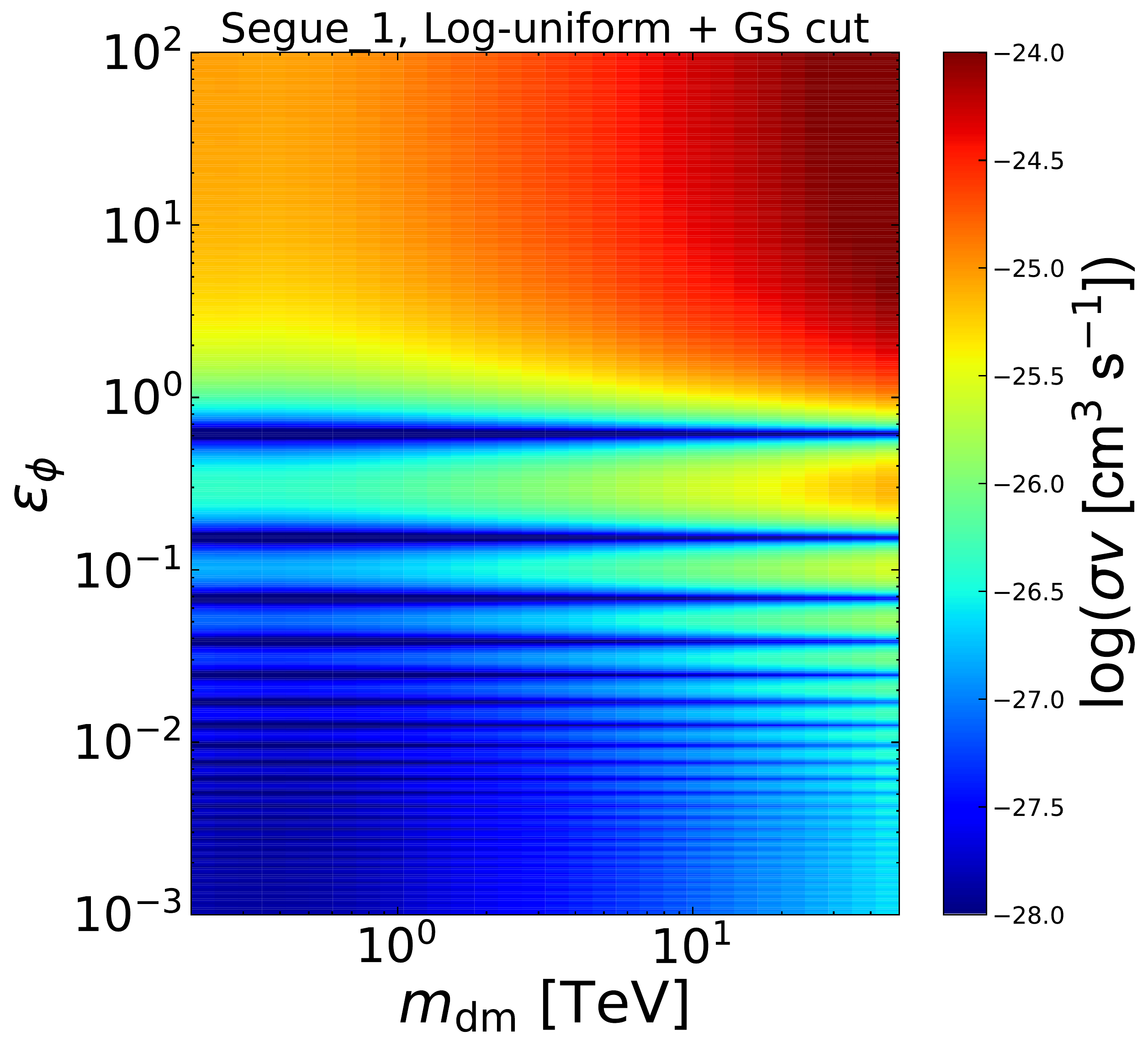}
    \includegraphics[width=4.9cm]{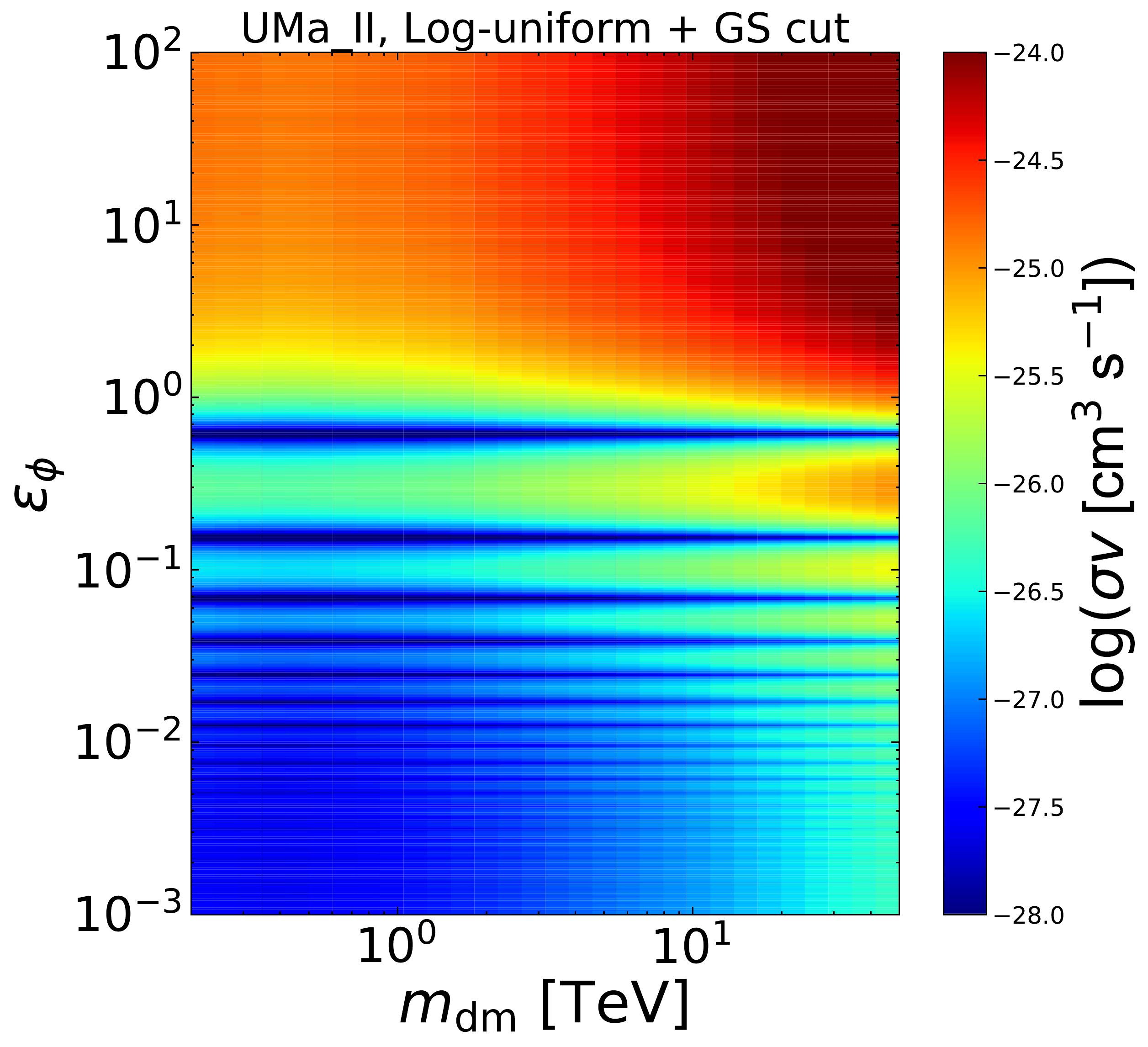}   

    \caption{Colored map of the upper limit of the annihilation cross
      section. Parameters are the same as
      Fig.\,\ref{fig:contour_ultrafaints_lightmediator}.}
  \label{fig:map_ultrafaints_lightmediator}
 \end{center}
\end{figure}

\begin{figure}
  \begin{center}
    \includegraphics[width=4.9cm]{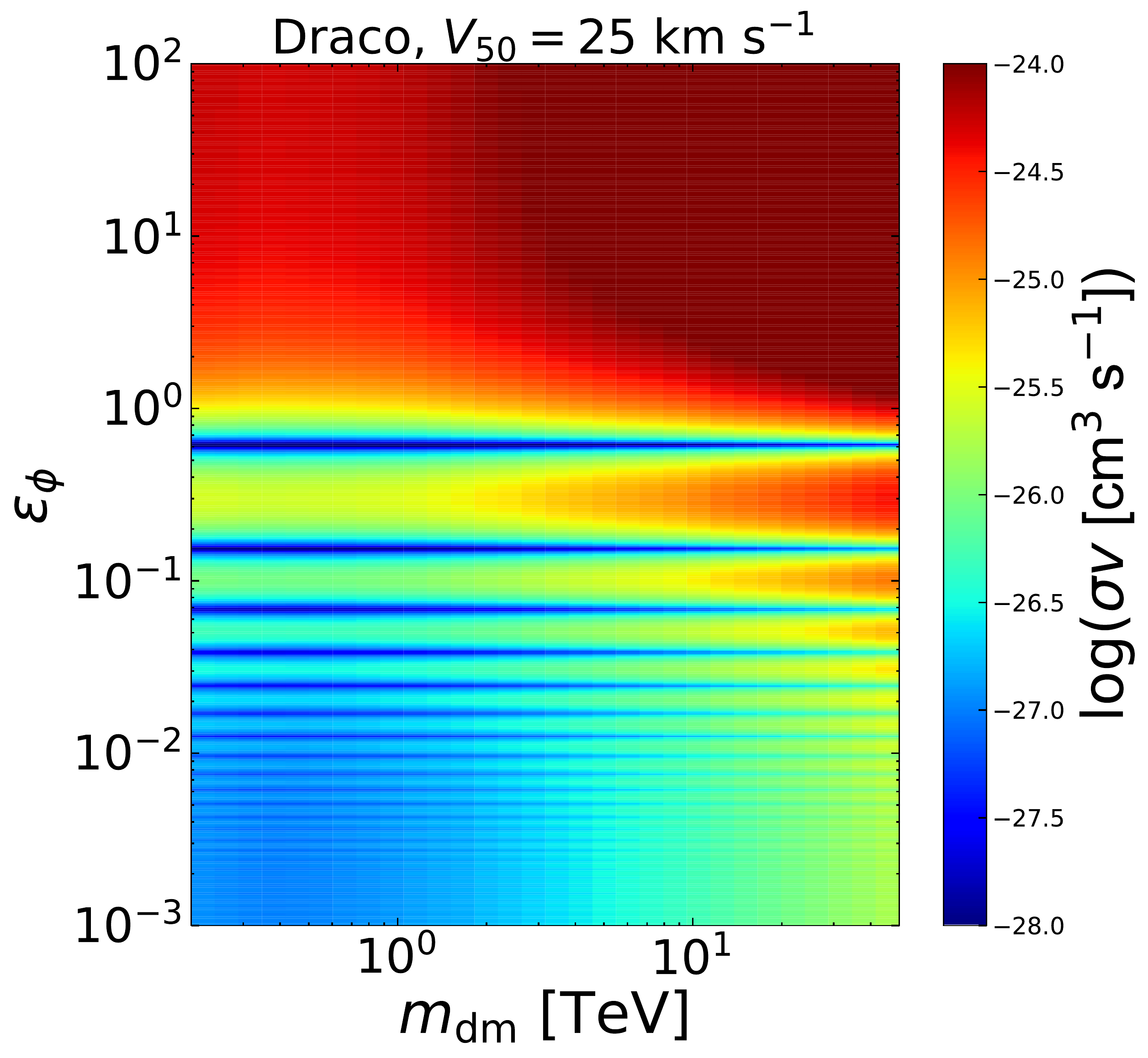}
    \includegraphics[width=4.9cm]{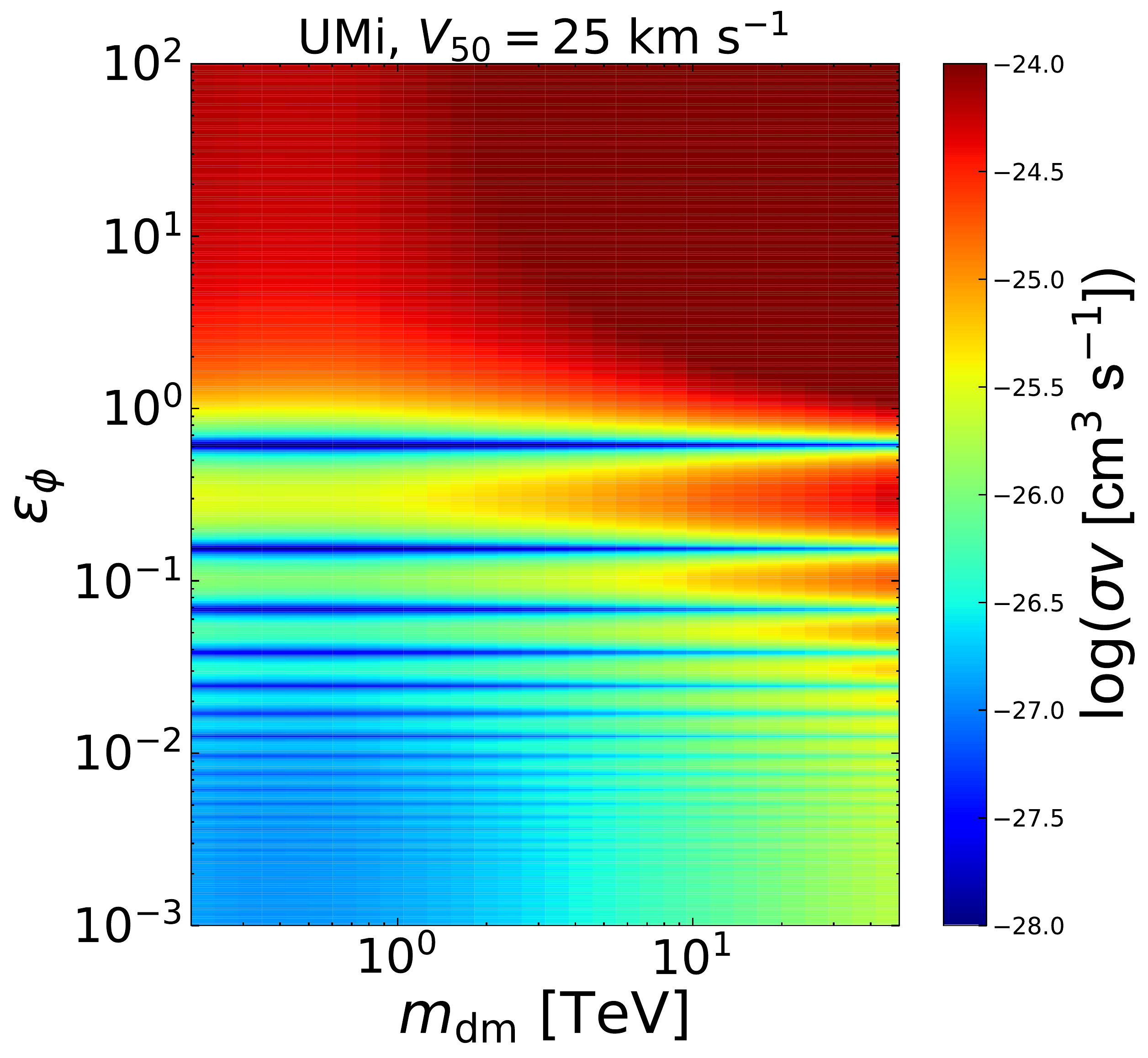}
    
    \includegraphics[width=4.9cm]{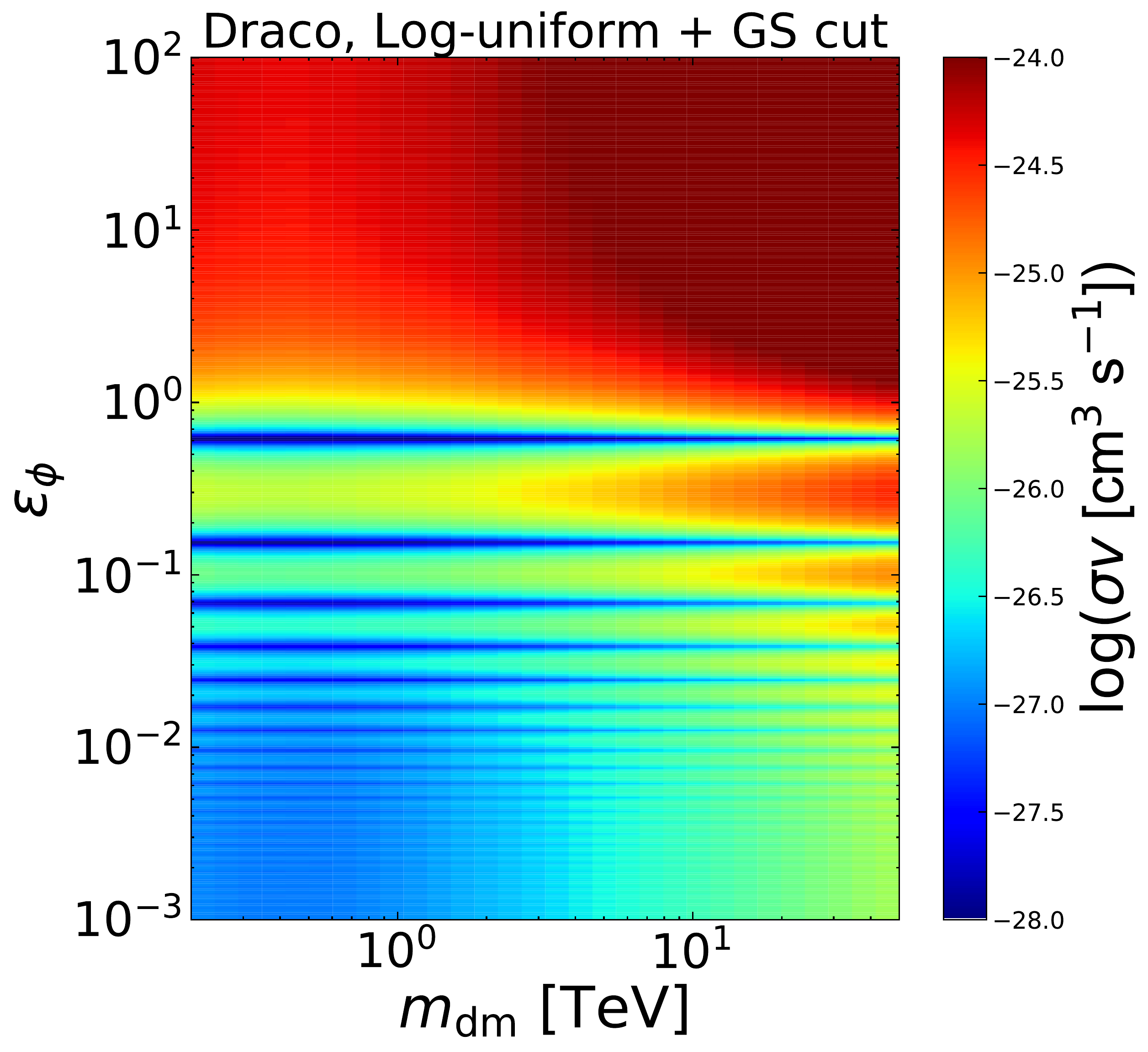}
    \includegraphics[width=4.9cm]{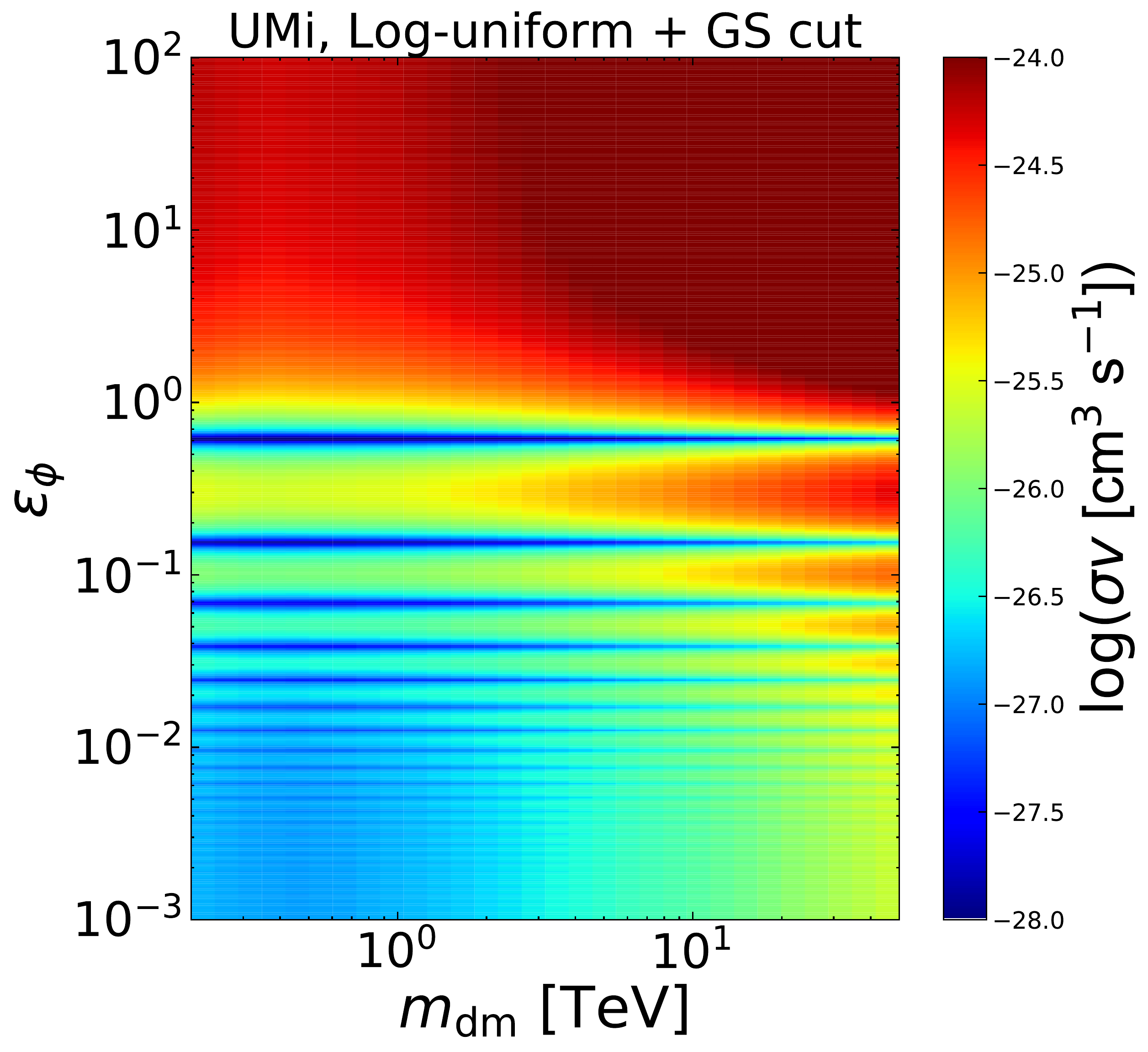}
    \caption{The same as Fig.\,\ref{fig:map_ultrafaints_lightmediator}
      but for Draco and Ursa Minor and the parameters are the same as
      Fig.\,\ref{fig:contour_classicals_lightmediator}.}
    \label{fig:map_classicals_lightmediator}
 \end{center}
\end{figure}

We are now ready to show the expected sensitivity of CTA North to the DM
annihilation that is enhanced by the Sommerfeld effect. In
Fig.\,\ref{fig:contour_ultrafaints_lightmediator}, the expected upper
limits using ultrafaint dSphs on the annihilation cross section in the
light mediator model is shown.  We assume that DM
annihilates into $b\bar{b}$. In the plot, we use median J-factor
computed in Sec.\,\ref{sec:scalarmed}.  As
in Fig.\,\ref{fig:J_ultrafaints_selected}, Reticulum II, Segue 1, and
Ursa Major II are used in the analysis where the satellite prior with
$V_{50}=$ 10.5\,km\,s$^{-1}$ and 18\,km\,s$^{-1}$ compared to log-uniform
+ GS15 cut prior. For all ultrafaint dSphs plotted here it is found
that the results with satellite prior (with $V_{50}=10.5$\,km\,s$^{-1}$)
gives weaker upper limits compared to the log-uniform + GS15 cut prior.
The most stringent bound is obtained from Segue 1. The result with
$V_{50}=10.5$ km s$^{-1}$ shows that DM annihilation with
$\sigma v=10^{-26}$ cm$^3$\,s$^{-1}$ can be detected in the
mass range $m_{\rm dm}<\order{10\,{\rm TeV}}$ ($\order{1\,{\rm TeV}}$)
for $\epsilon_\phi<\order{10^{-2}}$ ($\order{10^{-1}}$).  The result
with $V_{50}=18$\,km\,s$^{-1}$, on the other hand, gives a bit more
optimistic expectation; DM annihilation with $\sigma
  v=10^{-26}$ cm$^3$\,s$^{-1}$ can be detected in the mass range
$m_{\rm dm}<\order{10\,{\rm TeV}}$ ($\order{1\,{\rm TeV}}$) for
$\epsilon_\phi<\order{10^{-2}}$ ($\order{1}$).
Fig.\,\ref{fig:contour_classicals_lightmediator} shows the upper limits
from classical dSphs: Draco and Ursa Minor. It is found that the
constraints are comparable or weaker than Segue 1.  For full
information about the upper limits of the annihilation cross section,
we provide the colored map in
Figs.\,\ref{fig:map_ultrafaints_lightmediator} and
\ref{fig:map_classicals_lightmediator}.

A comment on J-factor estimates of Segue 1 is in order.  In order to
estimate its density profile parameters, we adopt a velocity
dispersion of the member stars based on Ref.\,\cite{Simon:2010ek}.
However, there is a large uncertainty on what stars are regarded as a
member of the system, and depending on their inclusion or exclusion,
the estimates of the J factor of Segue 1 can be smaller by orders of
magnitude~\cite{Bonnivard:2015xpq, Hayashi:2016kcy}.  We caution that
our estimates are on an optimistic side, in a similar spirit of
earlier paper by the Magic collaboration~\cite{Aleksic:2013xea}.
Unlike earlier work including Ref.\,\cite{Aleksic:2013xea}, however,
our new modeling adopting satellite priors systematically shifts the
best estimates of the J factors toward lower values, whereas our
results in the case of uninformative priors are consistent with the
results of the earlier studies.

\begin{figure}
  \begin{center}
    \includegraphics[width=4.8cm]{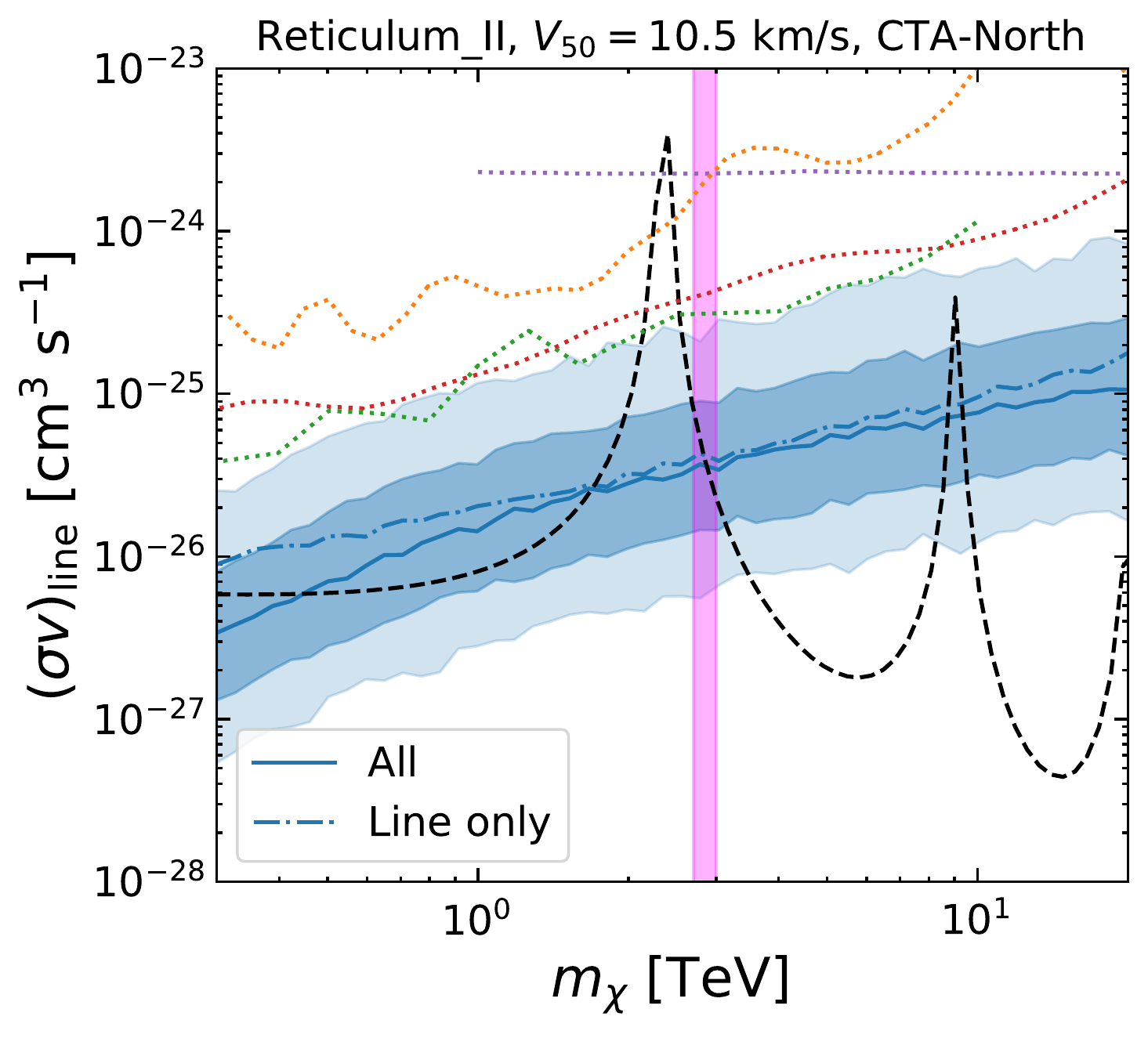}
    \includegraphics[width=4.8cm]{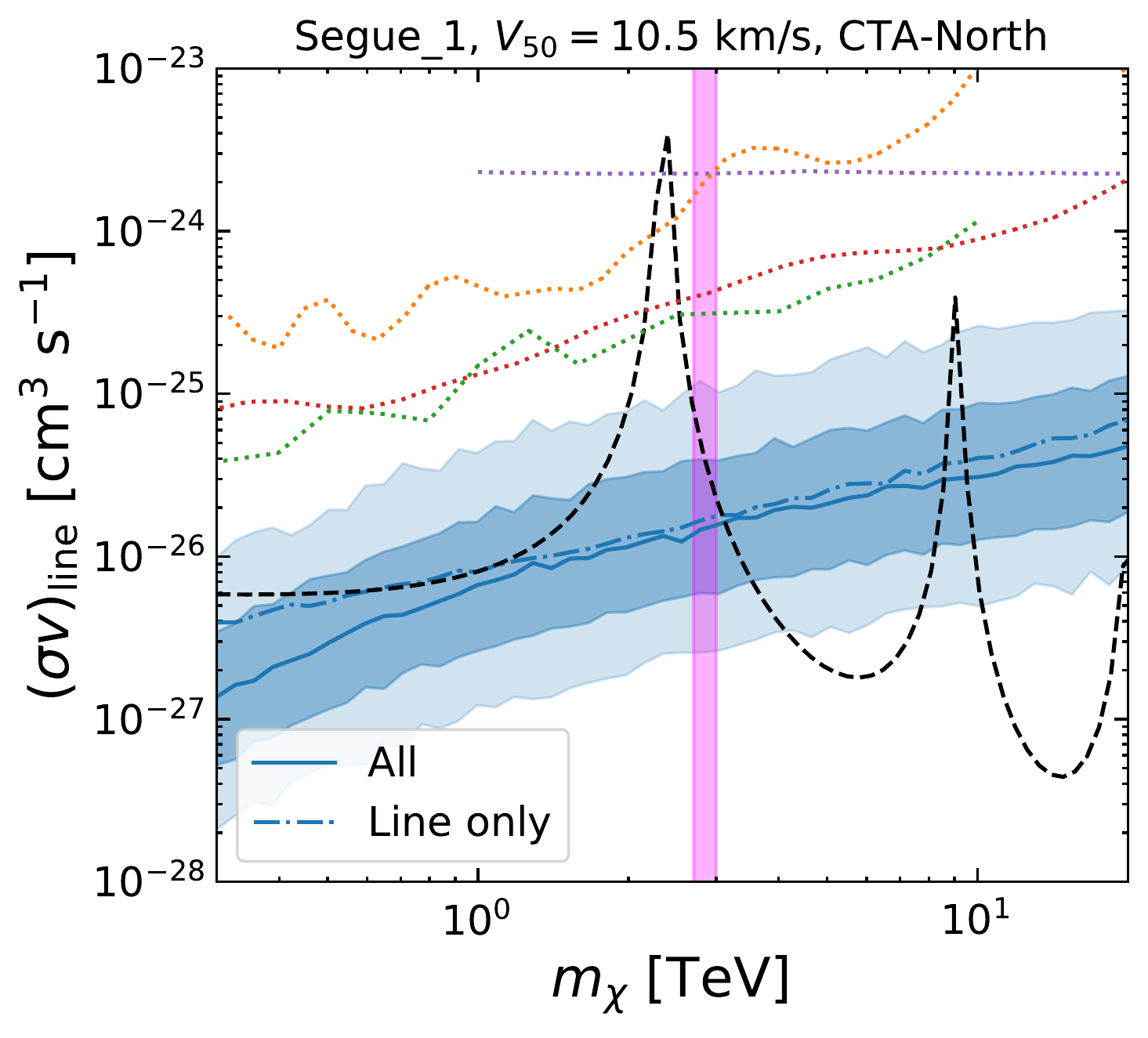}
    \includegraphics[width=4.8cm]{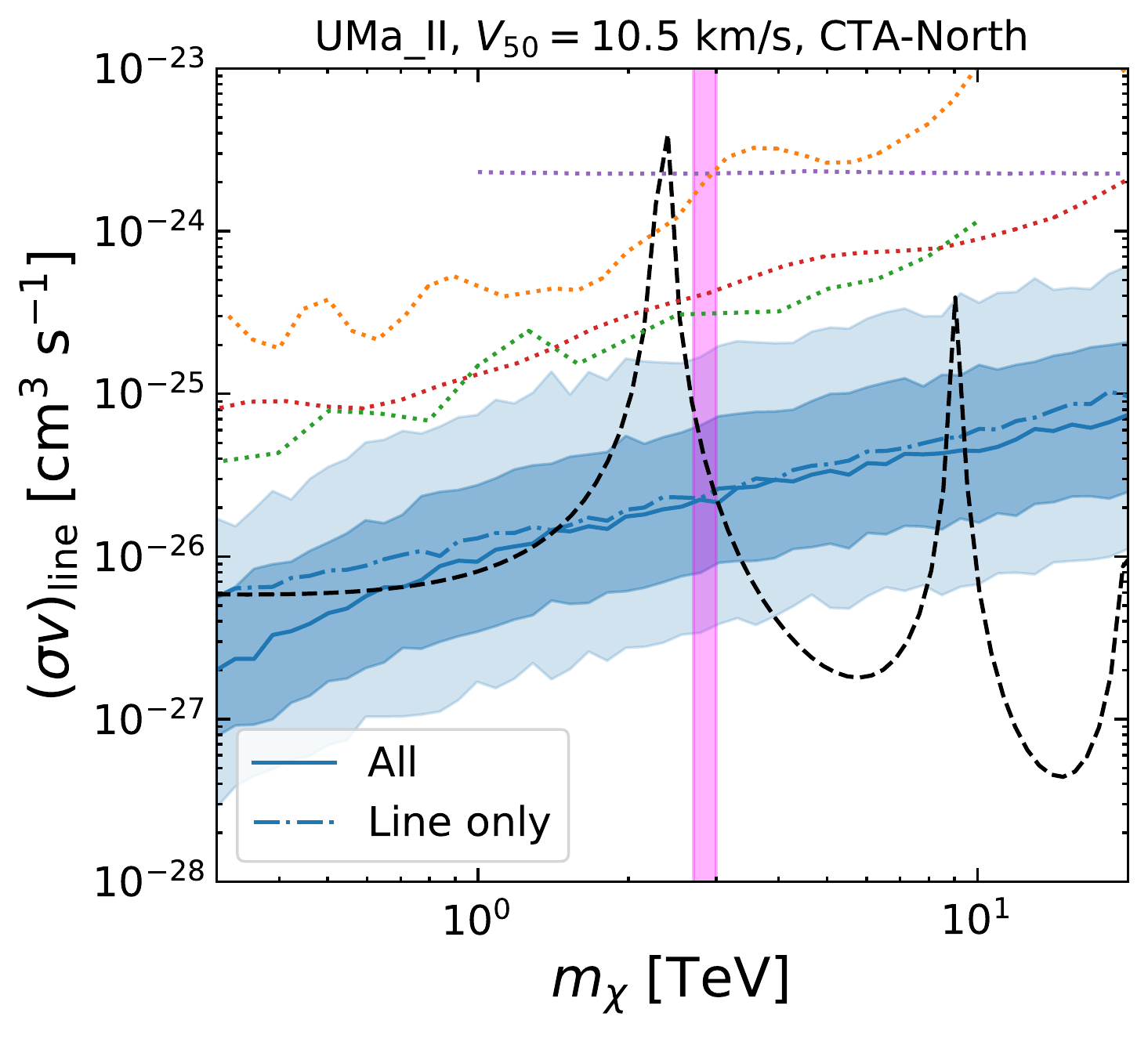}
    \includegraphics[width=4.8cm]{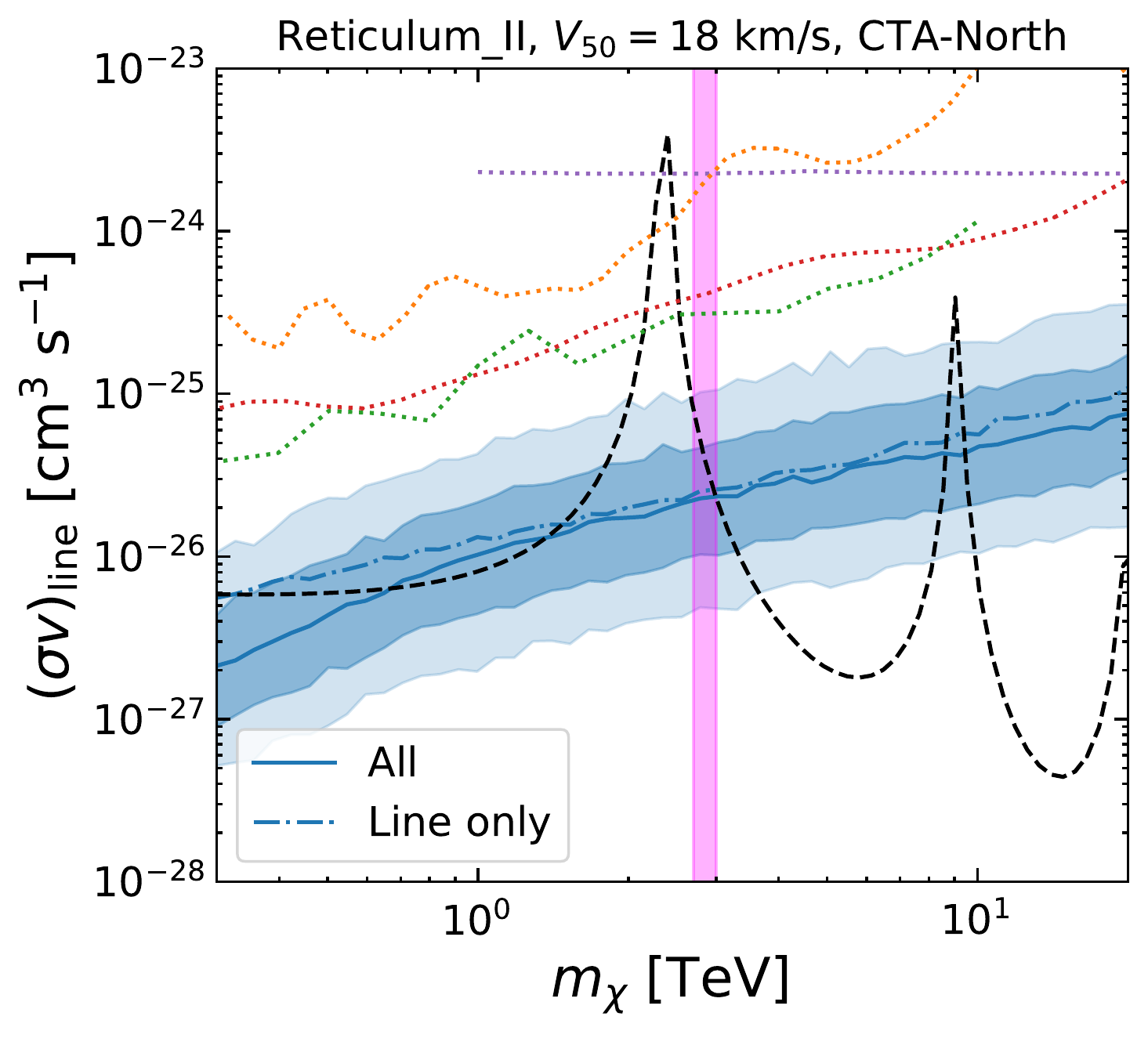}
    \includegraphics[width=4.8cm]{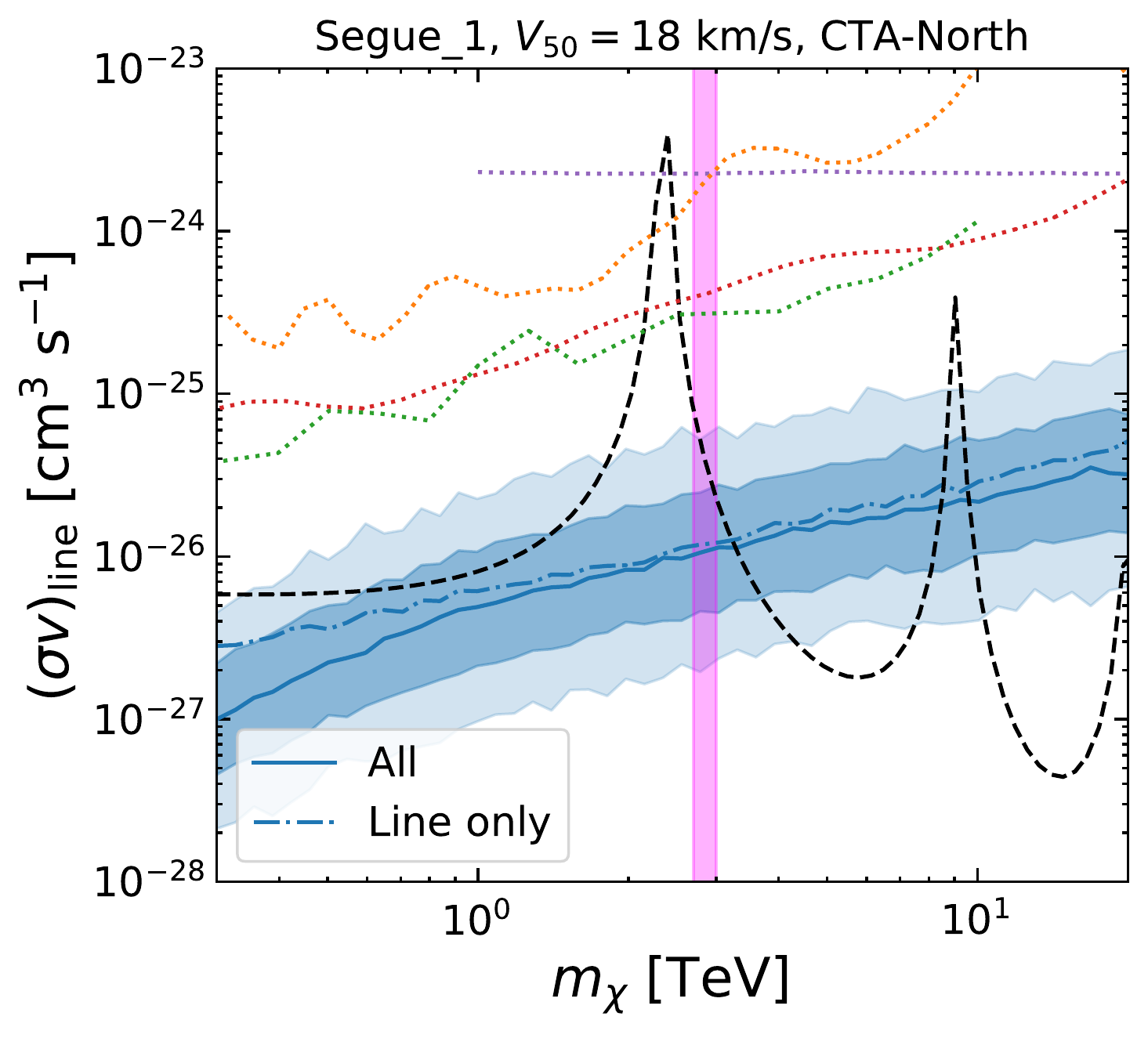}
    \includegraphics[width=4.8cm]{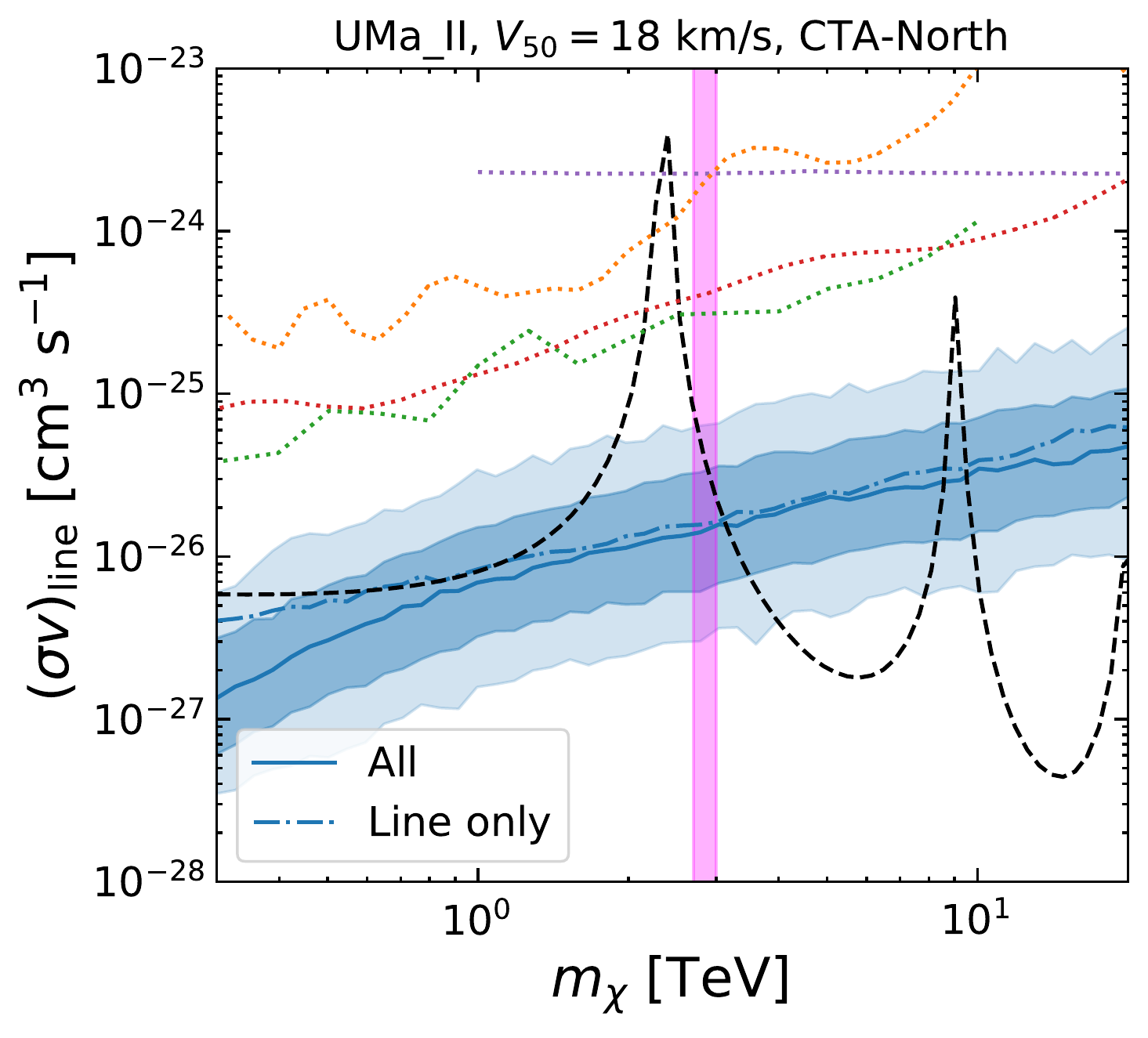}
    \caption{Expected 95\% upper limits on $\sigma v_{\rm line}$ by
      CTA North with 500 hours of exposure for the Wino DM, for the
      ultrafaint dSphs:Reticulum II (left), Segue 1 (middle), and
      Ursa Major II (right). Top and bottom panels show the results
      corresponding to different satellite priors with $V_{50} = 10.5$
      and 18~km~s$^{-1}$, respectively. The solid curve shows the
      expected median sensitivity at 95\% CL, while thick and thin
      bands are 68\% and 95\% containment regions,
      respectively. Dotted curves are existing upper limits by the
      current generation of telescopes: HESS~\cite{Abdalla:2018mve}
      (orange), MAGIC~\cite{Aleksic:2013xea} (red),
      VERITAS~\cite{Archambault:2017wyh} (green), and
      HAWC~\cite{HAWC:2019jvm} (purple). The dashed curve shows the
      expected Wino annihilation cross section with the Sommerfeld
      enhancement, whereas purple vertical region highlights the most
      likely region of the Wino mass, 2.7--3~TeV.}
    \label{fig:CTA_limits_ultrafaint}
 \end{center}
\end{figure}

\begin{figure}
  \begin{center}
    \includegraphics[width=4.9cm]{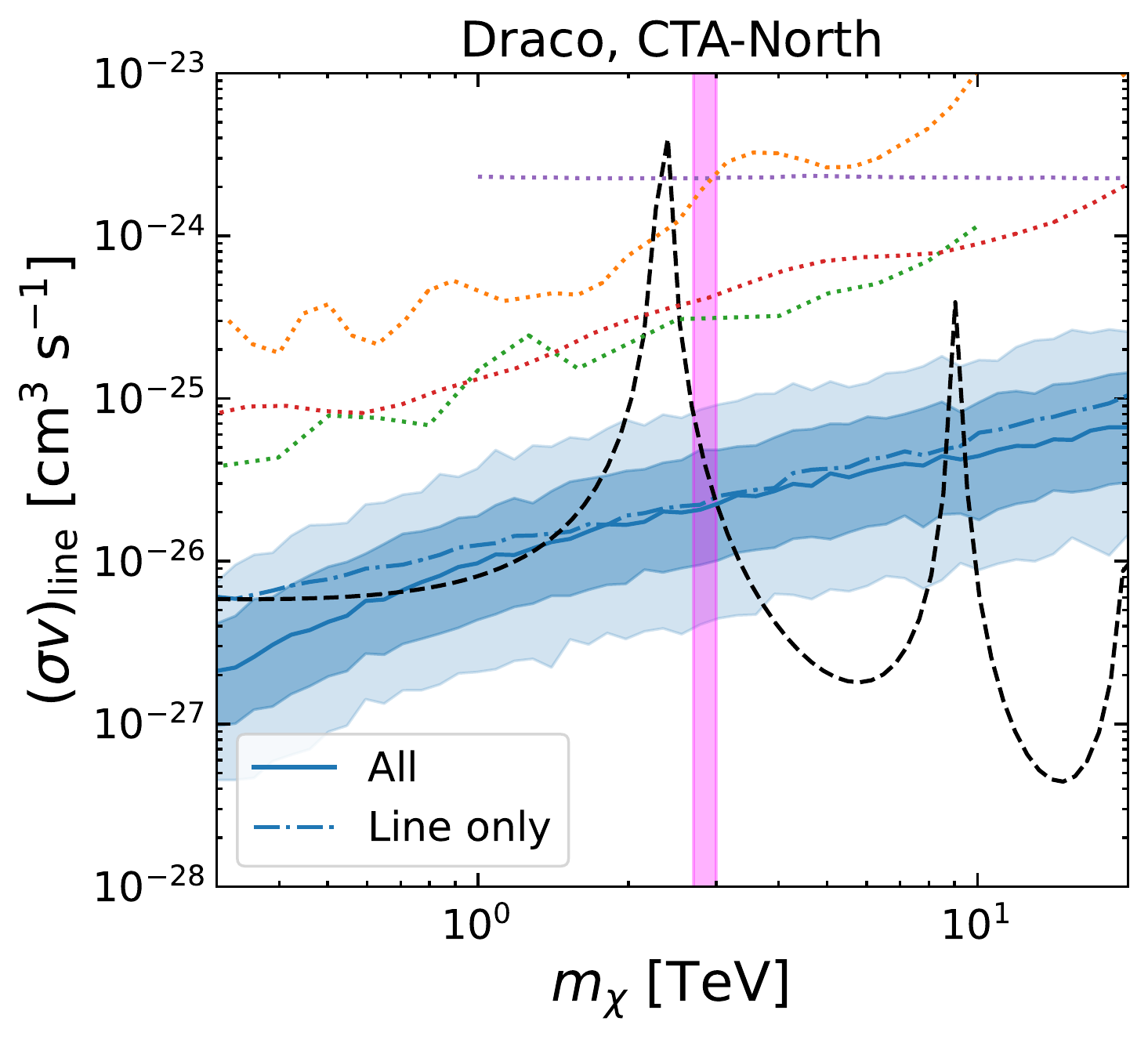}
    \includegraphics[width=4.9cm]{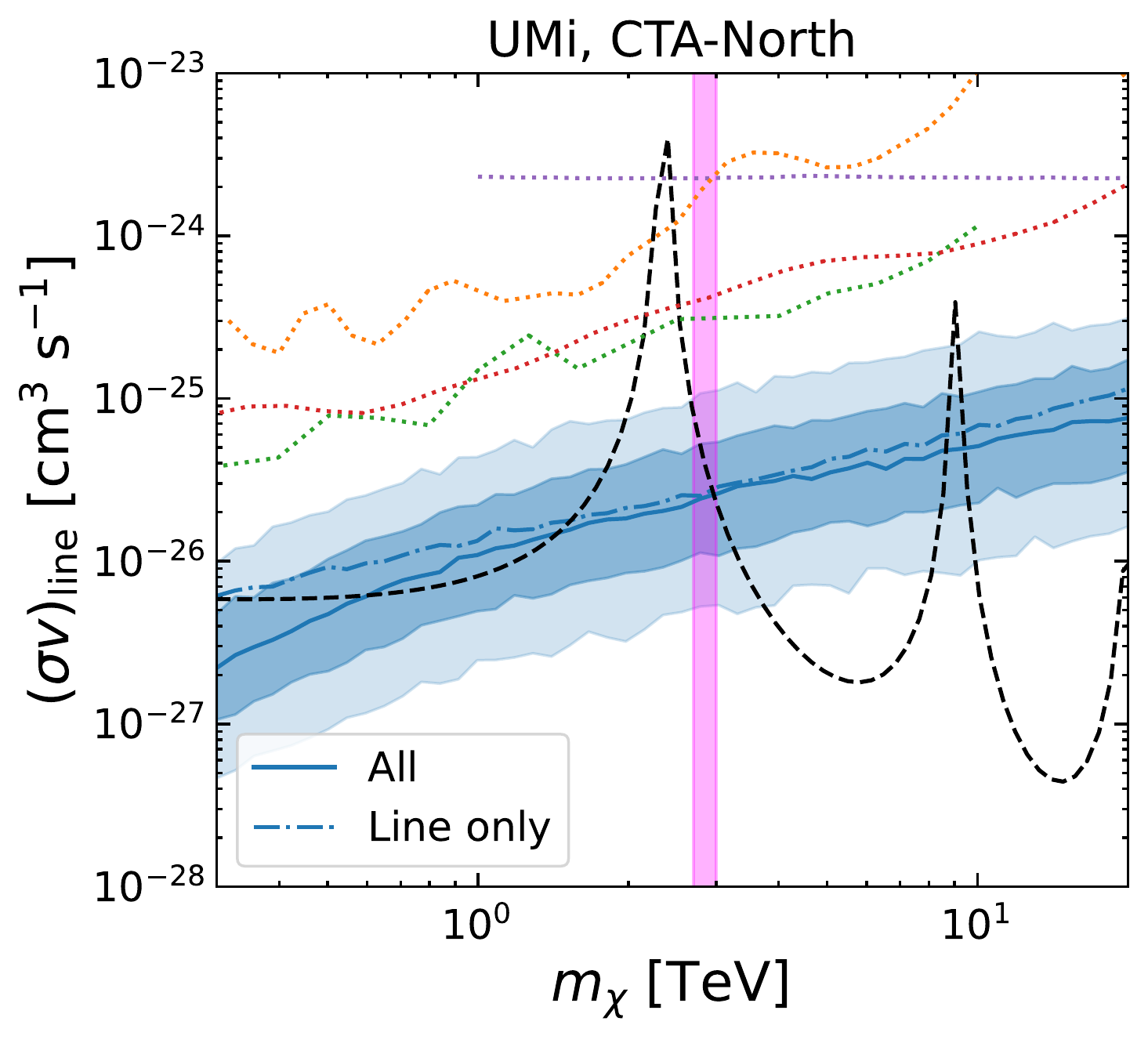}
    \caption{Same as Fig.~\ref{fig:CTA_limits_ultrafaint}, but for
      classical dSphs:Draco (left) and Ursa Minor (right).  Satellite
      prior with $V_{50}=$25 km~s$^{-1}$ is used. }
    \label{fig:CTA_limits_classical}
 \end{center}
\end{figure}

We now move on to the sensitivity to the Wino DM.
Fig.\,\ref{fig:CTA_limits_ultrafaint} shows the expected 95\% CL upper
limits on the $\sigma v_{\rm line}$ (defined in
Eq.\,\eqref{eq:sigmav_line}) by observing ultrafaint dSphs, Reticulum
II, Segue 1, and Ursa Major II.  It has been found that all selected
ultrafaint dSphs have the sensitivity to detect 2.7--3 TeV Wino DM. It
is worth noticing that this conclusion is independent of the prior
model. Namely in either model with $V_{50}=10.5$\,km\,s$^{-1}$ or
$18$\,km\,s$^{-1}$, the result shows that the CTA observing those
ultrafaint dSphs can detect the Wino DM with a mass of 2.7--3~TeV. The
same conclusion is derived by observing classical dSphs, Draco and
Ursa Minor (see Fig.\,\ref{fig:CTA_limits_classical}). They can be
promising for detection of the Wino DM. Therefore, it is concluded
that the observation of dSphs by CTA will provide another robust
avenue to look for the Wino DM in addition the Galactic center
region~\cite{Rinchiuso:2018ajn}.

\section{Conclusions}
\label{sec:conclusion}

In this paper we have studied the detection of DM whose annihilation
process is enhanced by the Sommerfeld effect. To this end, we focus on
observations of dSphs. In order to derive the current limits and the
future sensitivities on the annihilation cross section of DM, it is
crucial to determine the J-factor of the each dSph. Recently the
J-factors of the dSphs were calculated by using proper theoretical
priors and it was found that the J-factors were reduced by at least a
factor of a few for ultrafaint dSphs~\cite{Ando:2020yyk}. By applying
these satellite priors, we have calculated the J-factors of the dSphs
for DM annihilating into the standard model particles enhanced by the
Sommerfeld effect. We emphasize that most of the previous work on the
dSph density profile estimates adopted uninformative log-flat priors
for the density profile parameters such as $r_s$ and $\rho_s$. In this
paper, therefore, we focused on evaluating the impact of adopting the
new satellite priors in comparison with traditional uninformative
priors.  To be concrete, we have studied two models: (1) DM
annihilation boosted by a light scalar mediator via the Yukawa
interaction; and (2) Wino DM in supersymmetric models.

In the former case, the enhancement factor depends on the velocity of
DM that is determined by the profile.  Since the
priors give probability distribution of the profile parameters, the
prior dependence of the J-factor is non-trivial. By computing the
J-factor with the priors, we have found that although J-factor rapidly
oscillates as function of $\epsilon_\phi$, ratio of light mediator mass
to DM mass times the Yukawa coupling squared, the relative difference of
J-factors with different priors is less sensitive to $\epsilon_\phi$.
In addition, assuming that DM self-annihilates into
$b\bar{b}$ final state, we have calculated expected sensitivity limits
on the annihilation cross section by the CTA on $\epsilon_\phi$ and
DM mass plane by using different priors for ultrafaint and
classical dSphs.

In the latter case, a free parameter in the DM sector of the
Lagrangian is the Wino mass and the cross section is determined
uniquely for a given Wino mass. Besides, the Wino mass is determined
to be 2.7--3~TeV if the thermal freeze-out scenario is assumed. We
have computed the Wino annihilation cross section at the
next-to-leading log level following Ref.\,\cite{Baumgart:2018yed} and
derived the current limits on the annihilation cross section and the
future sensitivity by CTA observation.  It has been found that
measurements of one of the ultrafaint dSphs (Reticulum, Segue 1, and
Ursa Major II) and classical dSphs (Draco and Ursa Minor) will give us
sufficient sensitivity to detect the Wino DM with 2.7--3~TeV mass with
CTA observations for 500 hours. This conclusion is nearly independent
of the priors, thus it is a robust prediction. One should, however,
keep in mind relatively large uncertainties related to the density
profile estimates of Segue 1 ultrafaint dSph, when making a decision
on which of these dSphs should be chosen as the best target. We
conclude that observing dSphs is another powerful tool to detect the
Wino DM as well as the observation of the Galactic center region.

\section*{Acknowledgments}

We are grateful to Nicholas L.~Rodd, Timothy~Cohen, Kohei~Hayashi,
Shigeki Matsumoto, and Takashi Toma for valuable discussion and
comments.  KI was supported by JSPS KAKENHI Grant Numbers JP17K14278,
JP17H02875, JP18H05542, and JP20H01894, and SA by JSPS/MEXT KAKENHI
Grant Numbers JP17H04836, JP18H04340, JP20H05850, and JP20H05861.

\end{document}